\documentclass[12pt, twocolumn, numberedappendix, tighten]{aastex62}
\usepackage{color} 
\usepackage{graphicx}

\shortauthors{Xu et al.}

\begin{document}

\title{Exploring the pertubed Milky Way disk and the substructures of the outer disk}

\author[0000-0002-2459-3483]{Y.~Xu} 
\affiliation{CAS Key Laboratory of Optical Astronomy, National Astronomical Observatories, Chinese Academy of Sciences, Beijing, 100101, China}
\email{xuyan@nao.cas.cn}

\author[0000-0002-1802-6917]{C.~Liu}
\affiliation{Key Laboratory of Space Astronomy and Technology, 
National Astronomical Observatories, 
Chinese Academy of Sciences, 
Beijing 100101, PR China;}
\email{liuchao@nao.cas.cn}

\author[0000-0003-3347-7596]{H.~Tian}
\affiliation{Key Laboratory of Space Astronomy and Technology, 
National Astronomical Observatories, 
Chinese Academy of Sciences, 
Beijing 100101, PR China;}

\author{H.~J.~Newberg}
\affiliation{Department of Physics, Applied Physics and Astronomy, Rensselaer Polytechnic Institute, Troy, NY 12180, USA}

\author{C. F. P. ~Laporte}
\affiliation{Department of Physics \& Astronomy, University of Victoria, 3800 Finnerty Road, Victoria, BC V8P 5C2, Canada}
\affiliation{Kavli Institute for the Physics and Mathematics of the Universe (WPI), The University of Tokyo Institutes for Advanced Study (UTIAS), The University of Tokyo, Chiba 277-8583, Japan}

\author[0000-0002-6434-7201]{B.~Zhang}
\affil{Department of Astronomy, Beijing Normal University, Beijing 100875, PR China, LAMOST Fellow}

\author[0000-0001-8459-1036]{H.~F.~Wang}
\affiliation{South$-$Western Institute for Astronomy  Research, Yunnan University, Kunming, 650500, China, LAMOST Fellow }

\author{X.~Fu}
\affiliation{KIAA-The Kavli Institute for Astronomy and Astrophysics at Peking University, Beijing
100871, China}

\author[0000-0002-0786-7307]{J. ~Li}
\affiliation{Physics and Space Science College,China West Normal University,
1 ShiDa Road, Nanchong 637002 , China}
\affiliation{SHAO, Chinese Academy of Sciences, Nandan Road, Shanghai 200030; China, LAMOST Fellow}

\author[0000-0001-9073-9914]{L.~C.~Deng} 
\affiliation{CAS Key Laboratory of Optical Astronomy, National Astronomical Observatories, Chinese Academy of Sciences, Beijing, 100101, China}

\begin{abstract}

	The recent discovery of a spiral feature in $Z-V_Z$ phase plane in the solar neighborhood implies that the Galactic disk has been remarkably affected by a dwarf galaxy passing through it some hundreds of millions of years ago. Using 429,500 LAMOST K giants stars, we show that the spiral feature exits not only in the solar vicinity; it also extends to about 15 kpc from the Galactic center, and then disappears beyond this radius. Moreover, we find that when the spiral features in a plot of $V_\phi$ as a function of position in the $Z-V_Z$ plane, at various Galactocentric radii, are re-mapped to $R-Z$ plane, the spiral can explain well the observed asymmetric velocity substructures. This is evidence that the phase spiral features are the same as the bulk motions found in previous as well as this work. Test-particle simulations and N-body simulations show that an encounter with a dwarf galaxy a few hundred million years ago will induce a perturbation in the Galactic disk. In addition, we find that the last impact of Sgr dSph can also contribute to the flare. As a consequence of the encounter, the distribution function of disk stars at a large range of radii is imprinted by the gravitational perturbation. 
\end{abstract}


\section{INTRODUCTION}
 There is a great deal of evidence that the Milky Way disk is in a state of disequilibrium. Near the Sun, there are vertical oscillations both in density and bulk velocities of disk stars \citep{2012ApJ...750L..41W, 2013ApJ...777L...5C, 2013MNRAS.436..101W, 2018MNRAS.475.2679C}. There are kinematic substructures and a radial velocity gradient \citep{2011MNRAS.412.2026S, 2015ApJ...809..145T, 2017RAA....17..114T, 2018MNRAS.477.2858W, 2018A&A...616A..11G}. There are ring-like overdensites appearing alternately on the north and south sides of the disk \citep{2002ApJ...569..245N, 2015ApJ...801..105X,2004ApJ...615..732R,2015MNRAS.452..676P}, From analysis of Gaia DR2 data, \citet{2019A&A...621A..48L} also provides evidence that the kinematic distribution of disk stars in the Milky Way deviates from a ``simple stationary configuration in rotational equilibrium."

To explain the perturbations, both internal processes (perturbation and resonance of the bar and the spiral arms) and external ones (accretion and passage of satellites) have been proposed \citep{2015MNRAS.452..747M, 2015ApJ...800...83B, 2012MNRAS.425.2335S, 2014MNRAS.440.2564F, 2014MNRAS.443.2757K, 2011MNRAS.417..762Q, 2015MNRAS.453.1867G, 2017RAA....17..114T, 2014MNRAS.443L...1D, 2013MNRAS.429..159G, 2016MNRAS.456.2779G, 2014MNRAS.440.1971W, 2019A&A...622L...6K}. 

Some studies considered the coupled effects between the vertical perturbation and planar perturbation. Recent simulations show that the rotating bar and spiral can produce vertical perturbation \citep{2015MNRAS.452..747M} and the passage of satellite can also produce radial and azimuth perturbation \citep{2016ApJ...823....4D}. 
 
Some of the substructures of the outer disk are well studied, such as the famous Monoceros overdensity \citep{2002ApJ...569..245N, 2003MNRAS.340L..21I, 2003ApJ...588..824Y, 2003ApJ...594L.119C, 2005MNRAS.364L..13C, 2006MNRAS.367L..69M, 2008ApJ...684..287I, 2012ApJ...753..116M, 2012ApJ...757..151L}.  The ring-like Monoceros overdensity was originally thought to be tidal debris \citep{2004MNRAS.355L..33M, 2006MNRAS.366..865B,2005ApJ...626..128P} or part of warp or flare \citep{2004A&A...421L..29M, 2014A&A...567A.106L, 2018MNRAS.478.3367W}, but more recent simulations suggest the ring could be induced by the gravitational interaction of the Sagittarius dwarf galaxy with the disk \citep{2011Natur.477..301P, 2018MNRAS.481..286L}. 

Other studies show that the outer disk overdensities could be related. \citet{2018MNRAS.473.2428D} and \citet{2018MNRAS.473..647D} suggested that the Anti-Center Stream (ACS) and Eastern Banded Structure (EBS) \citep{2006ApJ...651L..29G,2009ApJ...693.1118G} are parts of Monoceros overdensity, even though they appear to be distinct structures in photometric data. From star ages, \citet{2020MNRAS.492L..61L} showed that thin structures like the ACS and EBS are tidal structures excited by a satellite interaction with the disk, while the Monoceros ring is a distinct structure formed through the gradual flaring of the disk over a more extended disk-satellite interaction. \citet{2015MNRAS.452..676P} suggested that the Triangulum Andromeda Overdensity (TriAnd) is a disk component, since the stellar population is more similar to the disk than to tidal debris from a dwarf galaxy. \citet{2018Natur.555..334B} show that TriAnd and A13 have similar chemical abundance patterns as the thin disk. \citet{2020MNRAS.492L..61L} shows that the Monoceros ring, ACS, and EBS have [Fe/H]$\sim$-0.7 with no vertical gradient in [Mg/H] vs. [Fe/H], which is consistent with chemistry of thin disk and not consistent with the ``knee" typically seen in [Mg/H] vs. [Fe/H] for dwarf spheroided galaxies \citep{2015ApJ...808..132H,2019IAUS..343...49T}.\citet{2017ApJ...844...74L} suggested that all of the substructure in the outer disk: the Monoceros overdensity, TriAnd1 \citep{2014ApJ...793...62S}, A13 \citep{2010ApJ...722..750S}, and TriAnd2; which appear alternately north and south of the Galactic plane with increasing distance from the Galactic center, are vertical oscillations of the disk. \citet{2015ApJ...801..105X} showed that the outer disk overdensities which appear at 2 kpc (north near structure) and 5 kpc (south middle structure) from the Sun can be fit with a star count toy model for a bending wave in the disk. Simulations show that the passage of the Sagittarius dwarf spheroidal galaxy (Sgr dSph) through the Milky Way halo could produce oscillations of the disk with ring-like structures similar to those which are observed \citep{2011Natur.477..301P, 2017MNRAS.465.3446G, 2018MNRAS.481..286L}.  

\citet{2018Natur.561..360A} found a $Z-V_Z$ phase space spiral, believed to result from non-equilibrium phase-mixing from the passage of a satellite through the disk some several hundred million years ago. \citet{2018MNRAS.481.1501B} and \citet{2019MNRAS.486.1167B} make test particle simulations that can produce a phase space spiral which is quite similar to that found in Gaia DR2 and GALAH data.  \citet{2019MNRAS.485.3134L} show four density wraps on $Z-V_Z$ phase space of Gaia DR2 data within 1 kpc from the Sun; their simulation of Milky Way disk interacting with Sgr dSph can reproduce many features revealed in Gaia data.  \citet{2019arXiv190210113K} show the ridge-like structures in the (R,$V_\phi$) phase space correspond to constant energy or constant angular momentum, and show that phase mixing can produce ridges of constant energy. \citet{2018MNRAS.480.4244C} suggest that the cumulative effect of interaction between the disk and the satellites in a clumpy halo can produce long-lived bending waves. 

All of the observations mentioned above were made in a volume of the disk that is limited to 3-4 kpc from the Sun. LAMOST K giants can provide far more distant data, up to at least 20 kpc from the Galactic center, to compare with the results of simulations. 
Through studying the LAMOST K giants, we find a strong connection between disk kinematic substructures and the phase space spiral. Our work provides new insight into the origin of the outer disk substructures.  
 
In Sections 2, 3, and 4, we describe our data selection, distance calibration, and calculation of the 3D velocities for the K giant stars. In Sections 5, we analyze the $V_\phi$ kinematic maps, projected onto the $(R,Z)$ plane; we explore the velocity distribution in phase space and the connection between substructures of K giants and the phase space spiral. Section6, 7 and 8 show the $V_R$, $V_Z$ and metallicity distribution in the $(R,Z)$ plane. In Section 9, the observational kinematic features are compared with the results of simulations. Section 10 and 11 present the discussion and conclusion.
 
\section{Sample selection}
K giant stars from LAMOST DR5 were selected using the selection criteria provided in \citet{2014ApJ...790..110L}. The stellar parameters for the K giants were obtained from the Stellar LAbel Machine (SLAM) procedure \citep{2020ApJS..246....9Z}. We then cross-matched the LAMOST DR5 K giants and {\it Gaia} DR2 stars using TOPCAT \citep{2005ASPC..347...29T}.  The high-precision line-of-sight velocities measured by LAMOST DR5 \citep{2012RAA....12.1197C, 2012RAA....12..735D, 2012RAA....12..723Z}, and the proper motions from Gaia DR2 \citep{2001A&A...369..339P, 2018ENews..49a..29P, 2018IAUS..330...13B}, are combined to obtain  3D velocities.

 We remove K giants with true metallicity [M/H]$<-1.2$ from the sample to eliminate halo stars. We also remove duplication,  and red clump stars that were identified in \citet{2018ApJ...858L...7T}. Red clump stars are removed because they have different absolute magnitudes from other K giants at the same color. We require high signal-to-noise in the g band ($snrg>15$) to assure the accuracy of derived stellar parameters, radial velocity errors of less than 10 km s$^{-1}$, and proper motion errors less than 0.2 mas/yr are required to assure the precision of the velocity calculation. Stars with $M_K>0$ are eliminated because they have systematic distance errors larger than 10\% compared to distances from Gaia DR2 \citep{2018AJ....156...58B}, possibly due to contamination from dwarf K stars.  Using all of these selection criteria, we construct a sample of 429,500 K giant stars.  
  
\section{Distance calibration}

Our results rest on creating accurate maps of the kinematic distribution of K giants as a function of position in the Milky Way.  To achieve this, we need accurate distances for the stars in our sample, which can be observed out to 20 kpc from the Galactic Center. We derive the distances to the K giants by applying the Bayesian estimation method from \citet{2015AJ....150....4C}, using the stellar parameters from the SLAM pipeline; the distance estimates from Bayesian estimation with parameters from SLAM were much closer to {\it Gaia} distances than those produced using the standard LAMOST pipeline stellar parameters. 

In \citet{2015AJ....150....4C}, they calibrated the distances derived from LAMOST data with those from the Hipparcos sample to determine that the distance estimates have an uncertainty of about 20\%. However, Hipparcos data could only validate distances within several hunderd pc from the Sun. The distances derived from {\it Gaia} DR2 parallaxes can validate the distances within 4 kpc from the Sun. So, we recalibrate the distance of LAMOST DR5 data with the best estimated distances provided by Gaia DR2 parallaxes as found by \citet{2018AJ....156...58B}. Stars from Gaia DR2 within 3 kpc of the Sun and with relative distance errors of less than 10\% are matched to our dataset of LAMOST K giant stars. The distance error is defined by $1\sigma$ of the Gaussian distribution of estimated distance probability \citep{2018AJ....156...58B}. But keep in mind that the distance obtained by \citet{2018AJ....156...58B} is not model free; the \citet{2018AJ....156...58B} distance is still influenced by the adopted length scale model. 
 
\begin{figure*}
    \includegraphics[width=7cm]{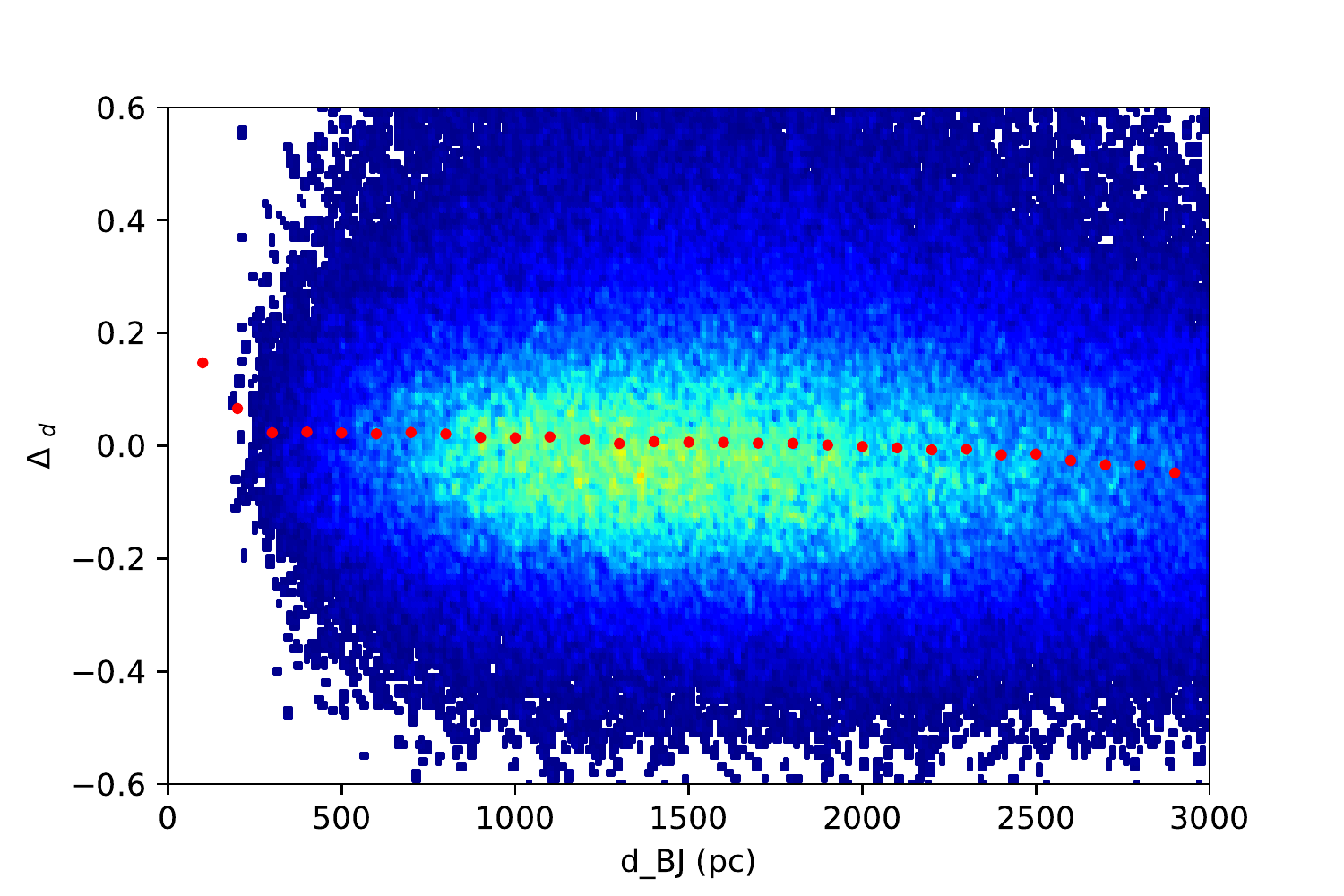}
    \includegraphics[width=7cm]{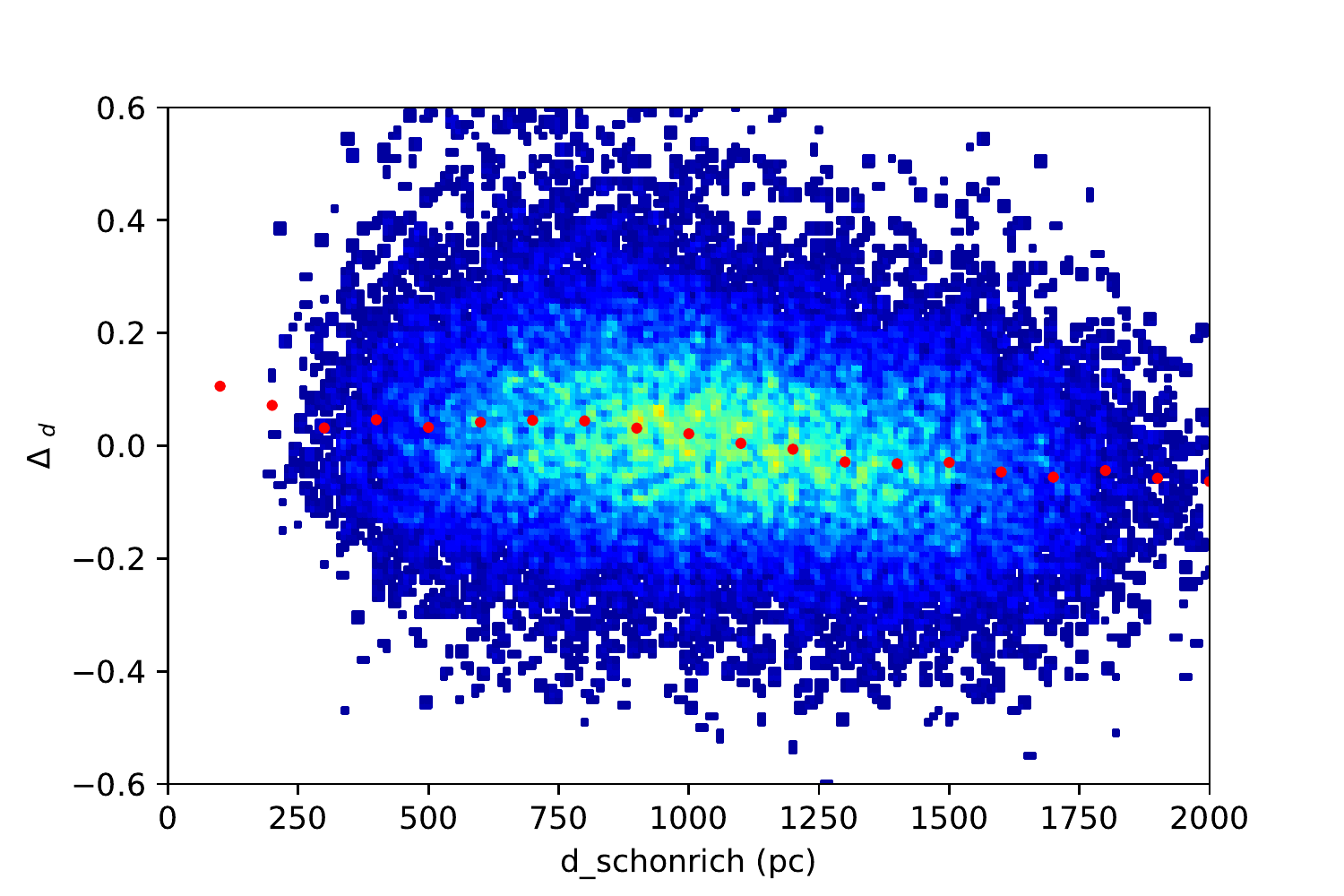}
	\caption{Comparison of the derived distances to LAMOST stars with parallax measurements from {\it Gaia}. The left panel shows the relative difference between LAMOST and {\it Gaia} distances ($\Delta_d$=$(d_{Carlin}-d_{BJ})/d_{BJ}$), as a function of the Gaia DR2 parallax distance ($d_{BJ}$), for stars with in 3 kpc of the Sun with distance errors smaller than 10\%. Red dots indicate the median offset in each bin of width 100 pc. The right panel gives the same information for the LAMOST distances compared with the corrected Gaia DR2 parallax for the ``entirely safe" sample from \citet{2019MNRAS.487.3568S}. These plots show that the systematic error in the distances we derive from LAMOST are small compared to the statistical errors.}
    \label{calibration}
\end{figure*}

Figure~\ref{calibration} shows the distribution of relative difference in distance for the two catalogs, $\Delta_d$, as a function of the Gaia DR2 distance for each star. The relative difference in distance is defined as $\Delta_d$$=(d_{Carlin}-d_{BJ})/d_{BJ}$.  $d_{Carlin}$ is the distance estimated by procedure outlined in \citet{2015AJ....150....4C}, applied to the LAMOST DR5 data. $d_{BJ}$ is distance obtained from {\it Gaia} DR2 parallaxes, as determined by \citet{2018AJ....156...58B}. The red dots indicate the median value of $\Delta_d$ in each distance bin, of width 100 pc.  The median value of median($\Delta_d$) is about 0.006, and the variation of  median ($\Delta_d$) of the calibration sample with distance is quite small. From this analysis, we determine that the systematic distance error can be ignored, since it is considerably smaller than the random error. 

We further checked our distances against a cleaner sample of {\it Gaia} parallaxes. \citet{2019MNRAS.487.3568S} found that the Gaia DR2 parallaxes have bias of 0.054 mas/yr, and produced a sample of corrected {\it Gaia} DR2 parallaxes \citep{2019MNRAS.487.3568S} that were free from this bias. This sample uses the ``entirely safe" criteria from Section 8 of \citet{2019MNRAS.487.3568S} to select stars from the Gaia sample, thus assuring that the estimated Gaia distances should be accurate. We crossmatch the LAMOST K giants sample with the selected sample. The right panel of Figure~\ref{calibration} shows the distribution of relative difference of distances, $\Delta_d$, vs. estimated distance from \citet{2019MNRAS.487.3568S}.  The median value of median($\Delta_d$) is about 0.004, confirming that the LAMOST distances are also consistent with estimated distance of \citet{2019MNRAS.487.3568S}, as expected. However, there appears to be a trend with distance; at larger distances, the Bayesian distance estimates are lower than the corrected Gaia measurements, with a maximum deviation of 5\%. 

\section{Cylindrical velocities: $V_r$, $V_\phi$, and $V_z$}
The 3D velocities are calculated with using line-of-sight velocities from LAMOST DR5, proper motions from Gaia DR2, and LAMOST distances calculated using the Baysian technique from \citep{2015AJ....150....4C}. A known systematic error of 4.4 km s$^{-1}$ in radial velocity is subtracted from the LAMOST catalog values \citet{2018MNRAS.477.2858W}. The 3D velocities in Galactic Cartesian coordinates ($U, V, W$) are calculated following the method in \citet{1987AJ.....93..864J}. Positive $U$, $V$, and $W$ are oriented towards the Galactic center, the direction of Galactic rotation, and the north Galactic pole, respectively. We adopt a distance from Sun to the Galactic center of $R_\odot=8.34$ kpc \citep{2014ApJ...783..130R}, the height of Sun above the Galactic plane of $Z_0=27$ pc \citep{2001ApJ...553..184C}, the peculiar velicity of the Sun with respect of the local standard of rest of $(U_\odot, V_\odot, W_\odot)=(9.58,10.52, 7.01)$ km s$^{-1}$ \citep{2015ApJ...809..145T}, and a circular velocity at the solar radius of $V_c=238$ km s$^{-1}$ \citep{2010MNRAS.403.1829S}. Then the 3D Cartesian Galactocentric velocities  can be obtained from $U=U+U_\odot$,  $V=V+V_\odot+V_c$, $W=W+W_\odot$.  The cylindrical velocities $V_R$, $V_\phi$, and $V_z$ are obtained following equations A13, A14, and A15 of \citet{2013MNRAS.436..101W}.  

\section{median $V_\phi$ distribution}
In this section, we study both the spatial (as a function of $R$ and $Z$) and phase space (as a function of $Z$ and $V_Z$) distribution of $V_\phi$. We also connect the bulk motion of disk stars, as a function of spatial position, with the phase space spiral. The spatial distribution of $V_\phi$ is illustrated in subsection 5.1. The phase space distribution of $V_\phi$ is illustrated in subsection 5.2. The relationship between the bulk motion and phase space spiral is illustrated in subsection 5.3.

\subsection{Median  $V_\phi$ as a function of (R,Z)}
\begin{figure*}
    \includegraphics[width=6cm]{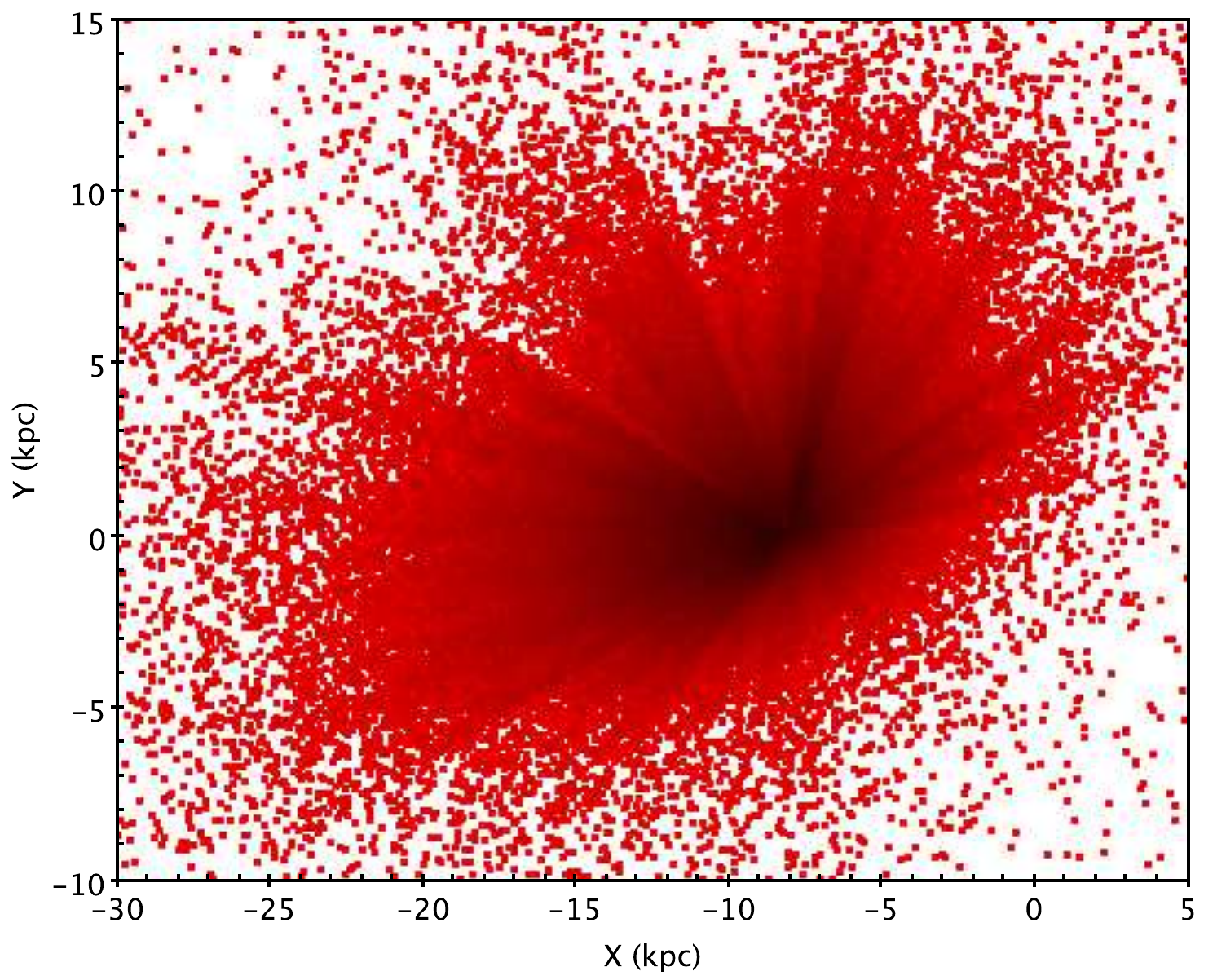}
  \includegraphics[width=6cm]{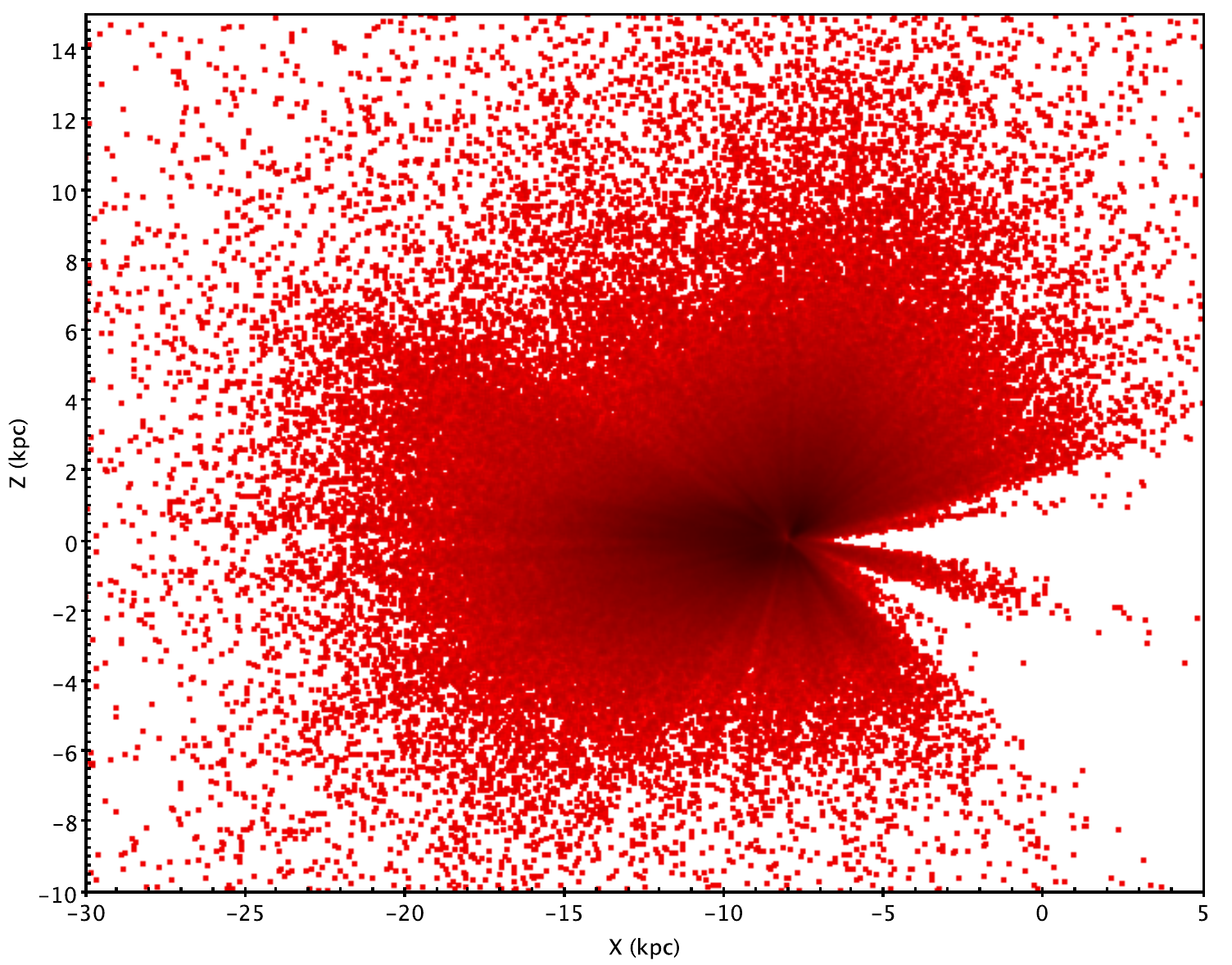}
  \includegraphics[width=6cm]{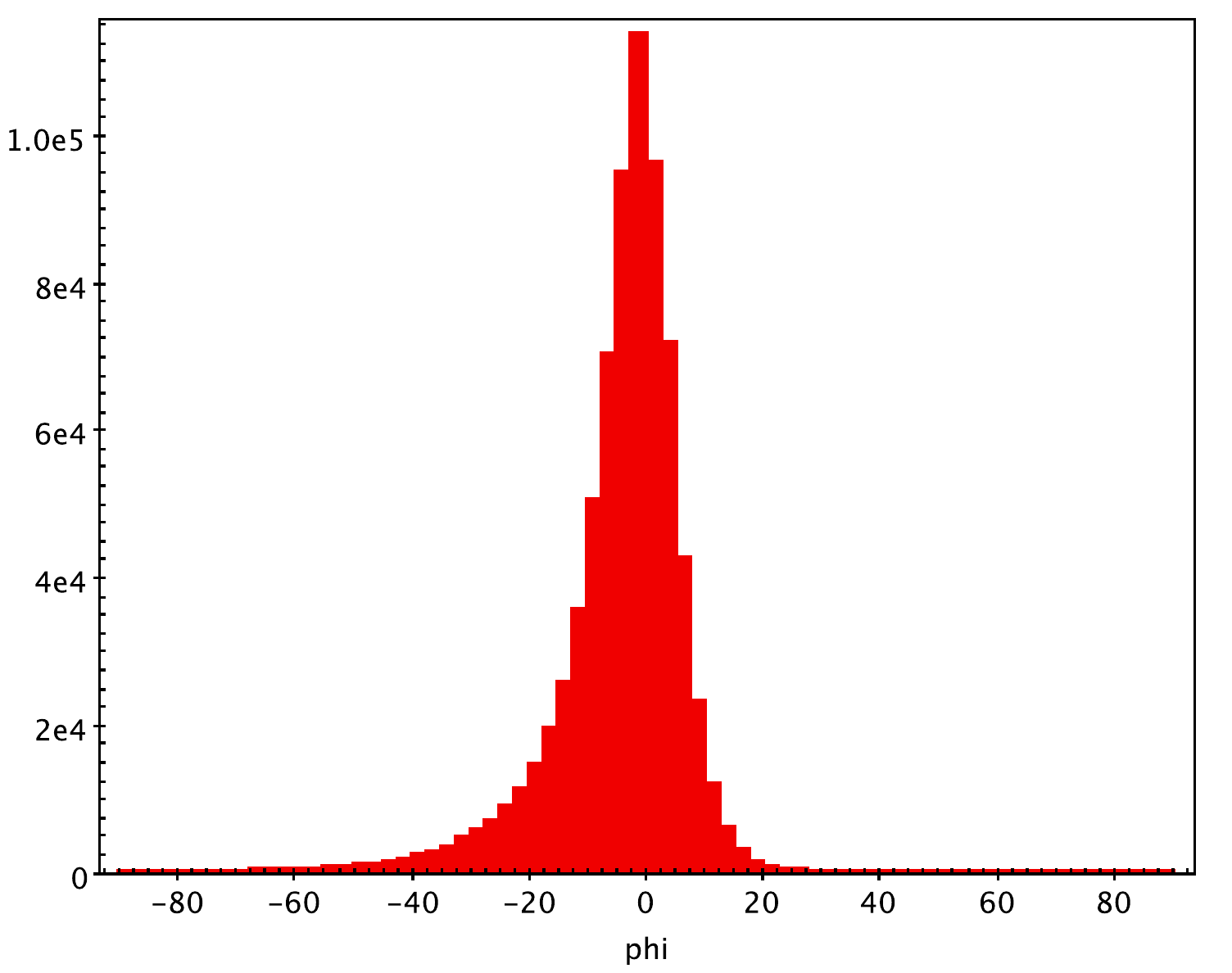}
      \caption{The spatial distribution of LAMOST K giant stars. The left panel shows Galactocentric $Y$ vs. $X$. The middle panel shows Galactocentric $Z$ vs. $X$.  The right panel shows that the stars are heavily concentrated towards the Galactic anticenter. Here, $phi=\arctan(Y/X)$, which is azimuthal angle from the anticenter directions, as seen from the Galactic center. Positive $\phi$ are found in the third quadrant, and negative $\phi$ are found in the second quadrant.}
    \label{spatialdistribution}
\end{figure*}

Figure~\ref{spatialdistribution} shows the spatial distribution of our selected LAMOST K giant stars. They  extend to more than 20 kpc from the Galactic center in the radial direction and extend several kpc from the Galactic plane in the $Z$ direction.  The right panel of Figure~\ref{spatialdistribution} shows that the LAMOST K giants concentrate towards the anti-center direction ($-20^\circ<\phi<10^\circ$, where $\phi$ is the Galactocentric azimuthal angle from the anticenter direction). 

\begin{figure*}
  \includegraphics[width=9cm]{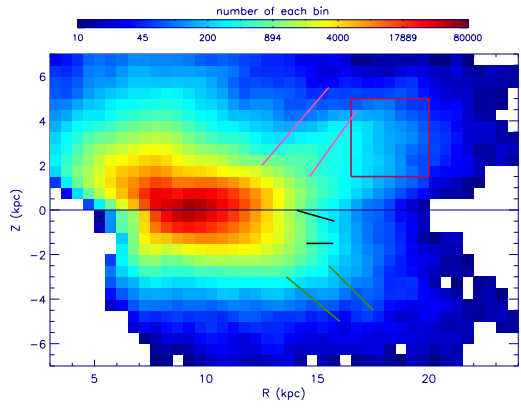}
      \caption{Heat map of the number of K giants in each bin of the R-Z map, using logarithmic scaling. Note that most of the stars are located closer to the Sun. The lines and square are for comparison with structures in Figures 4-7.}
    \label{num_RZ}
\end{figure*}

In this subsection, we will create and analyze two dimentional projections of the bulk velocities of these stars in $(R, Z)$ space. This is a convenient projection because K giants concentrate in a narrow range around the anti-center direction. Figure~\ref{num_RZ} shows the number of K giant stars in our sample as a function of $(R,Z)$. All of the bins contain at least 10 stars; within $R<20$ kpc, most bins have at least 50 stars.

Heat maps of the median $V_R$, $V_\phi$, $V_Z$ as function of $(R, Z)$ are now constructed to study the kinematic distribution of disk stars as a function of spatial position. The bin size in our maps is: $(\Delta R, \Delta Z)=(0,5, 0.5)$ kpc.  

\subsubsection{Disk Substructure in $V_\phi$}

\begin{figure*}
    \includegraphics[width=6.5cm]{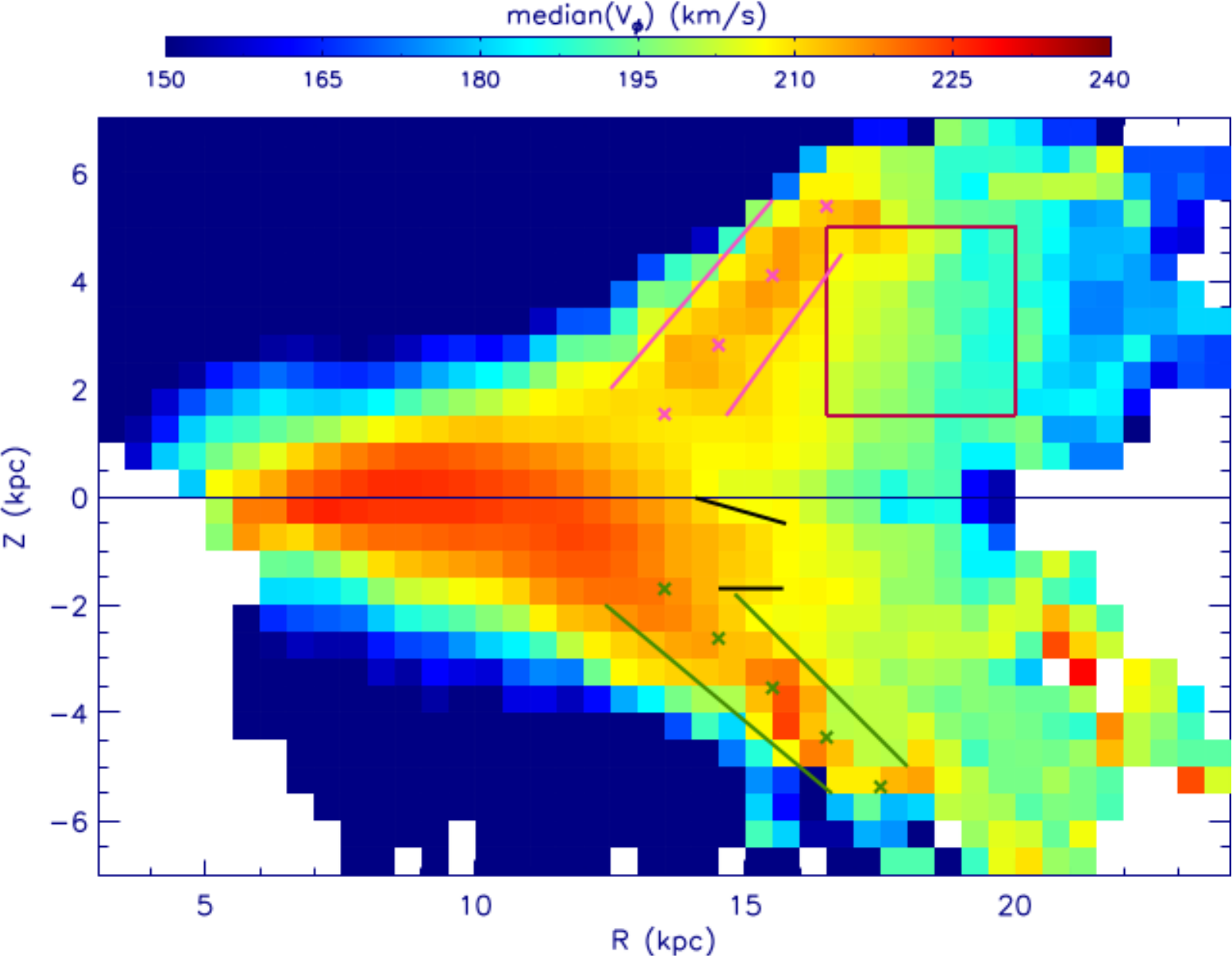}
  \includegraphics[width=6.5cm]{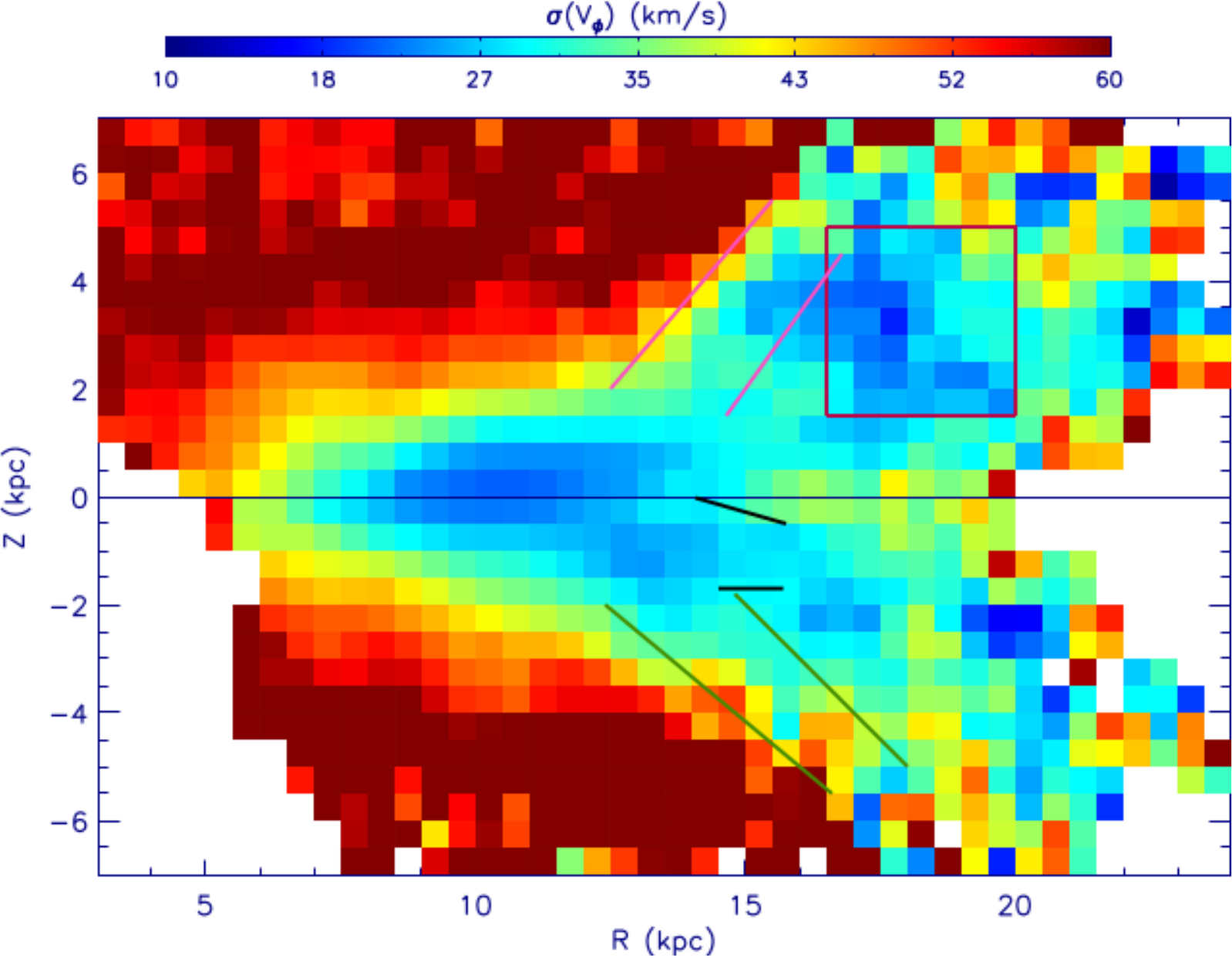}\\
      \includegraphics[width=6.5cm]{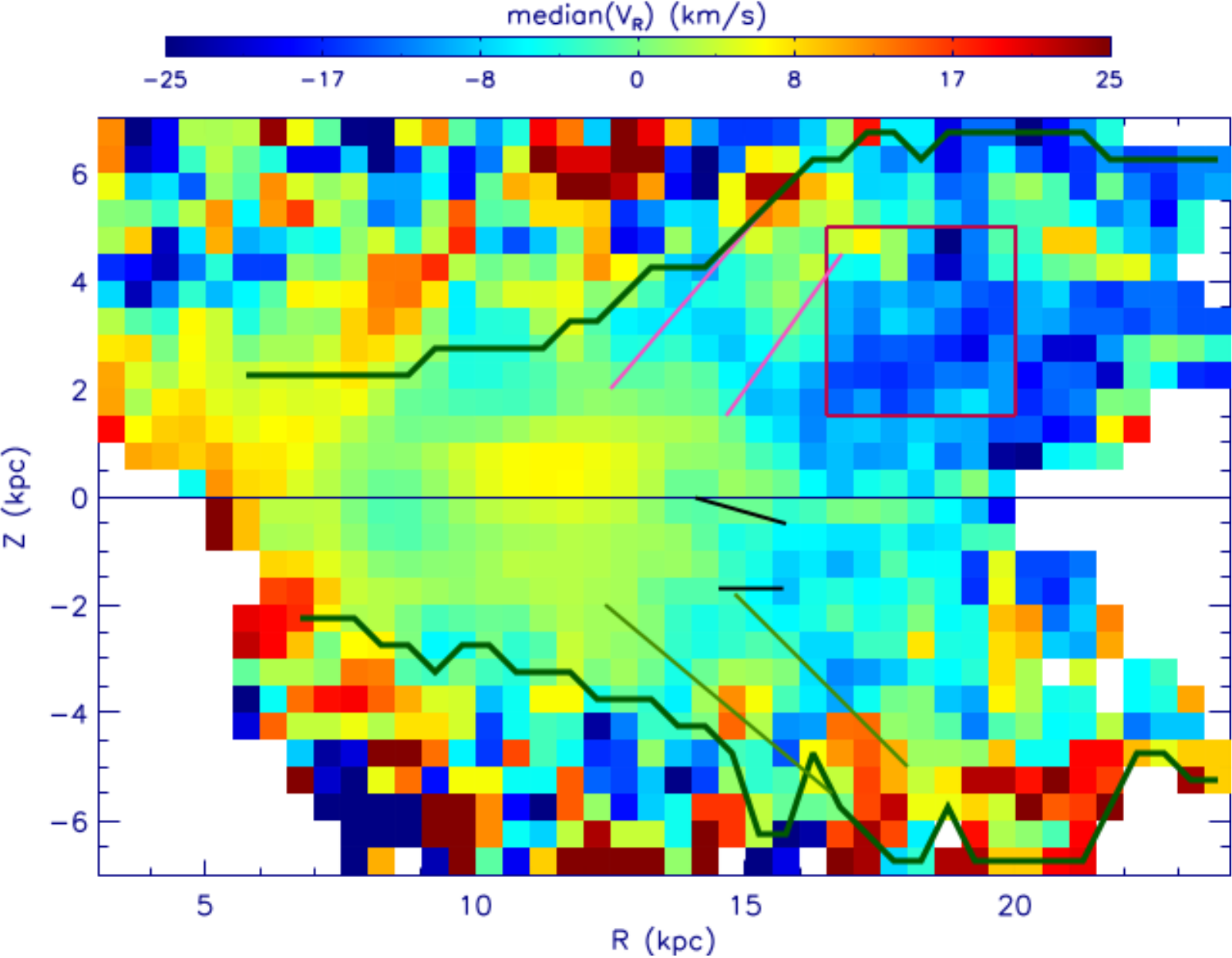}
    \includegraphics[width=6.5cm]{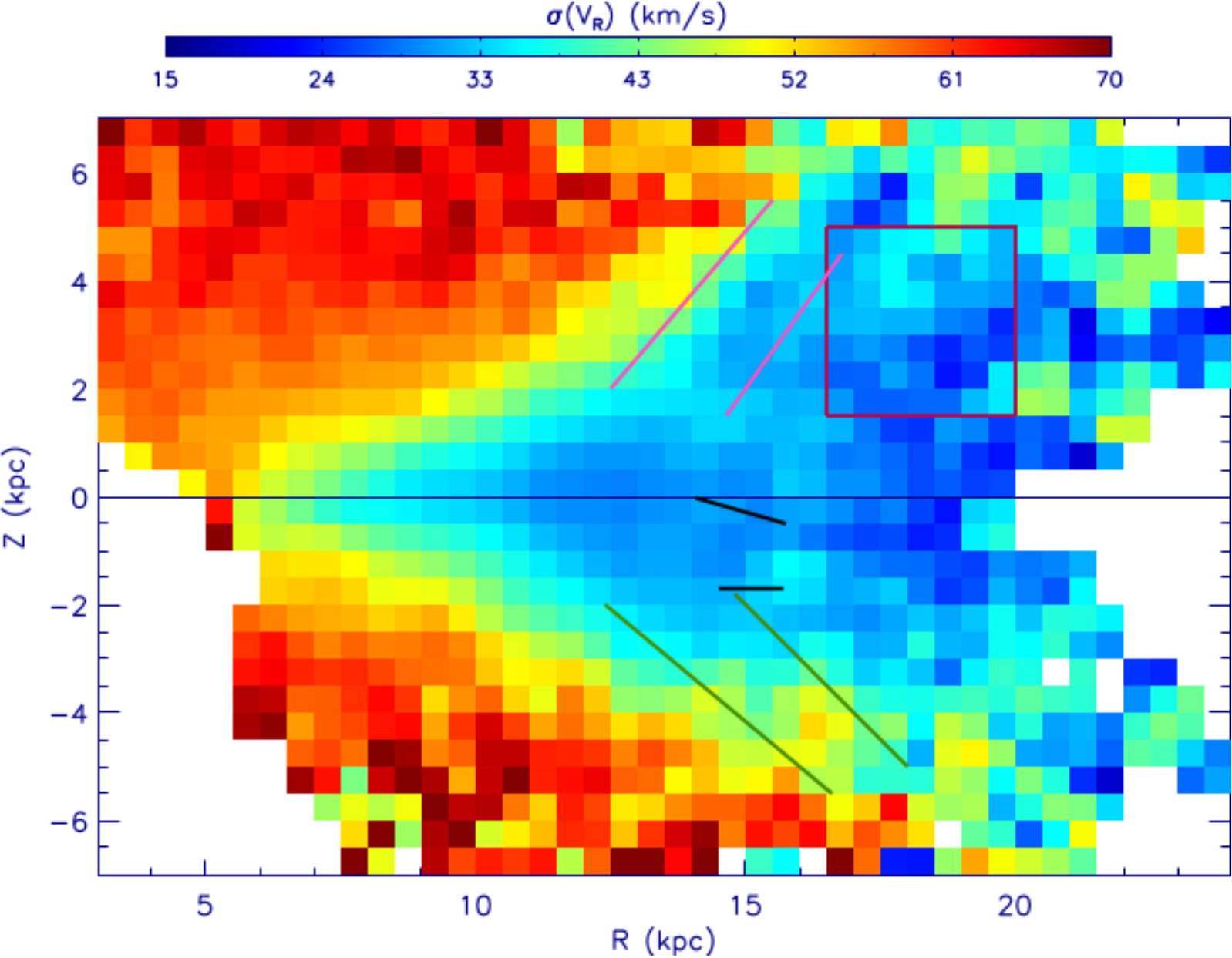}\\
        \includegraphics[width=6.5cm]{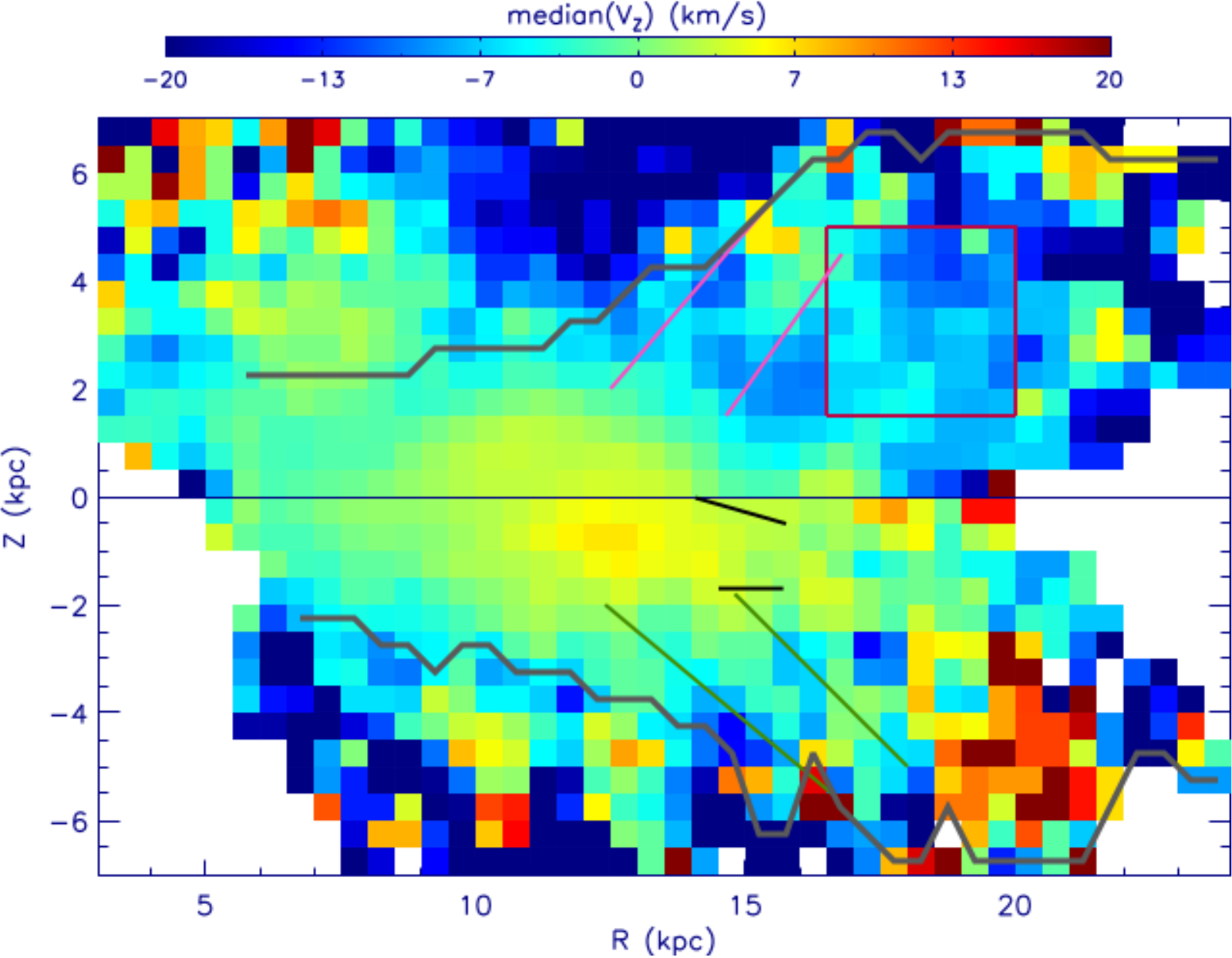}
    \includegraphics[width=6.5cm]{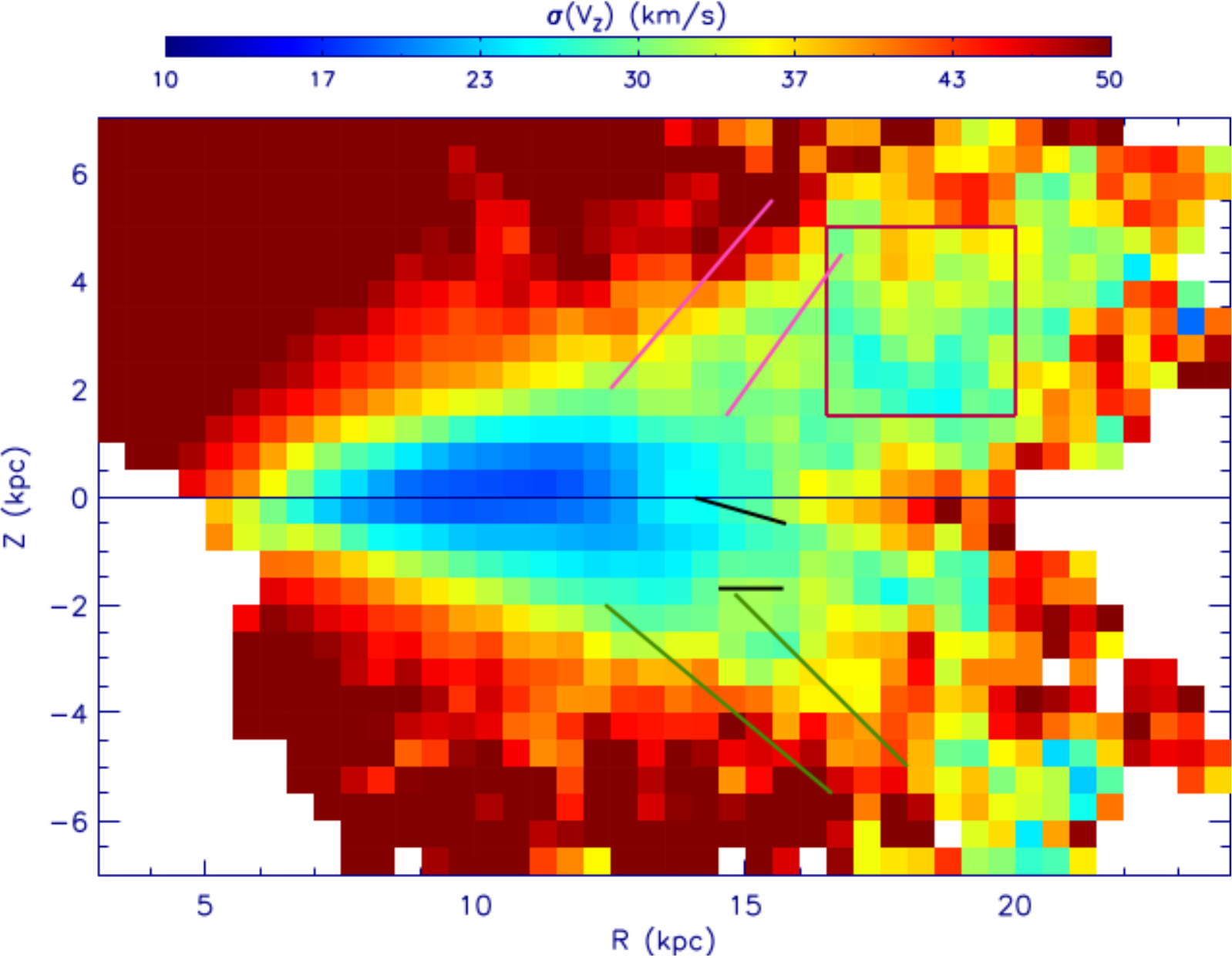}\\
        \includegraphics[width=6.5cm]{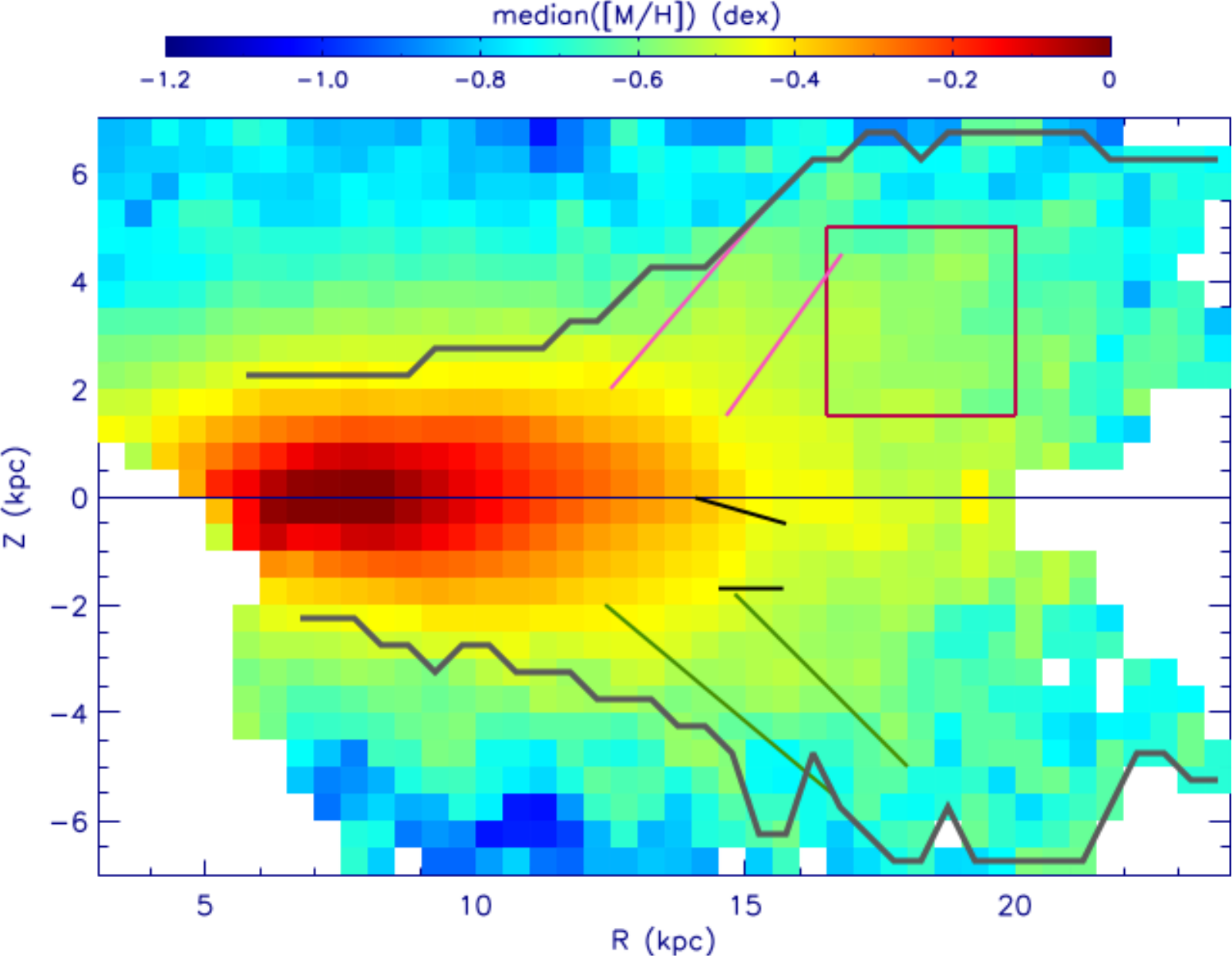}
    \includegraphics[width=6.5cm]{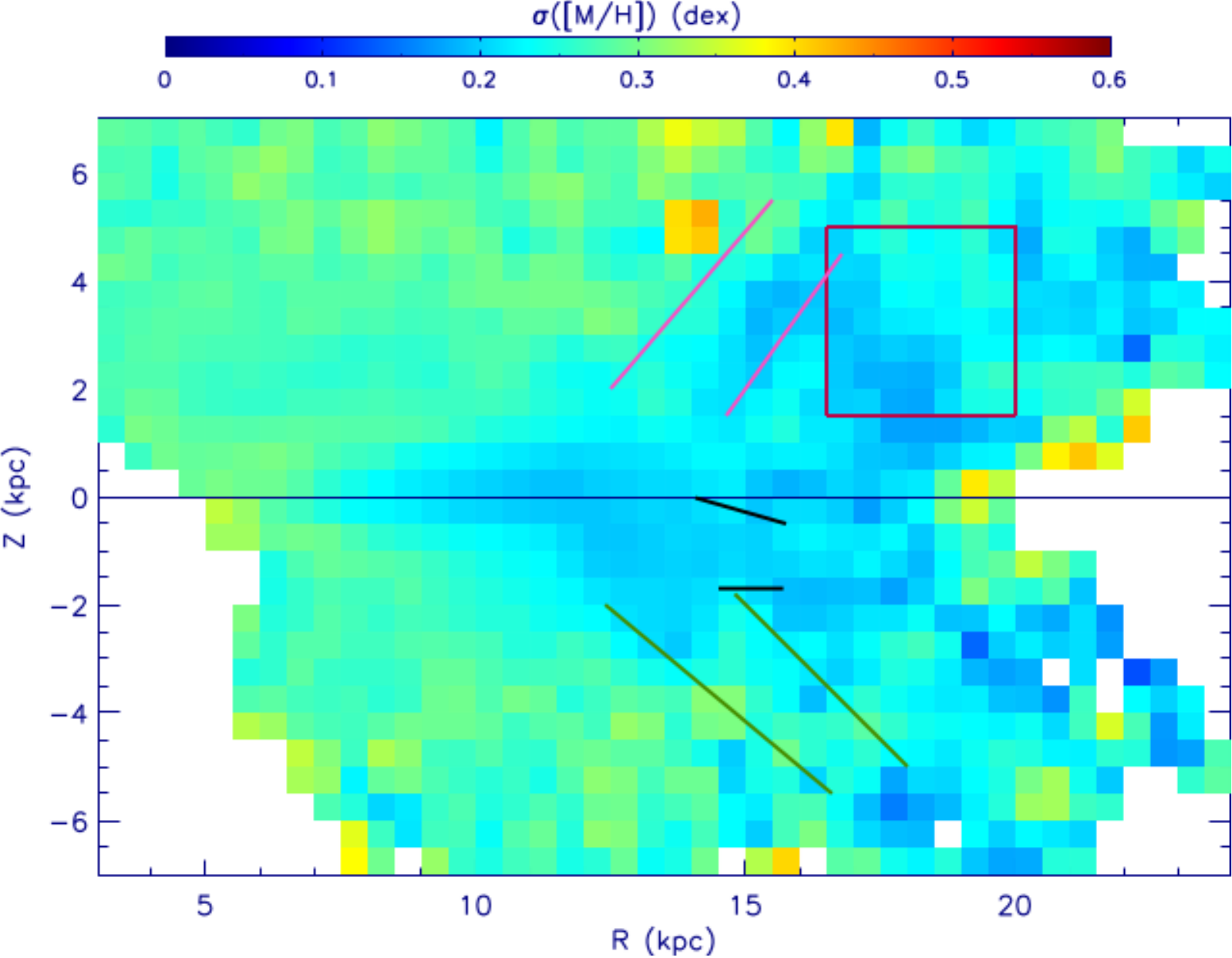}
      \caption{The first row: The median (left panel) and standard deviation (right panel) of $V_\phi$ for LAMOST K giants, as a function of $(R, Z)$. The bounds of the ``main branch," ``north branch" and ``south branch," which are identified as ridge lines in the left panel, are labeled by black lines, pink lines and green lines, respectively. The ridge line of the ``north branch" lies along $(R, Z)=(13.5,1.53), (14.5,2.81), (15.5,4.1), (16.5,5.38)$ kpc, as labeled in the left panel by pink crosses. The ridge line of the ``south branch" lies along $(R,Z)=(13.5, -1.7), (14.5, -2.625), (15.5, -3.54), (16.5, -4.46), (17.5, -5.375)$ kpc, as labeled in the left panel by green crosses. The ``Monoceros area," which is identified as a region of low standard deviation in the right panel, is labeled by a red square. The second, third and fourth row: The median (left panel) and standard deviation (right panel) of $V_R$, $V_Z$, and [Fe/H] for LAMOST K giants, as a function of $(R, Z)$. The bounds of the ``main branch," ``north branch," ``south branch," and ``Monoceros area" are labeled as in the first row of panels. In the left panels, the dark, jagged curves label the boundary of the region with median($V_\phi)>160$ km s$^{-1}$ in the top left. }
    \label{Vphi_RZ}
\end{figure*}

The first row of Figure~\ref{Vphi_RZ} \footnote{The image of Figure~\ref{Vphi_RZ} is smoothed to more clearly show the features. The median($V_\phi$), median($V_R$), median($V_Z$), median([Fe/H]) of each grid of (R, Z) with bin size ($R-1/2*\Delta R<R<R+1/2*\Delta R$, $Z-1/2*\Delta Z<Z<Z+1/2*\Delta Z$) is calculated from stars within the range of ($R-\Delta R<R<R+\Delta R$, $Z-\Delta Z<Z<Z+\Delta Z$). } shows the median($V_\phi$) and $\sigma$($V_\phi$) (the standard deviation of $V_\phi$) for our sample, as a function of $(R, Z)$. From this figure, the disk and halo populations can be clearly distinguished; the disk has a high $V_\phi$ and a low $\sigma$($V_\phi$).

The right panels of Figures ~\ref{Vphi_RZ} also illustrate the distribution of $\sigma$($V_R$), $\sigma$($V_Z$), and $\sigma$($[M/H]$), respectively. In these panels the disk-like stars also show the same characteristics; the disk-like stars are kinematically colder and have a smaller metallicity scatter compared to those of halo-like stars. The top right panel of Figure~\ref{Vphi_RZ} shows that in $\sigma$($V_\phi$) there is a smooth transition from the kinematically cold area with disk-like stars to the kinematically hot area with halo-like stars, especially before $R<13$ kpc. Beyond R=13 kpc, the transition is sharper.

If one looks at stars that are kinematically colder and have smaller metallicity scatter across the entire $(R,Z)$ plane, one finds that the distribution of median($V_\phi$) and $\sigma$($V_\phi$) of disk-like stars is horn-shaped. The scale height of disk stars grows quickly with $R$ in the outer disk, characteristic of a disk flare. The flare is consistent with the one seen in blue stragglers in the Canada-France Imaging Survey, crossmatched to Gaia DR2 and SEGUE and LAMOST \citep{2019MNRAS.483.3119T}.

In the solar neighborhood, the $V_\phi$ distribution is quite similar to the known disk kinematics, which are ``dominated by rotation with a smooth vertical gradient" \citep{2010ApJ...716....1B}. 

 Along the mid-plane, the median $V_\phi$ decreases with increasing $R$, which is consistent with the result of \citet{2016IAUS..317..354T}. In the range $R<11$ kpc, the median $V_\phi$ for bins in the mid-plane of the disk reaches to 230 km s$^{-1}$.  The median $V_\phi$ in the disk midplane decreases to around 210 km s$^{-1}$ when $R=15$ kpc.
 
 Beyond $R=11$ kpc, the most significant structure in disk kinematics is that the high $V_\phi$ stars are split into three branches, which we call the ``main branch," ``north branch," and ``south branch." The ``main branch" bends southward; the ridge line of the ``main branch" is distributed roughly along the line from $(R,Z)=(12,-0.5)$ kpc to $(R,Z)=(14,-1)$ kpc.  The ``north branch" is found along the slope from $(R,Z)=(13, 1.5)$ kpc to $(17, 5)$ kpc.  The ``south branch" is found along the line from $(R,Z)=(11,-1)$ kpc to $(R,Z)=(16,-4)$ kpc. The ``south branch" has higher median $V_\phi$ than that of ``north branch." Both of the north and south branches stretch along the boundary of the flare.

It is apparent from the $\sigma(V_\phi)$ panel of Figure~\ref{Vphi_RZ}, in the range of $R<13$ kpc, that the ridge of minimum standard deviation in $V_\phi$ is located slightly above the midplane, roughly at $Z=0.25$ kpc.  When $R>13$ kpc, the ridgeline of minimum velocity dispersion skews towards the south, following the main branch. This is consistent with a vertical displacement of the midplane of the disk. We will discuss this in more detail in Section 5.1.2.

Also apparent in the $\sigma(V_\phi)$ panel of Figure~\ref{Vphi_RZ} is an area with quite small $\sigma$($V_\phi$) in the range of $16.5<R<20$ kpc , $1.5<Z<5$ kpc. This location is consistent with famous substructure called the ``Monoceros overdensity." The location of the ``main branch," ``north branch" and ``south branch" kinematic structures, identified from the $median(V_\phi)$ panel of Figure~\ref{Vphi_RZ}, as well as the ``Monoceros area'' identified from the $\sigma(V_\phi)$ panel of Figure~\ref{Vphi_RZ}, are labeled by lines and a square in each panel the figure.

Estimates of the kinematic characteristics of ``north branch," ``south branch," and ``Monoceros area" from Figure~\ref{Vphi_RZ} are summarized in Table 1. To measure these properties, we selected stars most likely associated with the substructures, in the regions of parameter space delineated in Figure~\ref{Vphi_RZ}. The measured properties are the median over the the selected bins in $V_\phi$, $V_R$ and $V_Z$.

\subsubsection{The oscillating disk traced by kinematic features}
Using SDSS main sequence star counts, \citet{2015ApJ...801..105X} found that the midplane of the disk stars appears to oscillate vertically across the Galactic plane, as a function of distance from the Galactic center. 
The oscillating disk asymmetries appear at $D=2$ kpc, 5 kpc, 10 kpc, and 15 kpc, where $D$ is the distance from the Sun. The two nearer oscillation asymmetries can be approximated by model with an oscillating disk, in which the disk is offset up by 70 pc at a distance of 2 kpc from the Sun towards the anticenter, and down by 170 pc at a distance of 5 kpc from the Sun towards the anticenter. 

In this work, the mid-plane is traced by the locus of minimum standard deviation in $V_\phi$, and minimum standard deviation in $V_Z$. We slice the data into bins of different Galactocentric radius, $R$, and find the $Z$ location of the minimum standard deviation for each bin. We use a bin size of $\Delta R=0.2$ kpc for $R<12$ kpc and 0.5 kpc for $R>12 kpc$, due to fewer stars and larger statistical noise in the measurement of the minimum at large radius.   

In  Figure~\ref{waveonedisk}, the red triangles and the blue triangles show the locations of the minimum standard deviation of $V_\phi$ and $V_Z$, respectively. The $Z$ position of the minimum standard deviation in $V_\phi$ and $V_Z$ follow each other very closely.
The mid-plane shifts north in the range of $R=10-13$ kpc with amplitude of up to 300 pc, and then again shifts south. The black curve is the ``oscillation" derived from fitting disk main sequence star counts in \citet{2015ApJ...801..105X}. The trend found in the oscillation model obtained by fitting SDSS  main sequence stars is consistent with that of the mid-plane oscillation identified kinematically using LAMOST K giants. However, the oscillation in K giant stars suggests a larger oscillation amplitude. \citet{2018MNRAS.478.3367W} identified the South middle structure from LAMOST K giant star counts. They clarified that the South middle Structure defined by \citet{2015ApJ...801..105X} is an  extended structure from $(R,Z)=(10,-0.5)$ kpc to $(15,-3)$ kpc, and found that the asymmetry across the plane can be erased by shifting the mid-plane south by about 300 pc at $R=14$ kpc. This suggested a mid-plane offset is indicated in Figure~\ref{waveonedisk} with a green plus sign. 

We traced the mid-plane using the location of minimum standard deviation of $V_\phi$ and $V_Z$ for $R<14.5$ kpc. After 14.5 kpc, the locus of minimum standard deviation in $V_\phi$ is dominated by substructure with high $Z$. For example, the Monoceros area has small standard deviation in $V_\phi$ with median $Z$ of about 3 kpc, which is more consistent with substructure kicked out of the disk than it is with an oscillation.

\begin{deluxetable*}{l*{7}{c}}
\tablewidth{0pt}
\tabletypesize{\footnotesize}
\renewcommand{\arraystretch}{0.85}
\tablecaption{Characteristics of kinematic substructures of disk stars}
\tablehead{
\colhead{kinematic features} &
\colhead{$median(V_\phi)$} &
\colhead{$\sigma(V_\phi)$} & 
\colhead{$median(V_R)$} &
\colhead{$\sigma(V_R)$} &
\colhead{$median(V_Z)$} &
\colhead{$\sigma(V_Z)$}\\
&
 (km/s) &
 (km/s) &
 (km/s) & 
 (km/s) &
 (km/s) &
 (km/s) &
    }

\startdata
(i) north branch   & 213  & 32.5 & -1.6 & 35 &-4.3 & 33.9\\
(ii) south branch & 213 & 40.7  & 0.9 & 42 & -2.5 & 39.3\\
(iii) Monoceros area  &195.5 & 26.4 & -12.8 & 30 &  -7 & 31.\\
 \enddata
\label{Table:compareobssim}
\end{deluxetable*} 

\subsection{The $V_\phi$ phase space spiral and the outer disk substructures}

\citet{2018Natur.561..360A} discovered an impressive snail-like spiral in $Z$ vs. $V_Z$ phase space, providing direct evidence of phase mixing produced by a strong perturbation of the disk. The Sgr dSph was identified as the most likely Galactic intruder to produce this perturbation, due to the match between the very recent impact time and the phase mixing time scale. 

In this section, we will show the variation of the $V_\phi$ phase space spiral with Galactocentric radius, and show the connection between the $V_\phi$ phase spiral and the outer disk kinematic features.

 \subsubsection{Comparison of the $V_\phi$ phase space spiral in LAMOST K giants with that of Gaia DR2 data in the Solar neighborhood}

Before we construct the phase spiral maps for different ranges of Galactocentric radius, it is necessary to compare the $V_\phi$ phase spiral seen in LAMOST K giants with that of {\it Gaia} DR2 data in the solar neighborhood. 

The upper panel of Figure~\ref{gaiaspiral} shows the median $V_\phi$ distribution in $(Z-V_Z)$ phase space for 0.62 million stars crossmatched between LAMOST DR5 and {\it Gaia} DR2, that have Galactocentric distances in the $8.24<R<8.44$ kpc range. We will refer to it as the ``total sample" in this subsection. The upper panel of Figure~\ref{gaiaspiral} perfectly reproduces the two and half circles of phase spiral discovered by \citet{2018Natur.561..360A}. To produce this panel we used distances from \citet{2018AJ....156...58B}. 

The upper right panel of Figure~\ref{gaiaspiral} shows the 12,870 K giants selected from the ``total sample" with the additional selection criteria of Section 2. Using these stars we can only barely see the phase spiral. The K giants in the disk might not be quite as good as main sequence stars for illustrating the phase spiral, because they are a kinematically hotter disk population with a larger scale height and therefore less influenced by the Sgr dSph impact. Also, there are many fewer stars in the K giant sample. K giant stars comprise only one twentieth of the total sample within $8.24<R<8.44$ kpc; the subset of K giants that meet the criteria outlined in Section 2 are only one fiftieth of the ``total sample'' in this radial slice.

To determine the main reason for the difference in clarity between the upper panels of Figure~\ref{gaiaspiral},
we tried randomly extracting one fiftieth of the stars from the total sample and building the $Z-V_Z$ phase space map.
The histograms in the second row of Figure~\ref{gaiaspiral} show the distribution in $Z$ of the ``total sample" and K giant stars. The K giants are more spread out in the $Z$ direction, and less concentrated right at the solar position. This is because they are intrinsically bright and can be observed to further distances than many of the stars in the ``total sample." To determine whether the wider range of distances is important, we subsampled 12,870 stars from the ``total sample" in two ways. In the left panel of the third row, the stars are randomly sampled. In the right panel of the third row, the stars are extracted from the ``total sample" with a distribution in $Z$ that matches the K giant stars' $Z$ distribution. The two panels in the third row are also produced with  distances from \citet{2018AJ....156...58B}.

Both of the phase spirals of random samples in the third row of Figure~\ref{gaiaspiral} are similar, and only a little clearer than that of the K giant sample in the upper right panel.  Extracting randomly from the ``total sample" or extracting based on a broader distribution in $Z$ didn't make much difference, but the lower star number did make a big difference. From this experiment, we illustrate the importance of sample size in tracing the phase space spiral. Since K giants are much more sparsely distributed in space than main sequence stars, we cannot gather as many K giants as main sequence stars in a thin slice. Because the phase space spiral changes with Galactocentric distance, there is a tradeoff between using a thicker slice that will include more stars, and a thinner slice that will sample a narrower range of spiral properties; the phase spiral will be blurred in thicker slices of $R$. We choose 1 kpc bins in $R$ as a trade-off between gathering more stars and cutting the slice as thinly as possible.

The lower panels of Figure~\ref{gaiaspiral} show the K giant phase spiral for $8<R<9$ kpc. The sample size is about one twelfth of the total sample in the upper left panel. The lower left panel is constructed with distances from \citet{2015AJ....150....4C}; it shows a big central ``bulge" (centroid) which is the blurred inner circle of the phase spiral and also shows the wide strong, tail of the phase space spiral. In Figure 1 of \citet{2018Natur.561..360A}, the width of the inner circle is about $\Delta Z=0.1-0.2$ kpc and the outer circle width is about $0.2-0.3$ kpc. Because the inner portion of the spiral has a finer structure, 
the inner phase spiral can't be distinguished from the LAMOST K giant sample in solar neighborhood when the sample size is too small. The lower right panel shows the same data as the lower left panel, but constructed instead with the {\it Gaia} DR2 distances \citep{2018AJ....156...58B}, which is more precise than the LAMOST distances \citep{2015AJ....150....4C} in the solar neighborhood. In the lower right panel, we see that the outer portion of the spiral more closely matches the position of the outer spiral in the upper left plot. But the inner portion is not sampled, as can be understood from a comparison of the histograms in the second row of the figure; the central region of the spiral is heavily populated with a large number of intrinsically fainter stars.

Given that the LAMOST K giants are much sparser tracers but sample a much larger volume, we can explore the phase spiral over a larger portion of the Milky Way, but with lower resolution than was seen in local {\it Gaia} DR2 data. We will use LAMOST K giants to determine whether there is a phase space spiral in volumes within 20 kpc of the Galactic center and trace the outline of any observed spiral, but we will not be able to observe the phase space spiral's detail.

\subsubsection{The median$(V_\phi)$ phase spiral as a function of distance from the Galactic center}
We are now in a position to look for variations in the phase space spiral as a function of position in the Galaxy. First, we separate the disk stars into 1 kpc bins along the projected Galactic radius, $R$, in the range $6<R<20$ kpc. In Figure~\ref{ZVz_Vphi}, the stars in each bin are plotted in (Z, $V_Z$) phase space, color coded by median($V_\phi$). 

The phase spiral appears in each bin from 7 kpc to 15 kpc; after that, the spiral disappears.

 At first glance, the phase space spiral in Figure~\ref{ZVz_Vphi} appears to be sketched by the high median($V_\phi$) stars. In the gap of high median($V_\phi$) spiral, there are relatively low $V_\phi$ stars. From 7 kpc to 15 kpc, each phase spiral map is similar to the last one, with some small but significant change. The $R=14$ to 15 kpc bin is the last bin in which we can see a full circle of the phase spiral. There is still a high-$V_\phi$ spiral standing out from the lower $V_\phi$ background, though it is blurry and not smooth.  After $R=15$ kpc, there are only a few high $V_\phi$ segments in the $R=15$ to 16 kpc bins.
 
In the caption of Figure~\ref{ZVz_Vphi}, the sample size of each bin is labeled. It is noticeable that sample size is only 5 thousand in the $R=14$ to 15 kpc bin. In the upper right panel of Figure~\ref{gaiaspiral}, the phase spiral can only barely be seen when the sample size decreases to around 12 thousand. However, the phase spiral can still be seen in the panels of Figure~\ref{ZVz_Vphi} in the bin R=14 to 15 kpc. This is because the phase spiral is in the range of $-0.5< Z < 0.5$ kpc in the $R=8$ kpc bin, and the interval between the different phase spiral circles is narrow so the phase spiral is quite easily blurred. In the $R>12$ kpc bin, the phase spiral is spread over a wider distance from the Galactic plane, $-3<Z<3$ kpc, with a different distribution. The spiral is still visible because the K giants are more spread in $Z$ when $R$ is larger.

From 7 kpc to 15 kpc, the winding of the phase spirals becomes looser with increasing $R$.  
Moreover, in the $7<R<15$ kpc range, the phase space gets narrower (compression) in the $V_Z$ direction and wider (stretching) in $Z$ direction with each increasing $R$ bin. For example, in the bins from $10<R<11$ kpc to $13<R<14$ kpc, the high $V_\phi$ spiral spreads from 2 kpc extent to 3.5 kpc in the $Z$ dimension. This is consistent with simulation of \citet{2019MNRAS.485.3134L} and \citet{2019MNRAS.486.1167B} ; who explained this behavior as due to the weaker potential with increasing $R$. As $R$ increases, the stars travel further in the Z direction and move more slowly. 

\subsubsection{Comparing the $V_\phi$ phase spiral in LAMOST K giants with previous studies}
With Figure ~\ref{ZVz_Vphi}, we showed that the LAMOST K giants can be used to trace the global features of the phase spiral, which change with increasing $R$. The global features of the phase spirals are consistent with the theoretical predictions, but the detailed features of the phase spiral cannot be fully explored with LAMOST K giants data. 
We are here trying to corroborate the characteristics of $V_\phi$ phase spiral in LAMOST K giants by comparing with Gaia DR2 data within $R<10$ kpc.

\citet{2019MNRAS.485.3134L} construct the $V_\phi$ phase spiral at $R=6,$ 8, and 10 kpc with {\it Gaia} DR2 stars in Figure 14 in their paper. Comparing their phase spiral with that of the $R=8$ to 11 kpc bin of LAMOST K giants, the shape of the outline is similar and we also can see the outer circle tail from our data, but the inner circle data is also blurred into big centroid, and it is clear that the phase of the phase spiral in the Gaia DR2 data changes with Galactocentric radius.  The number of phase spiral circles also changes with Galactic radius. 

\citet{2019ApJ...877L...7W} study the shape of the phase spiral as function of $R$ using stars in common between the LAMOST and {\it Gaia} samples, in the range $R<12$ kpc. They also see  different phases of spiral shapes with different R. Before $R=9.34$ kpc, there are two and half spirals and after that radius there is only one circle in \citet{2019ApJ...877L...7W}. From Figure~\ref{ZVz_Vphi} of this work,  we also can see the phase of spirals changing with $R$, though the phase spiral is blurry. In the first three bins, with $7<R<10$ kpc, the spiral warp leads towards $(Z,V_Z)=(0 {\rm~kpc},0 {\rm~km s}^{-1})$, where we see a centroid with a strong and wide tail protruding in the clockwise direction in the first and second quadrants.  In the $10<R<11$ kpc bin, the spiral is composed of a big centroid and a long, narrow ``tail" in the second and third quadrant. The centroid fades with increasing $R$, and the centroid deviates from $(Z,V_Z)=(0 {\rm~kpc},0 {\rm~km ~s}^{-1})$ to the location of $Z<0$ kpc.

From Figure 14 of \citet{2019MNRAS.485.3134L}, the tail of the phase spiral splits when $R>8$ kpc. For example, the lower middle panel of Figure 14 of \citet{2019MNRAS.485.3134L} shows a branching of spiral in the range $-1<Z<1$ kpc, $30<V_Z<60$ km s$^{-1}$ for stars in {\it Gaia} DR2 data that are within 1 kpc of the Sun. We see similar branching in the $8<R<9$ kpc bin of Figure~\ref{ZVz_Vphi}; at the position of $(Z,V_Z)=(0.7 {\rm~kpc},30{\rm~km~s}^{-1})$, the spiral separates into several branches. We also see that the tail of phase spiral bifurcating in the third quadrant in the bin $10<R<11$ kpc. We barely see that the tail of phase spiral bifurcating in the fourth quadrant in the $11<R<14$ kpc bin. This means that either $Z, V_Z$ space is not the correct projection space in which to count the wraps \citep{2019MNRAS.485.3134L}, or the simple harmonic oscillation becomes chaotic, or it is not the simple harmonic oscillation but has higher modes.

\subsubsection{Cautions}
The phase spirals observed in LAMOST K giants are blurry and not smooth. The smaller sample size and larger distance error compared with the {\it Gaia} DR2 data explains part of the blurring, but there is another reason for this blurring effect. When we construct the phase spiral, the adopted bin in $(R,Z)$ space is $\Delta R, \Delta Z = (1{\rm~kpc}, 0.125{\rm~kpc})$ when $R<12$ kpc,  $\Delta R, \Delta Z = (1{\rm~kpc}, 0.15{\rm~kpc})$ when $12<R<15$ kpc,   $\Delta R, \Delta Z = (1{\rm~kpc}, 0.2{\rm~kpc})$ when $15<R<16$ kpc,  $\Delta R, \Delta Z = (1{\rm~kpc}, 0.25{\rm~kpc})$ when $16<R<18$ kpc,  $\Delta R, \Delta Z = (1{\rm~kpc}, 0.3{\rm~kpc})$ when $18<R<19$ kpc. The $\Delta Z$ bin size is larger with increasing $R$ in order to get enough stars in the sample, since the number of stars per radial increment decreases with increasing $R$. When we construct the phase spiral, there is an underlying assuption that there is no selection effect inside the $(R,Z)$ bin and the sampled $V_\phi$ distributions are not biased.  This assumption deviates more from the real situation when the bin size is larger. In the $R<10$ kpc bin, the phase spiral from LAMOST K giants reproduces most of the features of phase spiral from {\it Gaia} DR2, but the properties of the phase spiral beyond this range still need to be confirmed by a larger and more precise dataset.
 
\subsection{Connection between the $V_\phi$ phase spiral and kinematic substructures}
In order to study the relevance between the $V_\phi$ kinematic feature in $R-Z$ and phase space, we rotate the panels from Figure~\ref{ZVz_Vphi} clockwise and arrange them horizontally in order of increasing $R$ (Figure~\ref{ZVz_Vphi_vertical}). Then, we compare Figure~\ref{ZVz_Vphi_vertical} with the median($V_\phi$) panel of Figure~\ref{Vphi_RZ}:

i) In Figure~\ref{ZVz_Vphi_vertical}, the centroid of the phase spiral becomes less and less pronounced with increasing $R$. This is partially due to the smaller number of K giants with increasing R, and partially because of the lower restoring force which causes the phase spiral to be more spatially extended with larger $R$ \citep{2019MNRAS.485.3134L}.

\begin{figure*}
    \includegraphics[width=9cm]{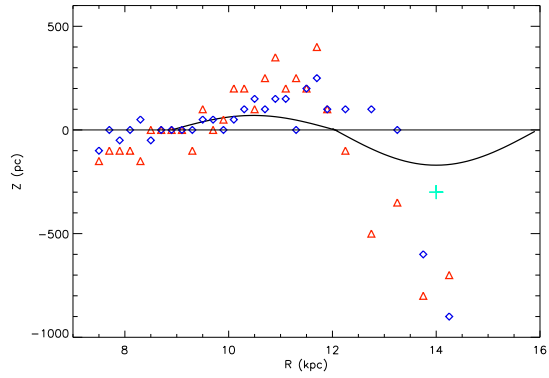}
        \caption{The red triangles and blue diamonds trace the location peak of minimum standard deviation of $V_\phi$ and $V_Z$ in $R-Z$ plane of Figures~\ref{Vphi_RZ} with bins of size ($\Delta$ R, $\Delta$ Z)=(0.2, 0.1) kpc. The black curve is  the ``oscillating" model fit to star counts of SDSS K dwarf stars \citep{2015ApJ...801..105X}. The green plus shows the location of mid-plane shifting at $R=14$ kpc, estimated from LAMOST K giant star counts \citep{2018MNRAS.478.3367W}. }
    \label{waveonedisk}
\end{figure*}

\begin{figure*}
    \includegraphics[width=11cm]{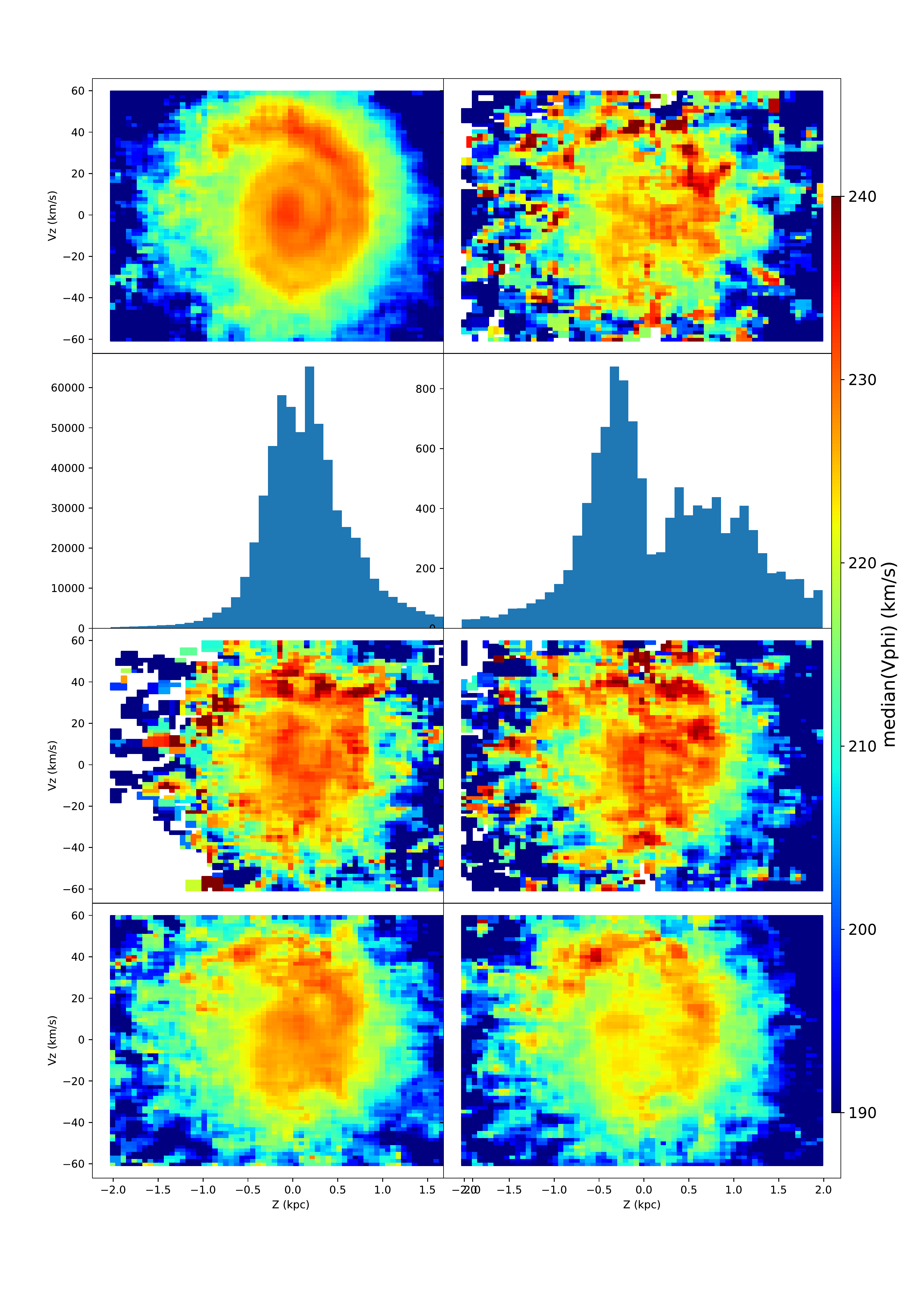}        
	\caption{The distribution of $V_\phi$ in $(Z-V_Z)$ phase space for stars in common between {\it Gaia} DR2 and LAMOST DR5. The upper left panel shows 0.62 million stars (the ``total sample") within the range $8.24<R<8.44$ kpc from the Galactic center. The upper right panel shows 12,870 K giants included in the ``total sample'' which satisfy the selection criteria of Section 2. The panels in the second row show the distribution of ``total sample" stars in the $Z$ direction (left panel of the second row) and K giants sample of the first row (right panel of the second row). It is obvious that the K giants sample spread wider in $Z$ direction. The left panel of third row shows the random 12,870 stars drawn from the ``total sample." The right panel of the third row shows the 12,870 stars drawn from the ``total sample" with the $Z$ distribution of K giants illustrated in right panel of second row. The Gaia DR2 distances \citep{2018AJ....156...58B} are adopted to make the all of above plots. The fourth row of panels of shows the distribution of $V_\phi$ in $(Z-V_Z)$ phase space for 77 thousand LAMOST K giants within the range of $8<R<9$ kpc. The two panels on the fourth row differ only in the method used to calculate the distances to the stars; distances are calculated from the LAMOST spectra \citep{2015AJ....150....4C} in the left panel and from {\it Gaia} parallaxes \citep{2018AJ....156...58B} in the right panel.}
    \label{gaiaspiral}
\end{figure*}

\begin{figure*}
    \includegraphics[width=16cm]{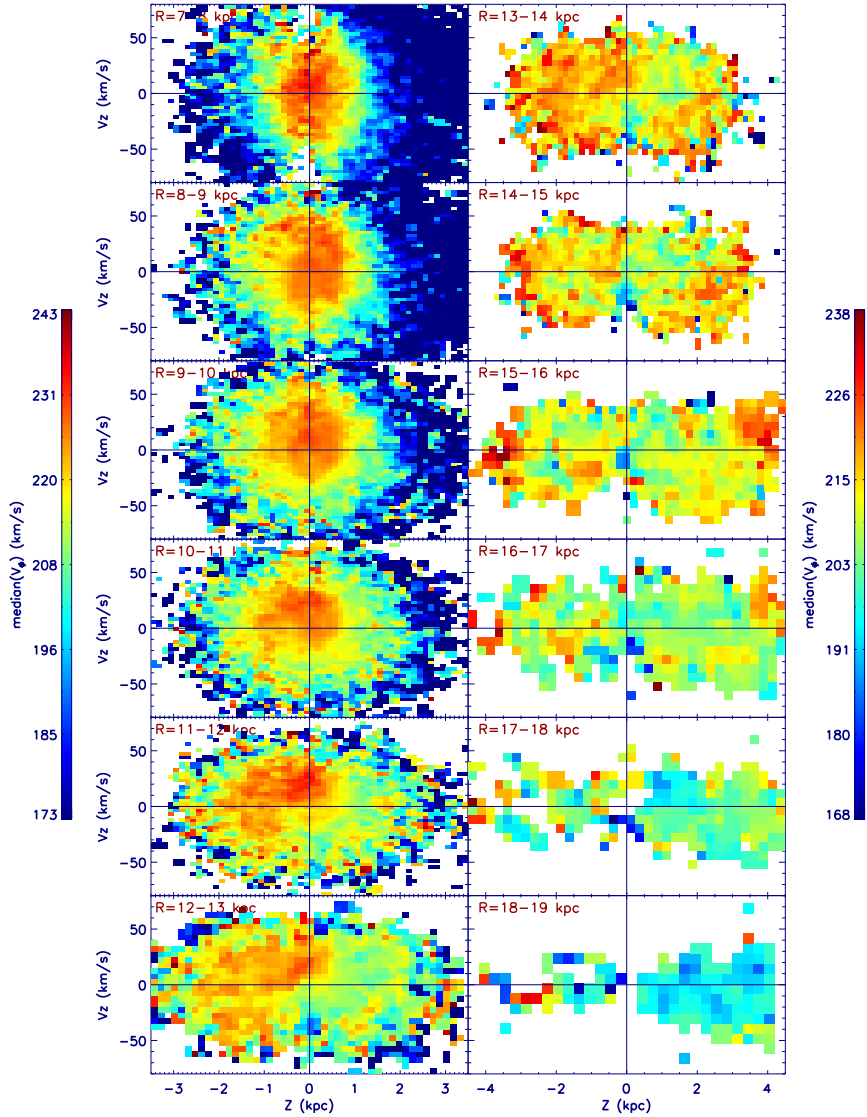}
        \caption{$V_\phi$ distribution in $Z-V_Z$ phase space, as a function of radial distance from the Galactic center, $R$. Each panel shows a 1 kpc range of $R$. In the left panels, from the top to bottom, $R$ changes from $7<R<8$ kpc to $12<R<13$ kpc. In the right panels, $R$ changes from $13<R<14$ kpc to $18<R<19$ kpc. The $V_\phi$ range of the color bar for the left panels is from 173 to 243 km s$^{-1}$, and the range is 168 to 238 km s$^{-1}$ for the right panels.  The $Z$-axis range in the left panels is $-3.5<Z<3.5$ kpc and it is $-4.5<Z<4.5$ kpc for the right panels. The number of stars in each pixel of each map is larger than 5. The total number of stars of each R bin is 61,571 ($7<R<8$ kpc); 77,335 ($8<R<9$ kpc); 100,519 ($9<R<10$ kpc); 72,030 ($10<R<11$ kpc); 41,380 ($11<R<12$ kpc); 20,470 ($12<R<13$ kpc); 9,679 ($13<R<14$ kpc); 5,139 ($14<R<15$ kpc); 3,190 ($15<R<16$ kpc); 2,114 ($16<R<17$ kpc); 1,479 ($17<R<18$ kpc); and 978 ($18<R<19$ kpc). The bin size in phase space is ($\Delta Z, \Delta Vz$) = (0.125 kpc, 3 km s$^{-1}$) when $R<12$ kpc,  ($\Delta Z, \Delta Vz$) = (0.15 kpc, 5 km s$^{-1}$) when $12<R<15$ kpc, ($\Delta Z, \Delta Vz$) = (0.2 kpc, 7 km s$^{-1}$) when $15<R<16$ kpc, ($\Delta Z, \Delta Vz$) = (0.25 kpc, 8 km s$^{-1}$) when $16<R<18$ kpc, and ($\Delta Z, \Delta Vz$) = (0.3 kpc, 9 km s$^{-1}$) when $18<R<19$ kpc.}
    \label{ZVz_Vphi}
\end{figure*}

The phase spiral is centered at $(Z, V_Z)=(0 {\rm~kpc},0 {\rm~km~s}^{-1})$ when $R=8$ to 9 kpc and $(Z, V_Z)=(-0.5{\rm~kpc}, 10 {\rm~km~s}^{-1})$ when $R=14$ to 15 kpc. In Figure~\ref{waveonedisk}, the mid-plane of the disk as identified by the position of the minimum standard deviation of $V_\phi$ and $V_Z$. The mean values of the location of minimum $V_\phi$ standard deviation from Figure~\ref{waveonedisk} in each 1 kpc R bin are also shown as black triangles in Figure~\ref{ZVz_Vphi_vertical}. We found that the black triangles trace the centroid of the phase spiral. We can connect this with the observation that the median($V_\phi$) of the``main branch" decreases with $R$ and bends south after $R>13$ kpc in the median($V_\phi$) panel of Figure~\ref{Vphi_RZ}.  That is consistent with the prediction that the phase spiral traces the bending wave\citep{2019MNRAS.485.3134L}. 

ii) The ``north branch" and ``south branch" of Figure~\ref{Vphi_RZ} have higher $V_\phi$ than the adjacent disk-like stars. The location of ``north branch" and ``south branch" are just at the location where the high median($V_\phi$) phase space spiral passes through this projection.  

For example, in the R=13 to 14 kpc bin the segment of the phase spiral within $-50<V_Z<50, Z<0$ kpc has a small $Z$ range; after integrating along $V_Z$ it corresponds to a pronounced high $V_\phi$ locus at $(R,Z)$=(13.5, -2) kpc. If we imagine compressing each $R$ bin in the $V_Z$ direction, we see that the kinematic substructure of the ``south branch" in the $median(V_\phi)$ panel of Figure~\ref{Vphi_RZ}, which stretches from $(R,Z)$=(12,-1.5) kpc to (15,-4) kpc, is equivalant to the location of the high median($V_\phi$) spiral at each radius. 

Similarly, the integration of the segment of the spiral with $-50<V_Z<0$ km s$^{-1}$ and $Z>0$ kpc, in bins in the range $12<R<16$ kpc, corresponds to the ``north branch" in the median($V_\phi$) panel of Figure~\ref{Vphi_RZ}. From this comparison, we know the  ``north branch" and ``south branch" kinematic substructures are a projection of the phase space spiral into the $R-Z$ plane. 

The peak lines of  ``north branch" and  ``south branch" in the $median(V_\phi)$ panel of Figure~\ref{Vphi_RZ} are overplotted on Figure~\ref{ZVz_Vphi_vertical} by crosses to show the corresponding relationship.

iii) There are different theories for how a flare is produced, such as secular evolution \citep{2012A&A...548A.127M,2015ApJ...804L...9M,2002A&A...394...89N,2012A&A...548A.127M} or the cumulative effect of interactions with passing dwarf galaxies \citep{2008ApJ...688..254K,2008MNRAS.391.1806V,2018MNRAS.481..286L}. 

In Figure~\ref{Vphi_RZ}, the disk-like stars flare with larger $R$. In Figure~\ref{ZVz_Vphi_vertical}, the edge of the flare aligns with the location of outer circle of phase space spiral, especially after $11<R<15$ kpc. Since the phase spiral is most likely induced by a recent passage of the Sgr dSph through the disk, it is reasonable to infer that the last passage of the Sgr dSph contributed stars to the flare.


\section{MEDIAN $V_R$ distribution}

The second row of Figure~\ref{Vphi_RZ} shows the distribution of the median and standard deviation of $V_R$ as a function of position in the $(R-Z)$ plane. In the first row of Figure~\ref{Vphi_RZ}, both the maps of median($V_\phi$) and that of $\sigma$($V_\phi$) show the distinction between disk-like stars and halo-like stars. There is no significant distinction between  disk-like stars and  halo-like stars in the $V_R$ map. So in the $median(V_R)$ panel of Figure~\ref{Vphi_RZ}, the boundary within which the bins of $median(V_\phi)$ panel of Figure~\ref{Vphi_RZ} have a median($V_\phi$) $>160$ km $s^{-1}$ is labeled to guide the eyes. The region of disk-like stars defined by median($V_\phi$) lies between these dark green lines. 

Observing the disk-like stars, we see the median($V_R$) is positive when $R<8$ kpc. The median($V_R$) is slightly lower than 0 km s$^{-1}$ at $R=9$ kpc. The median($V_R$) is positive in the range of $10<R<13.5$ kpc, $|Z|<3$ kpc. Then the median($V_R$) is negative again. To show it more clearly, the radial variations of $V_R$ within $|Z|<1$ kpc and $|Z|<3$ kpc are plotted in Figure~\ref{Vr_R}.
The median $V_R$ of the two positive $V_R$ areas in the range of $R<8$ kpc, $|Z|<3$ and $10<R<13.5 $ kpc, $|Z|<3$ kpc is about 5 km s$^{-1}$.

These $V_R$ features are consistent previous studies.    
 Using RAVE data, \citet{2011MNRAS.412.2026S} were the first to find a negative $V_R$ gradient in the range $6<R<9$ kpc; the negative gradient is consistent with that of the median($V_R$) of disk-like LAMOST K giants; it is positive in the range of $R<8$ kpc and negative in the range of $8<R<9$ kpc near the Galactic plane.
 
\citet{2017RAA....17..114T} and Liu et al. (2017) observed the same negative $V_R$ gradient when $R<9$ kpc and another positive gradient after $R>9$ kpc using RAVE-TGAS data, LAMOST red clump data, and LAMOST RGB data around the anticenter direction.  The positive gradient after $R>9$ kpc is consistent with the positive $V_R$ area in the range of $10<R<13.5$ kpc and $|Z|<3$ with LAMOST K giant tracers.
    
\citet{2018A&A...616A..11G} showed the $X-Y$ distribution of disk radial velocities in the range $4<R<12$ kpc using giant stars from {\it Gaia} DR2. 
Figure 10 of \citet{2018A&A...616A..11G} shows that there are two  positive $V_R$ regions at $4<R<8$ kpc and $R>10$ kpc, which is also consistent with our result.

\citet{2019AJ....157...26L} showed that the $V_R$ velocity of APOGEE stars increased towards the anti-center in the $R=9-13$ kpc range, reaching a maximum of 6 km s$^{-1}$.  $V_R$ decreased at larger $R$, crossing from positive $V_R$ to negative $V_R$  around $R=15$ kpc, which is the position of the Outer spiral arm.
   
  The $V_R$ gradients have variously been explained as a perturbation due to spiral arms \citep{2018A&A...616A..11G, 2019AJ....157...26L} or a resonance with the bar \citep{2017RAA....17..114T}.  Cheng et al. (2019)  reexamines the $V_R$ distribution in the $X-Y$ plane with LAMOST OB stars, over a larger range in $Y$. 
  They find that the two positive stripes are not aligned with the spiral arm, and therefore suggest that a satellite impact such as the Sgr dSph would better explain the $V_R$ ridges observed in the $X-Y$ plane.  
   
  We were not able to clearly trace the $V_R$ phase spiral, which is not as significant as the $V_\phi$ phase space spiral, using LAMOST K giants, so it is not included here; a larger amount data is needed to trace the $V_R$ phase spiral.

\section{median $V_Z$ distribution}
The third row of Figure~\ref{Vphi_RZ} shows the median and standard deviation of $V_Z$ as a function of $(R-Z)$. The stars around the anti-center direction are near the node of the warp, where $V_Z$ is expected to be positive. The mid-plane stars show positive median $V_Z$ of about 5 km s$^{-1}$, which is consistent with the presence of a warp as detected from {\it Gaia} DR2 kinematics \citep{2019gaia.confE..50P, 2020ApJ...897..119W}.

From the third row of Figure~\ref{Vphi_RZ}, we can see that the area of disk-like stars with  positive median($V_Z$) bends to the south after 12 kpc, following the ``main branch" as defined in Figure~\ref{Vphi_RZ}. The fact that both the locus of high median $V_\phi$ and the locus of positive $V_Z$ bend southward suggests that the disk mid-plane bends towards the south outside of 12 kpc from the Galactic center. 

\section{Distribution of metallicities abundance}
The fourth row of Figure~\ref{Vphi_RZ} shows the metallicity distribution, [M/H], as a function of $(R,Z)$ in our sample. [M/H] is obtained from LAMOST spectra using the SLAM pipeline \citep{2020ApJS..246....9Z}. The median([M/H]) panel in Figure~\ref{Vphi_RZ} shows that the value of [M/H] is maximum in the Galactic plane and decreases with increasing $|Z|$. Also, the flared area in the $(R-Z)$ map, where we find stars with the kinematics of disk stars, also has disk-like metallicity; the median([M/H]) value for stars in the ``north branch," ``south branch," and  ``Monoceros area" is  about -0.5 dex.
Not only do these stars have a high [M/H], but they also have a small spread in metallicity compared to the halo-like stars in our sample, as illustrated in the $\sigma[M/H]$ panel of Figure~\ref{Vphi_RZ}. 

\section{Comparison with simulations}
To better understand the observational data, we compare our results using LAMOST K giants with the results of both a test particle simulation and an N-body simulation \citep{2018MNRAS.481..286L} of the Sgr dSph galaxy gravitationally interacting with a Milky Way disk. Test particle simulations have relatively high efficiency and less computational cost; because the model is highly simplified, it is easy to observe the direct  influence of the intruder to the disk. The N-body simulation (Laporte et al. 2018) is more realistic, because it includes self-gravity, which cannot be ignored in many situations (Darling et al. 2019). In addition, the N-body simulation includes the effects of the Sgr dSph passing through the disk more than once.  In this section, we will describe this test particle simulation and N-body simulation, and then compare observational data  with the results.

\subsection{Description of the test particle simulation}
We reproduced \citet{2018MNRAS.481.1501B}'s toy model with galpy \citep{2014ascl.soft11008B}.  In the test particle simulation, the intruder passes perpendicularly through the disk only once, and the gravitational effect of the dwarf galaxy is calculated as an impulse to each of the bodies in the disk.

The galpy potential MKPotential2014, which is a convenient approximation for the Milky Way potential, is adopted. The potential MKPotential2014 includes three parts: a bulge model with spherical potentials that are derived from power-law density models, a disk model that follows the Miyamoto-Nagai potential, and a halo model that follows the NFW potential.
The parameters and properties of MWPotential2014 are summarized in Table 1 of Bovy et al. (2014). The virial mass of the Galaxy is $0.8\times10^{12}M_\odot$. The mass of the disk is $6.8\times10^{10}M_\odot$. The scale length of the disk is 2.6 kpc. 

The galpy  distribution function (df.quasiisothermaldf) is adopted in this work. It is  an approximately isothermal distribution function based on action angle variables. 
The MKPotential2014 and df.quasiisothermaldf are basically self-consistent; the simulated disk is stable for several hundred million years of integration. 

The dwarf galaxy is modeled as a point mass, simulated with the galpy Kepler potential. The influence of the Milky Way on the passing dwarf galaxy is not considered. The point mass is $2\times10^{10 }M_\odot$, as in \citet{2018MNRAS.481.1501B}. It passes through the disk from north side of the disk, starting from $Z=10$ kpc. The speed of the point mass remains constant at $300$ km s$^{-1}$. The influence of the dwarf galaxy disappears after $66$ Myr, when it arrives on the south side of the disk at $z=-10$ kpc.  

The time at the end of the impact is defined as t=0.0 Gyr in our simulation. The impact starts at -0.066 Gyr and it ends at 0.0 Gyr.  The orbit integration, however, starts at -0.5 Gyr (434 Myr before the impact) and ends at 0.5 Gyr (500 Myr after the impact). 

The goal of this section is to search across the whole disk in the simulated result to look for similar kinematic distributions to those found in observational data in both $R-Z$ space and $Z-V_Z$ space during the orbit integration time.  We observe each snapshot at each time grid in the directions $\phi=0^\circ, 180^\circ, 45^\circ,135^\circ, 225^\circ,$ and $315^\circ$, with a bin size of $\pm20^\circ$.  Finally, the stars in the wedge with $315^\circ-20^\circ < \phi < 315^\circ+20^\circ$ are selected to show the detail of kinematic features, because the phase space spiral several hundred million years after the impact in this direction is most similar to the observations. We will compare the observational data with simulation results in subsection 9.3. If the simulation was an exact replica of the encounter that produced the phase space spiral, then the $\phi$ directon in the simulation that matches the data would tell us something about the place or time of the impact that caused the spiral. But since this is only a toy model, the correspondance is only expected to be approximate.

A detailed description of the results of the test particle simulation is available in the appendix.

\subsection{Description of \citet{2018MNRAS.481..286L}'s N-body simulation}
 The N-body simulation of \citet{2018MNRAS.481..286L} is a high precision simulation which considers the entire orbit of the Sgr dSph after falling into the virial radius ($R_{200}$) of the Milky Way, including all disk passages, focussing on the reaction and evolution of the disk to the Sgr dSph. The results of the \citet{2018MNRAS.481..286L} simulation reproduce the vertical density and kinematic oscillation observed in the solar neighborhood and also the ring-like structure in the outer disk. 

\citet{2018MNRAS.481..286L} built 4 kinds of N-body simulation models (L1, L2, H1, H2) with different initial mass and different radial profiles for the Sgr dSph. The L2 model can best reproduce the spatial distribution of stars in the outer disk and flaring \citep{2019MNRAS.483.3119T} as well as the amplitude of density residual in the solar neighbourhood. In \citet{2019MNRAS.485.3134L}, the L2 model is adopted to explain the observed disk oscillation and phase spiral. We will use the simulation results for the L2 model in this work to compare with the observational data.

 The L2 model has a progenitor mass of M200=$6\times10^{10} M_\odot$ and is twice as concentrated as the mean of the mass-concentration relation \citep{2008MNRAS.387..536G}, with c=28. In the L2 model, the Sgr dSph travels around or passes through the Galaxy 5 times from 5 Gyr ago to the present day. Each passing can produce a disk oscillation and ring-like structures. It can also erase the imprint of the previous passing to a large degree, depending on a tradeoff between the rest mass of the Sgr dSph and the Galactocentric distance of the impact position. Each passing, the Sgr dSph losses mass, reducing the influence of the satellite on the Galactic disk. At the same time, the Sgr dSph gets nearer to the Galactic center, increasing the influence. These two effects compete with each other. From the simulation, the disk passage occurring at present day has a trivial effect on the disk, because the current mass of main body of Sgr dSph is only $10^9 M_\odot$. While the last passage happened more than 0.5 Gyr ago, is important because it reset the velocity distribution that we observe today in both physical space and phase space.

In \citet{2019MNRAS.485.3134L}, the N-body simulation results are compared with 
{\it Gaia} data in the range $6<R<10$ kpc mainly in $X-Y$ space and $Z-V_Z$ space.  We will compare with the LAMOST K giants in larger volume of the Galaxy, $8<R<20$ kpc in $R-Z$ space and $Z-V_Z$ phase space.  

The time window of the last Sgr dSph impact is about 0.46 Gyr - 0.8 Gyr ago \citep{2019MNRAS.485.3134L}, so we observe snapshots at 0.46 Gyr, 0.63 Gyr, and 0.8 Gyr in the directions $\phi=0^\circ, 180^\circ, 45^\circ,135^\circ, 225^\circ,$ and $315^\circ$,  with a bin size of $\pm20^\circ$. For N-body simulation, the $V_\phi$ phase spiral is most similar to that of observational data in the wedge with $225^\circ-20^\circ<\phi<225^\circ+20^\circ$. 

The definition of coordinates used by \citet{2019MNRAS.485.3134L} is the same as the test particle simulation. The position of the Sun is $(X,Y)=(-8,0)$ kpc. The azimuth angle $\phi$ is 0 in the direction of the negative $X$-axis. $\phi$ increases in the direction of disk star rotation. 

\subsection{Comparison of the observational data to the results of simulations}

In subsection 9.3.1, we compare the observed phase spiral with both the test particle and the N-body simulations \citep{2018MNRAS.481..286L}; each simulation provides different insights into the dynamical interaction. In subsection 9.3.2, other observational kinematic features in the data are compared with the test particle simulation only.



\subsubsection{Comparison of the observed and simulated phase spirals}

Both simulations can qualitatively reproduce the $V_\phi$ phase space spiral. For the test particle simulation, the wedge with $315^\circ-20^\circ<\phi<315^\circ+20^\circ$ is selected as the most similar to the data. For the N-body simulation, the wedge with $225^\circ-20^\circ<\phi<225^\circ+20^\circ$ is selected as the most similar.
  
For the test particle simulation, in the direction of $315^\circ-20^\circ<\phi<315^\circ+20^\circ$, Figure~\ref{testpartical_ZVz_table} shows the median($V_\phi$) distribution in $Z-V_Z$ space  in three different Galactic radii and five different times between 120 Myr and 400 Myr after the impact.

Figure~\ref{testpartical_ZVz_table} shows that the $V_\phi$ phase spiral appears first at smaller Galactic radii, and moves to larger radii with time. In this simulation, the phase spiral starts to appear in the bin with $8<R<9$ kpc when $t=120$ Myr after the impact. In the panel with $(R, t)=$(8 kpc, 120 Myr), the phase of the modeled spiral is consistent with that of the observed spiral; the phase space spiral is centered at ($Z, V_Z$)=(0 kpc, 0 km s$^{-1}$) and trails in the counterclockwise direction. At $t=200$ Myr after the impact, the phase space spiral starts to appear at larger $R$, where $R=14-15$ kpc.

Figure~\ref{testpartical_ZVz_table} also shows that as the phase spiral appears at larger radii, it disappears at smaller radii. The phase spiral has already disappeared from the $8<R<12$ kpc region by $t=360$ Myr after the impact. The phase spiral is just fading from the $14<R<15$ kpc region when $t=400$ Myr.

The fact that the phase spiral moves to successively larger values of $R$ is consistent with a perturbation that is propogating outwards.
In Figure~\ref{testpartical_ZVz_table}, the low median($V_\phi$) spiral appears at $R=11-12$ kpc at $t=200$ Myr. Then the low median($V_\phi$) spiral appears at $R=14-15$ kpc at $t=280$ Myr. This is consistent with Figure~\ref{testpartical_XY_Vphi} which shows that in the direction of $315^\circ-20^\circ<\phi<315^\circ+20^\circ$, the low median($V_\phi$) ring propagates outwards from $R=10$ kpc to $R=24$ kpc as the time evolves from $t=180$ Myr to $t=400$ Myr after the impact. Similarly, the high median$(V_\phi)$ phase spiral appears at radius $R=11-12$ kpc at $t=280$ Myr, and then moves outward to $R=14-15$ kpc when $t=360$ Myr. 

Figure~\ref{testpartical_ZVz_Vphi0p29} shows the phase spiral distribution over the full range of Galactocentric radius at a point $t=0.21$ Gyr after the encounter with the dwarf galaxy. Note that the phase space spiral can be seen over a wide range of radius ($9-15$ kpc).
 
In the N-body simulation result, the phase space spiral matched the observational phase space spiral in the Solar neighborhood $0.4-0.8$ Gyr after the Sgr dSph impact \citep[see Figure 10 of ][]{2019MNRAS.485.3134L}. In this time period, the distance range in which the phase space spiral appears changes from $8<R<15$ kpc to $8<R<20$ kpc, depending on azimuth angle. Figure ~\ref{Laporte_ZVzmap_Vphi} shows an example of phase spiral of each $R$ bin of the azimuth slice with $225^\circ-20^\circ<\phi<225^\circ+20^\circ$ in the N-body simulation result when $t=0.8$ Gyr.

The phase space spirals produced by both the test particle simulation and the N-body simulation qualitatively match the observational phase spirals. Similar to the behavior of the observational phase space spiral, the phase space spirals in simulations also are more extended in the direction of $Z$ and more contracted in the direction of $V_Z$ as $R$ increases. The winding of  phase space spiral gets looser with increasing $R$.
 In Figure~\ref{testpartical_ZVz_Vphi0p29} and ~\ref{Laporte_ZVzmap_Vphi}, there are two different phase space spiral shapes, one before and one after $R=12$ kpc. This finding is also similar to the observational phase spirals. 
  
The most significant difference between the results of the test particle simulation and the N-body simulation is the persistence time scale of phase space spirals.  In the test particle simulation, the phase space spiral only exists for a short time; it survived only 400 Myr within the range of $R<15$ kpc. But in the \citet{2019MNRAS.485.3134L} N-body simulation (Figure~10 of their paper), the phase space spiral survived more than 800 Myr within the same distance range. This is consistent with the conclusion of \citet{2019MNRAS.484.1050D}, who compared the responses of model disks with and without self-gravity to the same pertubation; they found the bending wave can last 1 Gyr in the simulated disk with self-gravity (as is present in the N-body simulation) while the bending wave damps out within 500 Myr in the simulated disk without self gravity (similar to the test particle simulation).

At Galactocentric distances larger than 16 kpc, the complete phase space spiral is not seen in the result of N-body simulations in the direction of $225^\circ-20^\circ<\phi<225^\circ+20^\circ$; only ``hatched chunks'' are visible in $Z-V_Z$ space. This is because the phase space spiral has been refreshed by the last impact of Sgr dSph in the inner disk, while in the outer disk the phase space shows the result of perturbations from multiple impacts \citep{2019MNRAS.485.3134L}. The ``hatched chunks" are not seen in the results of the test particle simulation outside of 16 kpc. This is because there is only one time impact in the test particle simulation.
 

\subsubsection{Comparison of the other observed kinematic features with the test particle simulation}

(i) The test particle simulation reproduces vertical disk oscillations several hundred million years after the impact. 
For example, the Figure~\ref{XYmap_z_testparticle} shows the median $Z$ distribution in the $X-Y$ plane at time $t=0.21$ Gyr after the impact in the test particle simulation. One sees an oscillating disk around $\phi=270^\circ$ (the direction of the negative $Y$ axis). The disk bends to the south at $R=13$ kpc, then bends to the north at $R=17$ kpc, with an oscillation amplitude of about 300 pc.

(ii) The test particle simulations can reproduce $V_R$ ripples similar to those found by Cheng et al. (2019) or the $V_R$ ripple in $median(V_R)$ panel of  Figures~\ref{Vphi_RZ} of this work. Cheng et al. (2019) studied the kinematic distribution of O, B stars in $X-Y$ plane. They found a ripple pattern: $V_R$ is -8 km s$^{-1}$ at $R=9$ kpc, 0 km s$^{-1}$ at $R=12$ kpc, and -10 km s$^{-1}$ when $R>13$ kpc. This ripple is similar to the $V_R$ distribution in $R-Z$ space of Figure~\ref{Vphi_RZ} and $V_R$ distribution in $R$ of Figure~\ref{Vr_R} of our paper. In Figure~\ref{Vphi_RZ} and Figure~\ref{Vr_R}, we see that the median($V_R$) of disk-like stars is positive when $R<8$ kpc, dips lower at $R=9$ kpc, and is positive again when $R>10$ kpc.

In the test particle simulation wedge with $315^\circ-20^\circ<\phi<315^\circ+20^\circ$, the $V_R$ ripple pattern appears after $t=0.14$ Gyr. The upper right panel of  Figure~\ref{RZmap_Vphi_0p36_testparticle} shows an example 
$V_R$ distribution for $315^\circ-20^\circ<\phi<315^\circ+20^\circ$ in $R-Z$ space when t=0.24 Gyr. The $V_R$ ripple is apparent in this panel. The median($V_R$) is about 10 km s$^{-1}$ when $2.5<R<4$ kpc, about -15 km s$^{-1}$ when  $5<R<7.5$ kpc, and larger than 10 km s$^{-1}$ when $7.5<R<14$ kpc. This ripple is illustrated more clearly in Figure ~\ref{Vr_R_simulation}, which shows the $V_R$ variation of stars within $|Z|<1$ kpc as a function of Galactocentric radius. Comparing the Figure~\ref{Vr_R_simulation} with the Figure~\ref{Vr_R}, the two positive $V_R$ peaks in the test particle simulation result are quite similar to those in the observations.
The full sequence of snapshots of the $V_R$ distribution in $R-Z$ space is shown in  Figure~\ref{testpartical_RZ_Vr} of Appendix.

(iii) The test particle simulation can reproduce the high median $V_\phi$ substructure in $R-Z$ space, similar to the ``north branch" and ``south branch" defined in subsection 5.1. If we study the phase space spiral in the simulation results, we also find a correspondence between the projection of the phase space spiral in the $R-Z$ space and the high median $V_\phi$ structures. 

Figure~\ref{ZVzmap_vertical_Vphi_0p22_315} shows an example of the median $V_\phi$ distribution in the $R-Z$ plane (the upper panel) and the $V_\phi$ phase space spiral in the $Z-V_Z$ plane for each $R$ bin (the lower panel) of snapshot of the test particle simulation with $315^\circ-20^\circ<\phi<315^\circ+20^\circ$ and $t=0.21$ Gyr. The panels of the phase space spiral map are rotated clockwise and lined up in the direction of increasing $R$. The upper panel of Figure~\ref{ZVzmap_vertical_Vphi_0p22_315} shows the high median $V_\phi$ substructure in the range $10<R<13$ kpc has $Z$ around 1.5 kpc, which is quite similar with the ``north branch" of Figure~\ref{Vphi_RZ}. There is a less well populated ``south branch" with $Z \sim -2$ kpc in that same distance range. The lower panel of Figure~\ref{ZVzmap_vertical_Vphi_0p22_315} shows that the high median $V_\phi$ branch is equivalent to the projection of the phase space spirals in the $10<R<13$ kpc bins.  
 
(iv) The test particle simulation exhibits a transient flare that is excited by the Sgr dSph impact. 
The upper panel of Figure~\ref{ZVzmap_vertical_Vphi_0p22_315} shows an example of a flare excited by a dwarf galaxy impact in the test particle simulation. Comparing the upper and lower panels, we see that the  boundary of the flare is defined by the outer ring of the phase space spiral, just as in Figure~\ref{ZVz_Vphi_vertical}. Figure~\ref{testpartical_RZ_Vphi}, ~\ref{testpartical_RZ_Vr}, and ~\ref{testpartical_RZ_Vz} show that the flares grow as a function of $R$, with the oscillation propagating outward. From Figure~\ref{testpartical_RZ_Vphi}, we can see that there is no flare before the Sgr dSph impact.



(v) The test particle simulations produce Monoceros-like substructures.
 For test particle simulation, Figure~\ref{RZmap_Vphi_0p36_testparticle} shows kinematically cold substucture at $(R,Z)=(15, 2)$ kpc with median $V_\phi=160$ km s$^{-1}$, $\sigma$$V_\phi =15$ km s$^{-1}$, and median $V_R= 20$ km s$^{-1}$. Figure~\ref{testpartical_RZ_Vphi} shows that this substructure is puffer and has moved to higher radius at $t=0.46$ Gyr. 
 
We use the test particle simulation to explore the particle dynamics qualitatively, but the test particle simulation does not include self-gravity which is important in reducing substructure and maintaining waves over long periods of time\citep{2019MNRAS.484.1050D}. These concerns are mitigated by the fact that we are only looking at times within a few hundred million years of the impace, but the test particle results in this section should be verified in the future with full N-body simulations.

\subsubsection{Summary of simulation results}

Comparing observations with simulations, the Sgr dSph impact can qualitatively explain most of the observed disk substructure phenomena. From the test particle and N-body simulations, the influenced stars are rotated and dragged into rings, and show a phase space spiral several hundred Myr after the impact. The phase space spiral first appears at small $R$, and then gradually moves to larger $R$ because the azimuthal frequency decreases with $R$. Spirals wind up quickly after a few orbital periods. 

The passage of the Sgr dSph through the disk can produce the observed substructure. It can produce vertical and radial waves \citep{2013MNRAS.429..159G, 2016ApJ...823....4D, 2019MNRAS.485.3134L}; $V_R$, $V_\phi$ ripples \citep{2016ApJ...823....4D}; \citep{2018MNRAS.481..286L} phase space spirals \citep{2018MNRAS.481.1501B, 2019MNRAS.486.1167B, 2019MNRAS.485.3134L};  and stream-like rings like the Monoceros ring. 
In summary, the Sgr dSph passing through the disk is likely the reason for the disk substructures, as was reported in \citep{2018MNRAS.481..286L}. We show here in great detail how that substructure, seen in many different phase space projections, is related.

\section{DISCUSSION}

\subsection{Comparison with the previous studies}
6D phase space information from {\it Gaia} DR2 allowed the local sample of stars to be studied in many projections. \citet{2018Natur.561..360A} studied the $V_\phi$ and $V_r$ distributions in $Z-V_Z$ phase space and discovered the phase space spiral. \citet{2019MNRAS.485.3134L} find the number counts in the $Z-V_Z$ phase space spiral. And they find that the vertical density oscillation \citep{2012ApJ...750L..41W, 2013ApJ...777...91Y} is consistent with the 1D projection of the number counts for the spiral in the $Z$ direction. \citet{2018MNRAS.478.3809S} study the  $W$ vs. $L_Z$ distribution with {\it Gaia}-TAGS data, and find that a feature of the warp is that $W$ is positive in the direction of the anti-center. Also, they find that there is an oscillation in the increase of $W$ with $L_Z$. 
They find that the $W-V_\phi$ oscillation is just the projection of a spiral in $Z-V_Z$ phase space.  Cheng et al. (2019) study LAMOST K giants in the $X-Y$ plane. 
 From our observational data and previous study, we know the dominant features of the velocity field in $R-Z, X-Y , Z-V_Z$, and $R-V_\phi$ phase space; the density distribution in the solar neighborhood along $Z$, and the density distribution in $Z-V_Z$ phase space, are different projections of the same kinematic feature imprinted by the same perturbation.
  
From our observational data and previous study, the bar is ruled out as the main reason for phase space spirals. This is because i) The observational phase spiral in LAMOST K giants is apparent $7-15$ kpc from the Galactic center, while the phase spiral produced by bar buckling is apparent only $4-10$ kpc from the Galactic center, even 4 Gyr after the buckling \citep{2019A&A...622L...6K}.  
ii) Stars with different ages coexist in the same phase space spiral at the same $R$ \citep{2018ApJ...865L..19T, 2019MNRAS.485.3134L}. 
iii) From the simulations, the bar cannot produce the observed size of the bending wave \citep{2015MNRAS.452..747M}.  iv) A pattern speed of 60 km s$^{-1}$ kpc$^{-1}$ is required to fit the observed radial variation of the median($V_R$) in the test particle simulation of \citet{2018IAUS..334..109L}, which is large compared with recent measurements of $34-47$ km s$^{-1}$ kpc$^{-1}$ for the pattern speed \citep{2016ARA&A..54..529B}. 
  
Spiral structure is also ruled out as the main formation mechanism for the phase space spiral. If the heavy spiral arms produce ripples in the disk, the ripple should follow those spiral arms \citep{2014MNRAS.443L...1D, 2012MNRAS.425.2335S, 2014MNRAS.440.2564F, 2016MNRAS.457.2569M, 2018A&A...616A..11G}. However, in the larger volume of LAMOST data we see that the ripples do not follow the spiral arms \citep{2019ApJ...872L...1C}. 

In the Milky Way disk, there are still kinematic substructures which can't be explained in detail as a projection of the high-$V_\phi$ phase spiral, and which need more precise data and more research to explore. For example ``South middle opposite" found by \citet{2019ApJ...884..135W} is located in an apparent gap in the phase spiral. In addition, \citet{2019MNRAS.490..797C} found that there is a strong azimuthal gradient in $V_R$, which is not predicted by the phase space spiral.

\subsection{The Monoceros substructure}
The Monoceros overdensity is characterized by a velocity streaming feature. The stars in the ``Monoceros area" have disk-like metallicity ([M/H]$\approx$-0.5 dex). 

\citet{2016ApJ...825..140M} estimate that the mass of the stars on the north side of the disk in the Monoceros overdensity is $4\times10^6 M_\odot$, and the mass of stars in ``south Monoceros," which is identified as a distinct structure from the south middle structure of \citet{2015ApJ...801..105X}, is $4-4.8\times10^7M_\odot$.  The total stellar mass of the Monoceros overdensity is less than $5.2\times10^7M_\odot$, even when the ``south Monoceros" structure is included.  However, the simulation of  \citet{2005nfcd.conf....1G} demonstrates that a dwarf galaxy stellar mass should be at least $10^8M_\odot$ to produce [Fe/H] as high as -0.5 dex which is an index of the total metallicity. This metallicity estimate for LAMOST K giants is additional evidence pointing to the identification of the Monoceros overdensity being part of the disk rather than an accreted satellite. 

 \citet{2020MNRAS.492L..61L} also find that the Monoceros ring is an extension of the outer disk, and not accreted tidal debris, from a more detailed chemical composition of the stars. They detect the iron abundance and the $\alpha$-abundance of the Monoceros ring and ACS with APOGEE data, and find that both of the structures have abundances in the range $-0.8<$[Fe/H]$<-0.3$, $0<$[Mg/Fe]$<0.15$. Stars with these abundances are at the metal poor end of the thin disk branch in the map of [Fe/H] vs. [Mg/Fe] in Figure 6 of Laporte et al. (2019b).  

Li et al. (2019, in submition) show the same result using LAMOST K giant and M giant stars to trace the Galactic anticenter substructure(GASS), which includes Monoceros ring stars. They find the mean and dispersion of the GASS iron abundance are -0.56 dex and 0.5 dex, respectively. The $\alpha$ abundance is concentrated in the range of 0 to 0.2 dex. The lower panel of Figure 7 in Li et al. (2019, in submition) also shows a evidence that the dwarf galaxies of the Milky Way and GASS do not occupy the same region of the [$\alpha$/Fe] vs. [Fe/H] plane. 
In addition, the high metallicity of the Monoceros ring is inconsistent with a dwarf galaxy of the mass of Fornax; if one uses the mass-metallicity relation for dwarf spheroidal from \citet{2013ApJ...779..102K}, the implied stellar mass of a galaxy of metallicity [M/H]$\sim-0.5$ would be $10^{10} M_\odot$.

The metallicity suggests that the substructures are very likely connected to the disk rather than with debris from and accretion event.

\section{Conclusion}

\begin{figure*}
    \includegraphics[width=18cm]{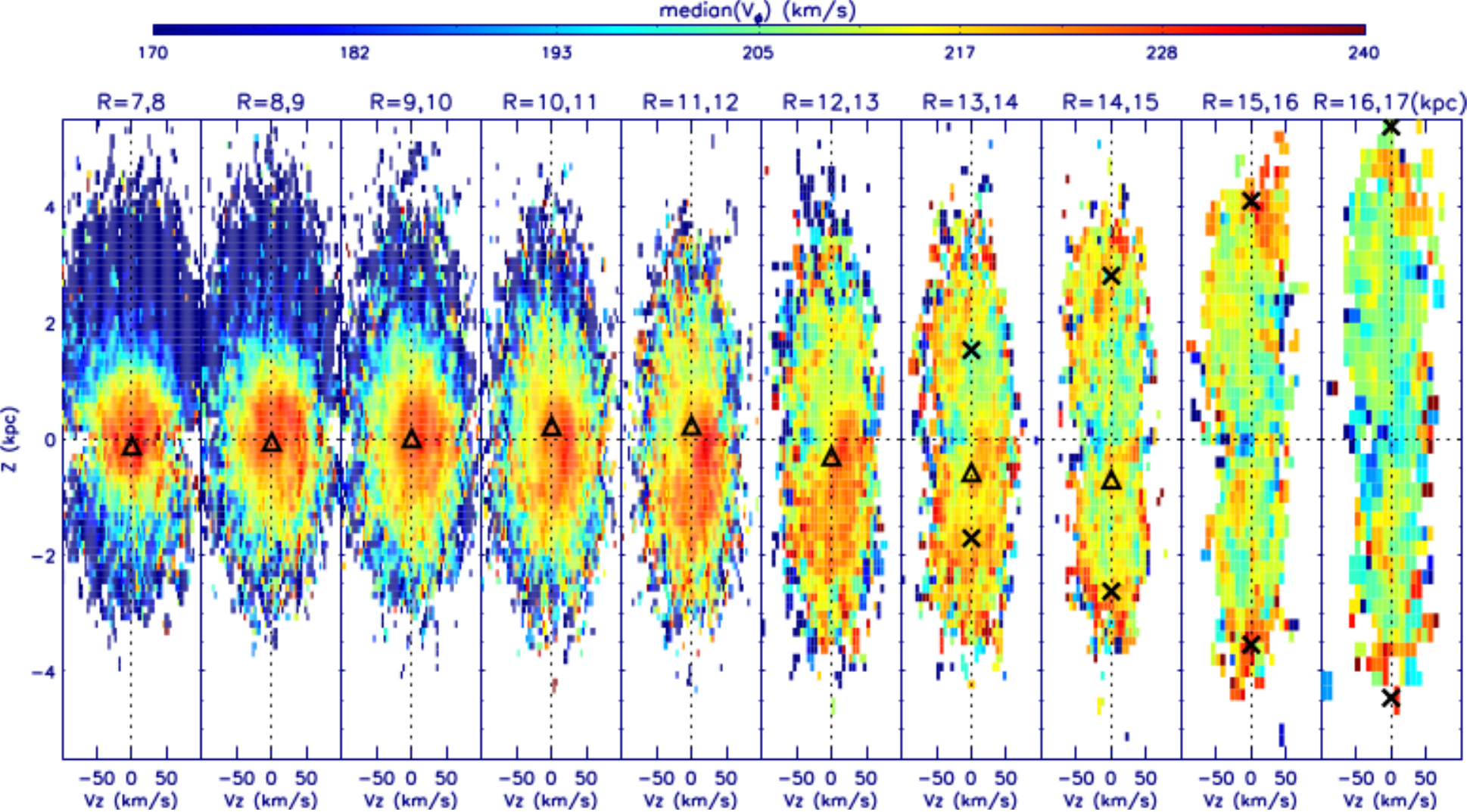}
    \caption{$V_\phi(V_Z, Z)$ as function of $R$. The panels are same as the panels of Figure~\ref{ZVz_Vphi}; they are rotated counterclockwise and ordered by increasing $R$. The locations of $Z$ of peak lines in each $R$ bin of the ``north branch" and the ``south branch," as seen in the left panel of Figure~\ref{Vphi_RZ}, are shown with crosses. The mean $Z$ value in each 1 kpc $R$ bin, at the location of the minimum standard deviation in $V_\phi$ as shown in Figure~\ref{waveonedisk}, is labeled with a triangle.}
    \label{ZVz_Vphi_vertical}
\end{figure*}

\begin{figure*}
  \includegraphics[width=9cm]{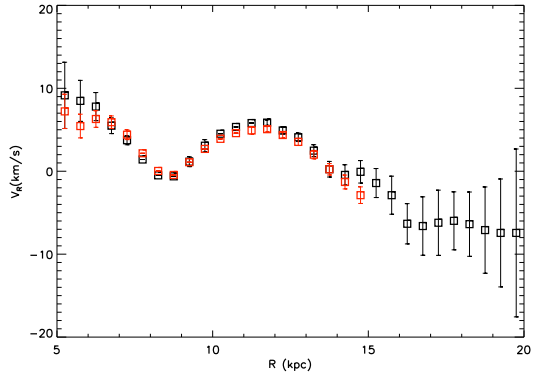}
	\caption{Radial variation of the median($V_R$). The black squares show radial variation of median($V_R$) within $|Z|<1$ kpc. The red squares show the radial variation of median($V_R$) within $|Z|<3$ kpc at $R<15$ kpc. From Figure~\ref{Vphi_RZ}, the median($V_R$) is seriously influenced by stars of the ``Monoceros area" after $R>15$ kpc, especially at higher $Z$, so the red squares are limited to $R<15$ kpc. The error bars show the standard deviation of the values of median($V_R$) for subsamples obtained by 1000 times bootstrap.}
    \label{Vr_R}
\end{figure*}

\begin{figure*}
    \includegraphics[width=19cm]{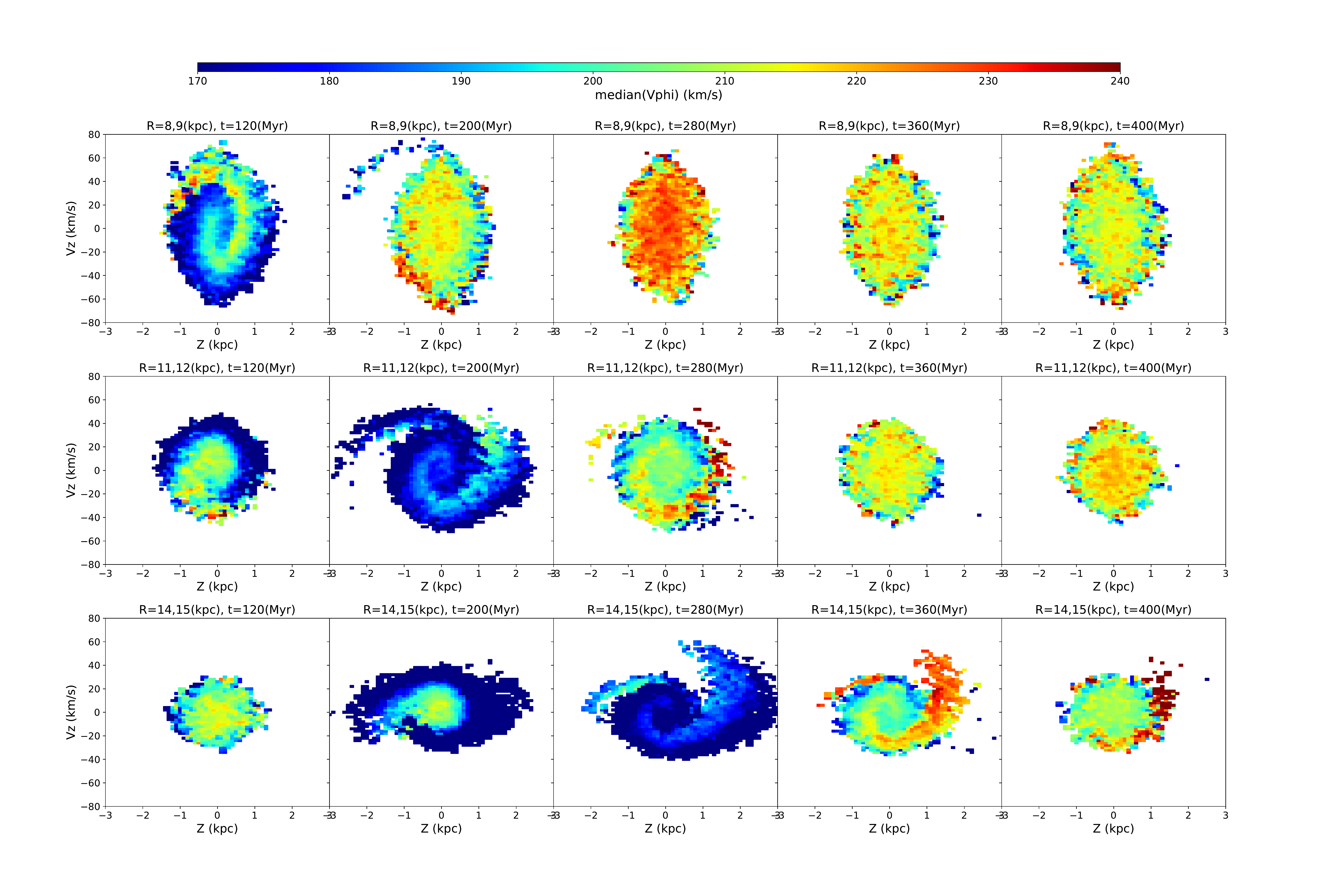}
    \caption{The $V_\phi$ distribution from the test particle simulation. Each panel shows the median($V_\phi$) distribution in $Z-V_z$ space for one time and one range of Galactocentric radius for one wedge of the test particle simulation with $315^\circ-20^\circ<\phi<315^\circ+20^\circ$. The top row shows data within $8<R<9$ kpc, the middle row shows $11<R<12$ kpc, and the bottom row shows $14<R<15$ kpc. From left to right the simulation times shown are $t=0.12$ Gyr, 0.2 Gyr, 0.28 Gyr, 0.36 Gyr, and 0.4 Gyr. This figure shows that the oscillation propagates outwards with increasing $R$. The phase spiral gradually moves from smaller $R$ to larger $R$ after the impact time.}
    \label{testpartical_ZVz_table}
\end{figure*}

\begin{figure*}
    \includegraphics[width=19cm]{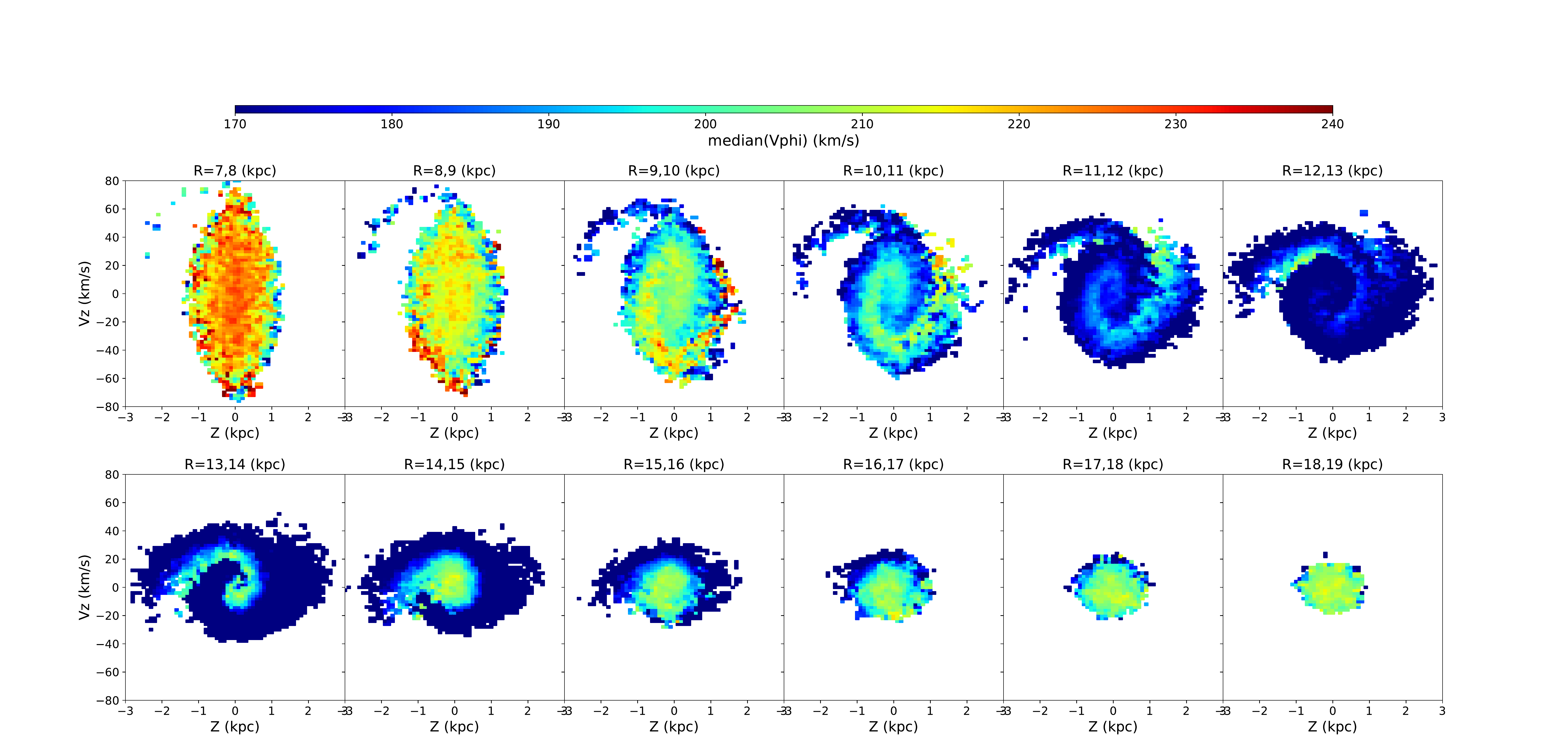}
    \caption{$V_\phi$ phase spirals at 210 Myr after impact and in the range $315^\circ-20^\circ<\phi<315^\circ+20^\circ$ over a full range of $R$. Note that the phase spirals appear in a large $R$ range.} 
    \label{testpartical_ZVz_Vphi0p29}
\end{figure*}

\begin{figure*}
    \includegraphics[width=19cm]{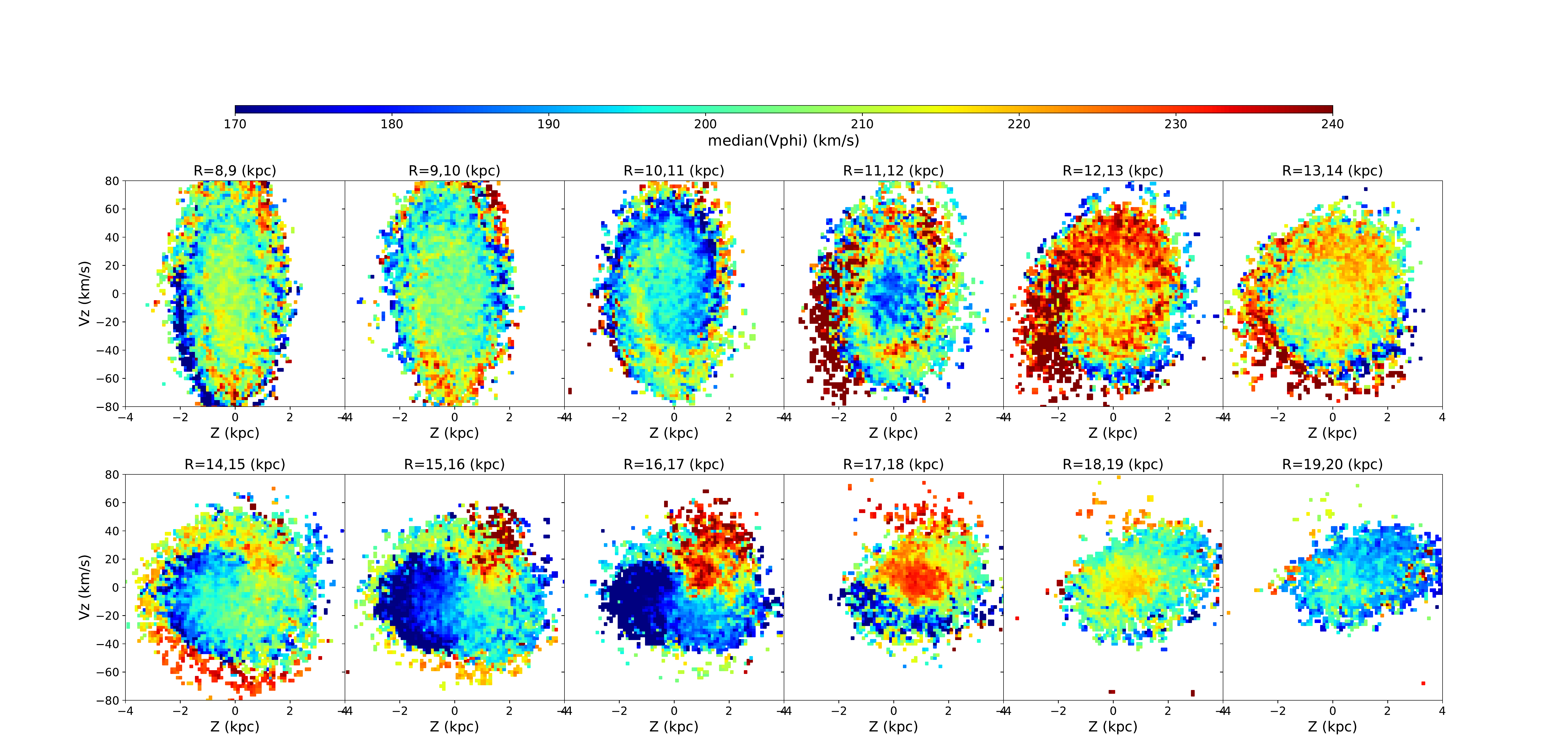}
    \caption{The result of N-body simulation \citep{2019MNRAS.485.3134L}. The $V_\phi$ distribution in $Z-V_Z$ phase space of stars in the range of $225^\circ-20^\circ<\phi<225^\circ-20^\circ$ at a time of 0.8 Gyr.}
    \label{Laporte_ZVzmap_Vphi}
\end{figure*}

\begin{figure*}
    \includegraphics[width=9cm]{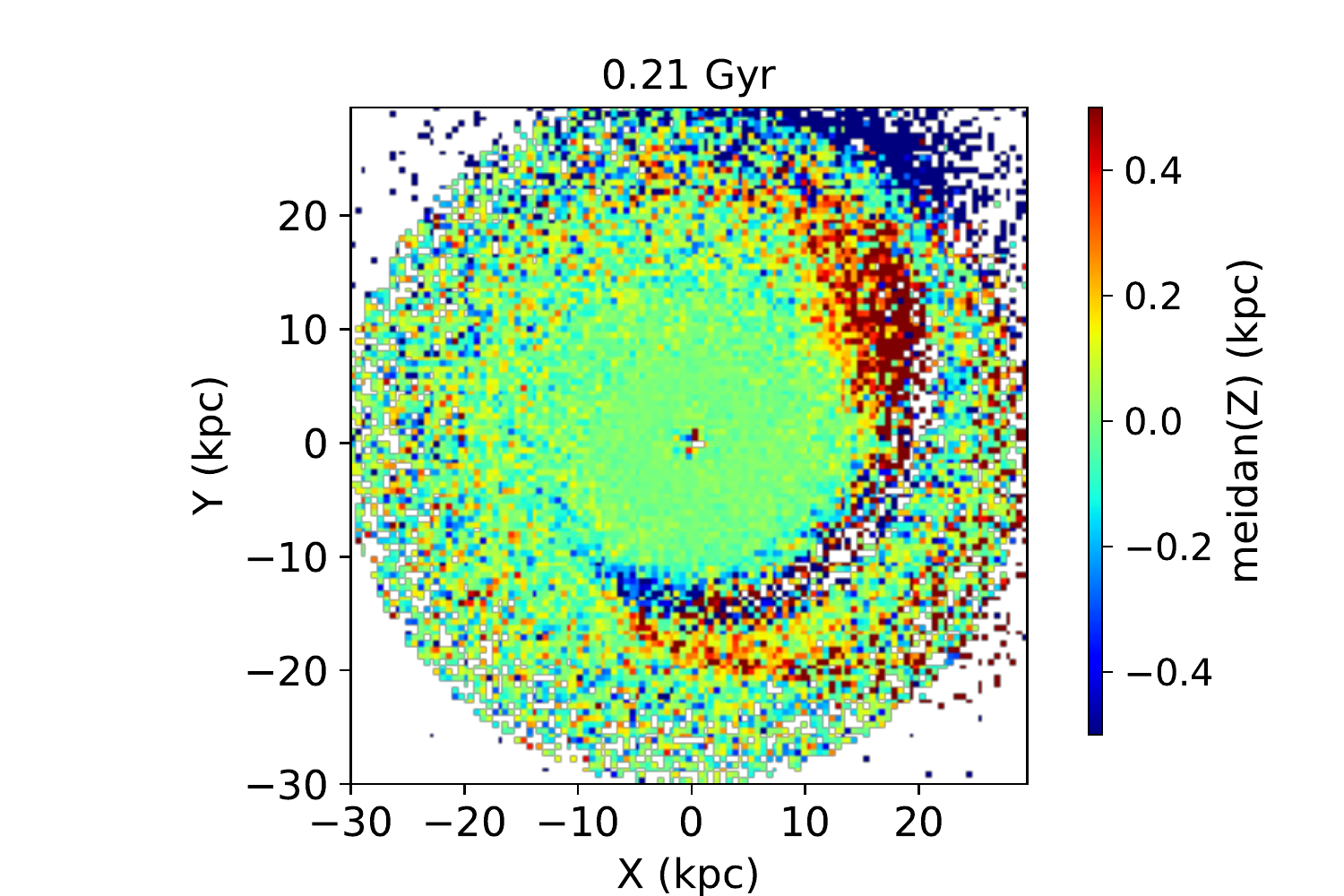}
    \caption{Median(Z) distribution in the $X-Y$ plane of test particle simulation when $t=0.21$ Gyr after the impact.}  
    \label{XYmap_z_testparticle}
\end{figure*}

\begin{figure*}
    \includegraphics[width=9cm]{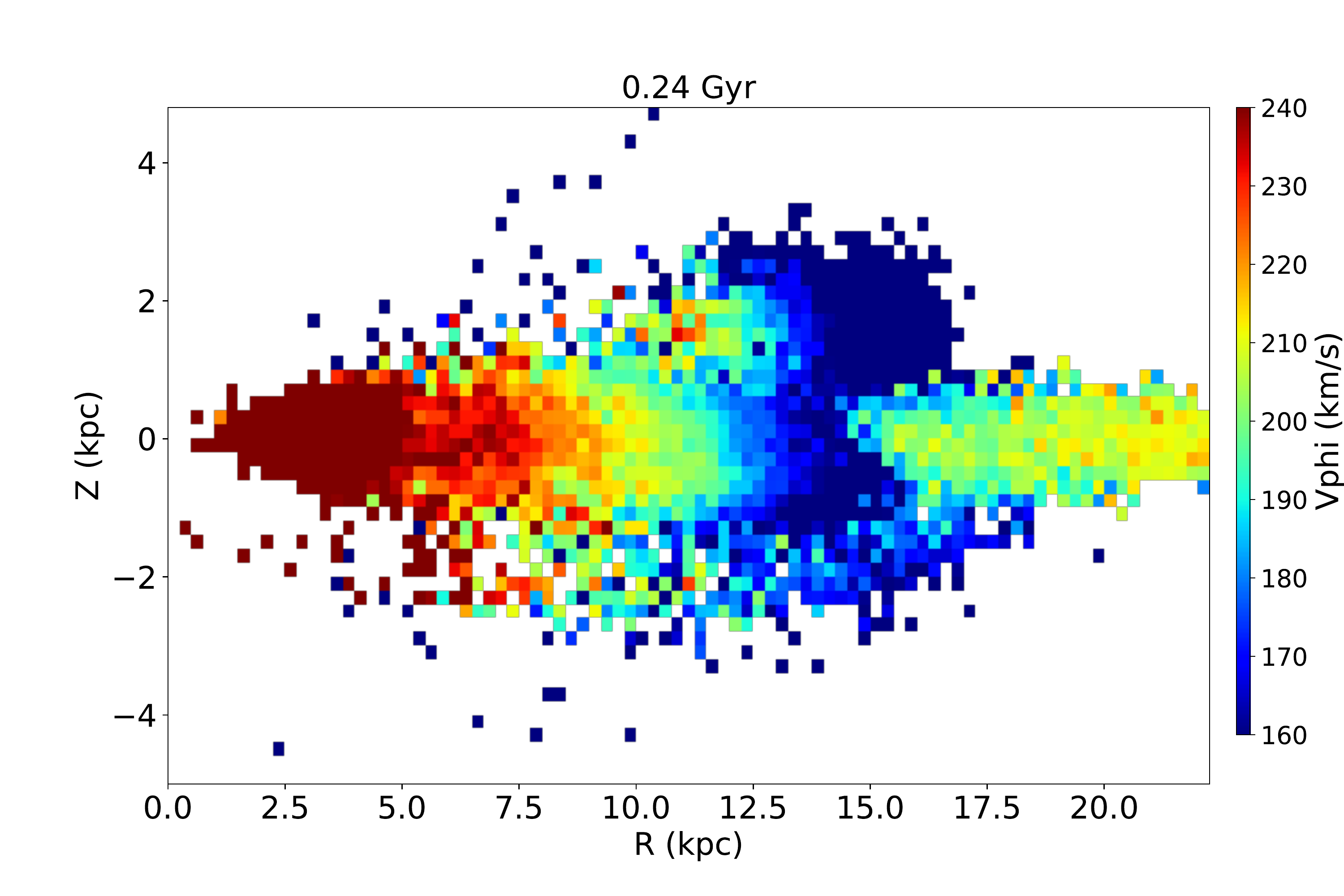}
    \includegraphics[width=9cm]{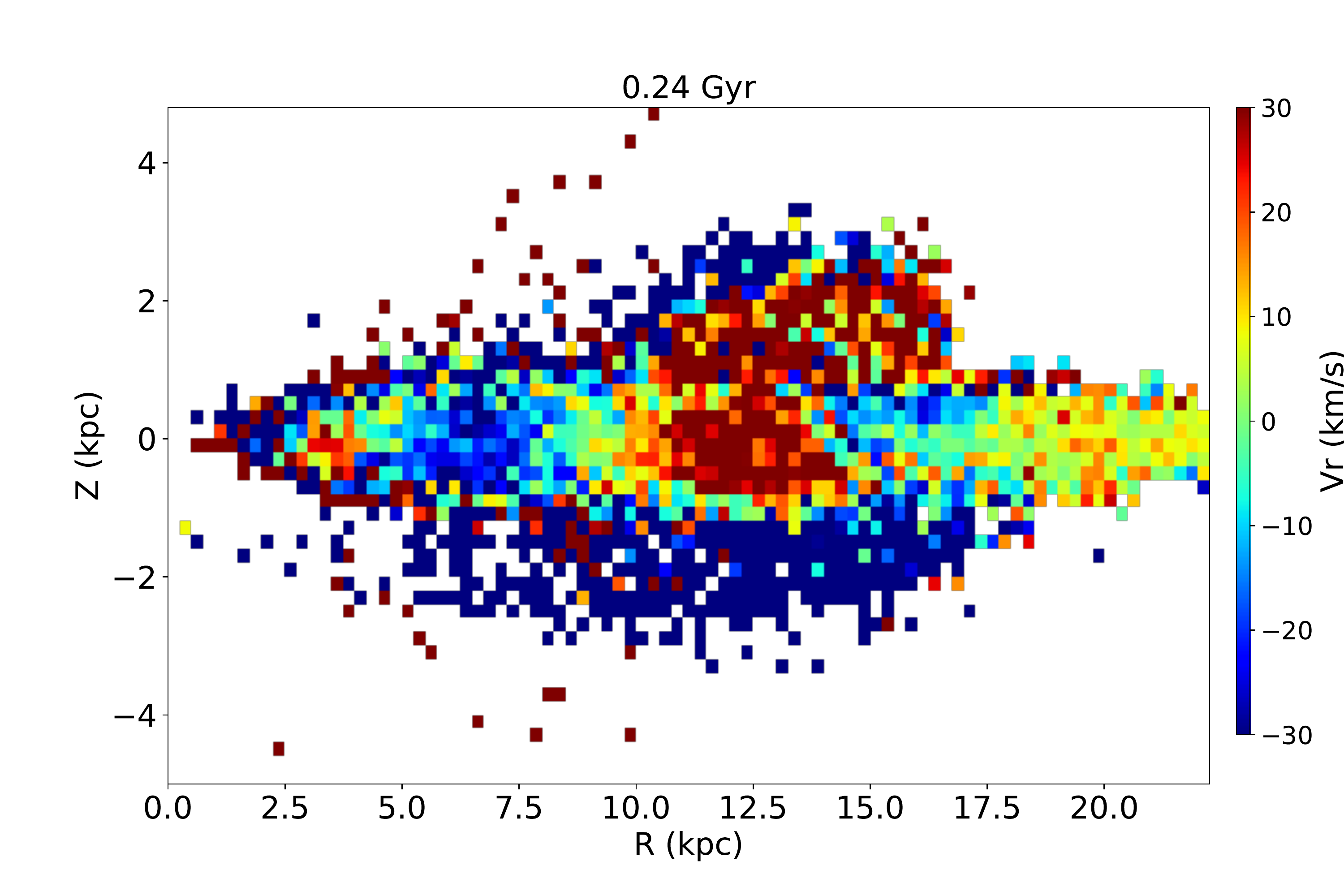}
    \includegraphics[width=9cm]{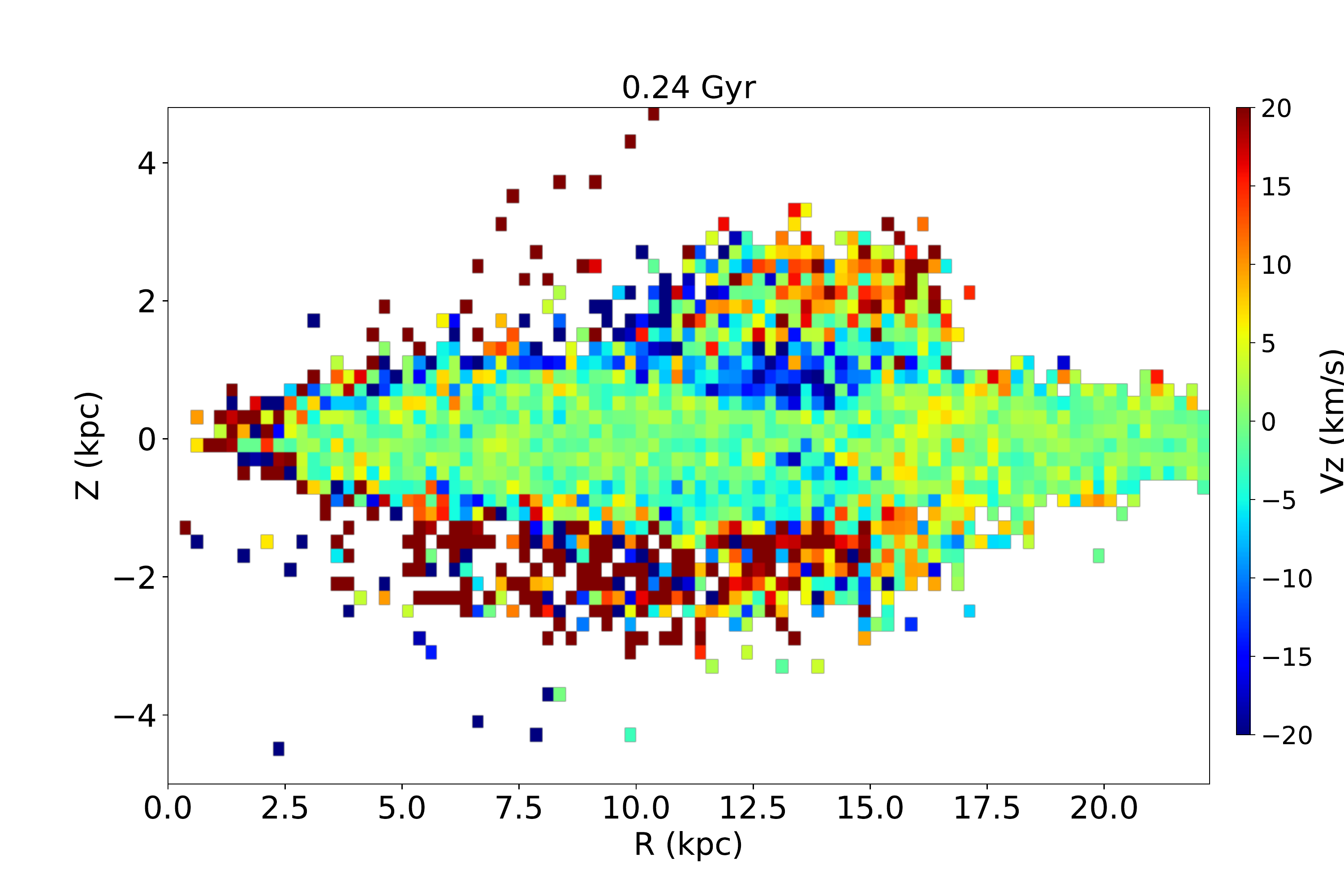}
    \includegraphics[width=9cm]{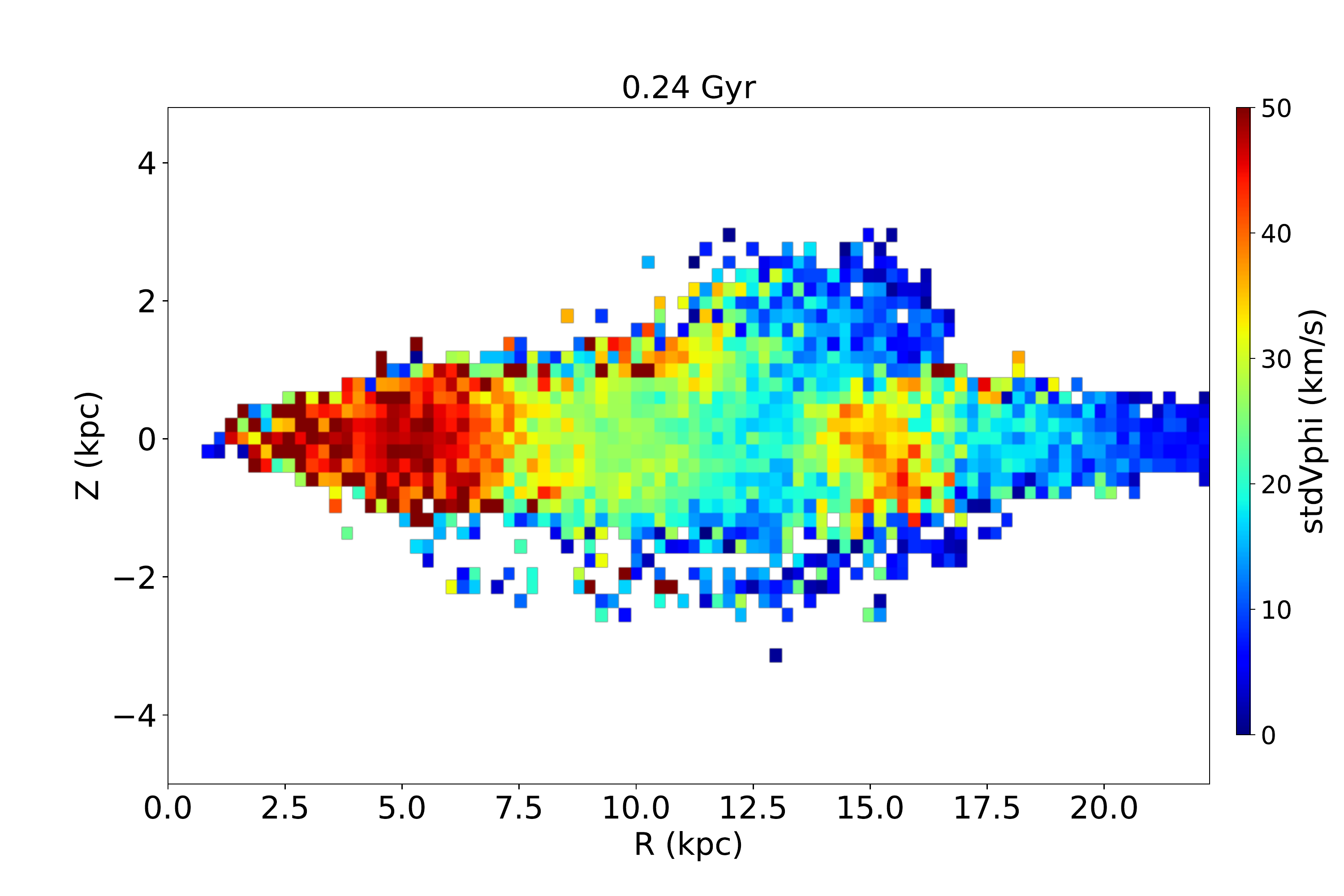}
            \caption{
            Sample median($V_\phi$), median($V_R$), median($V_Z$) and $\sigma$($V_\phi$) distributions in $R-Z$ space from the test particle simulation. This segment was selected at time $t=0.24$ Gyr in the simulation, and includes particles in the range of $315^\circ-20^\circ<\phi<315^\circ+20^\circ$. }  
    \label{RZmap_Vphi_0p36_testparticle}
\end{figure*}

\begin{figure*}
    \includegraphics[width=9cm]{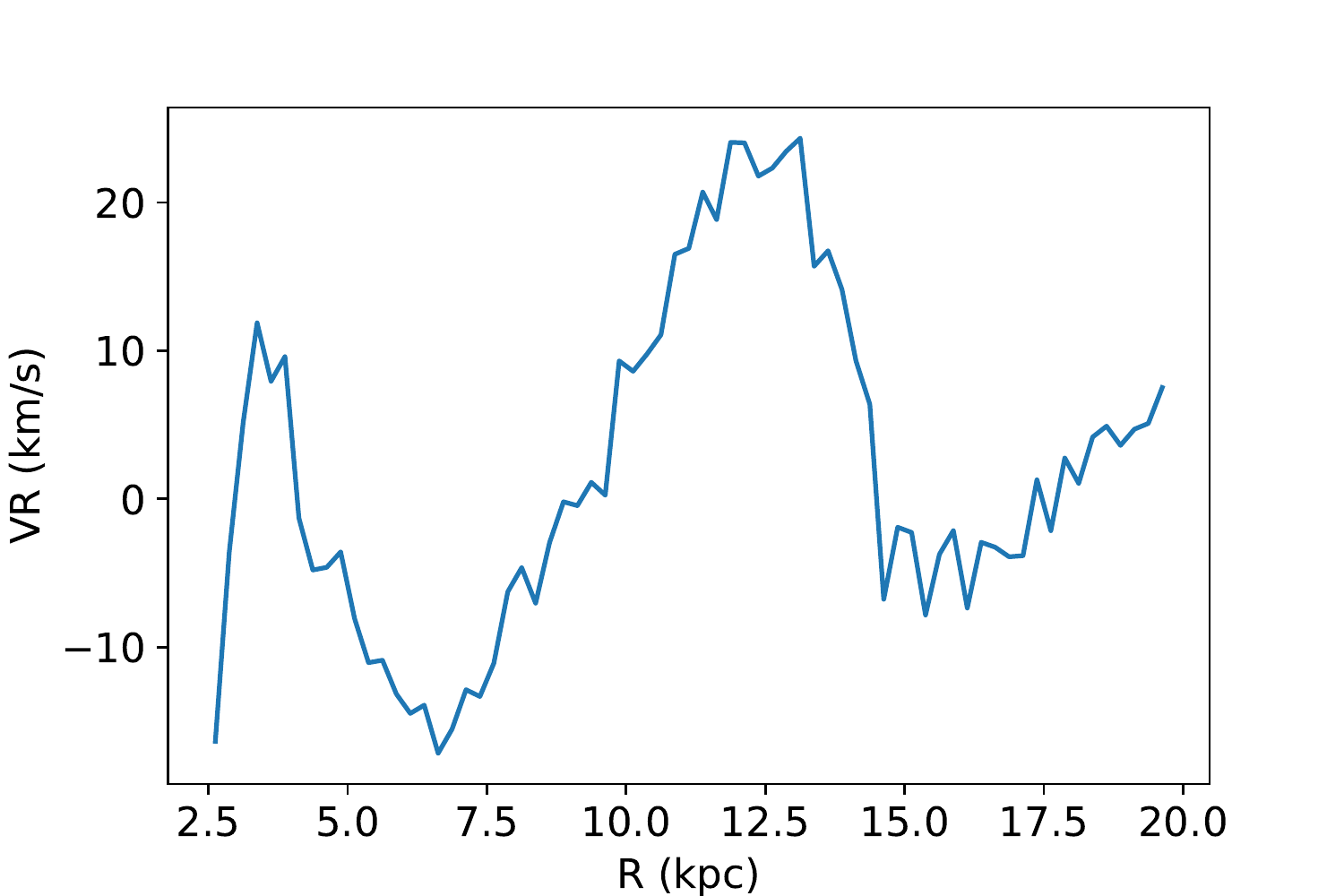}
             \caption{Ripple in median($V_R$) as a function of $R$ within $|Z|<1$ kpc and azimulthal angle $315^\circ-20^\circ<\phi<315^\circ+20^\circ$ for the test particle simulation when t=0.24 Gyr. This variation is comparable to the ripple in the median($V_R$) in the data, as shown in Figure~\ref{Vr_R}.}
    \label{Vr_R_simulation}
\end{figure*}

\begin{figure*}
     \includegraphics[width=7cm]{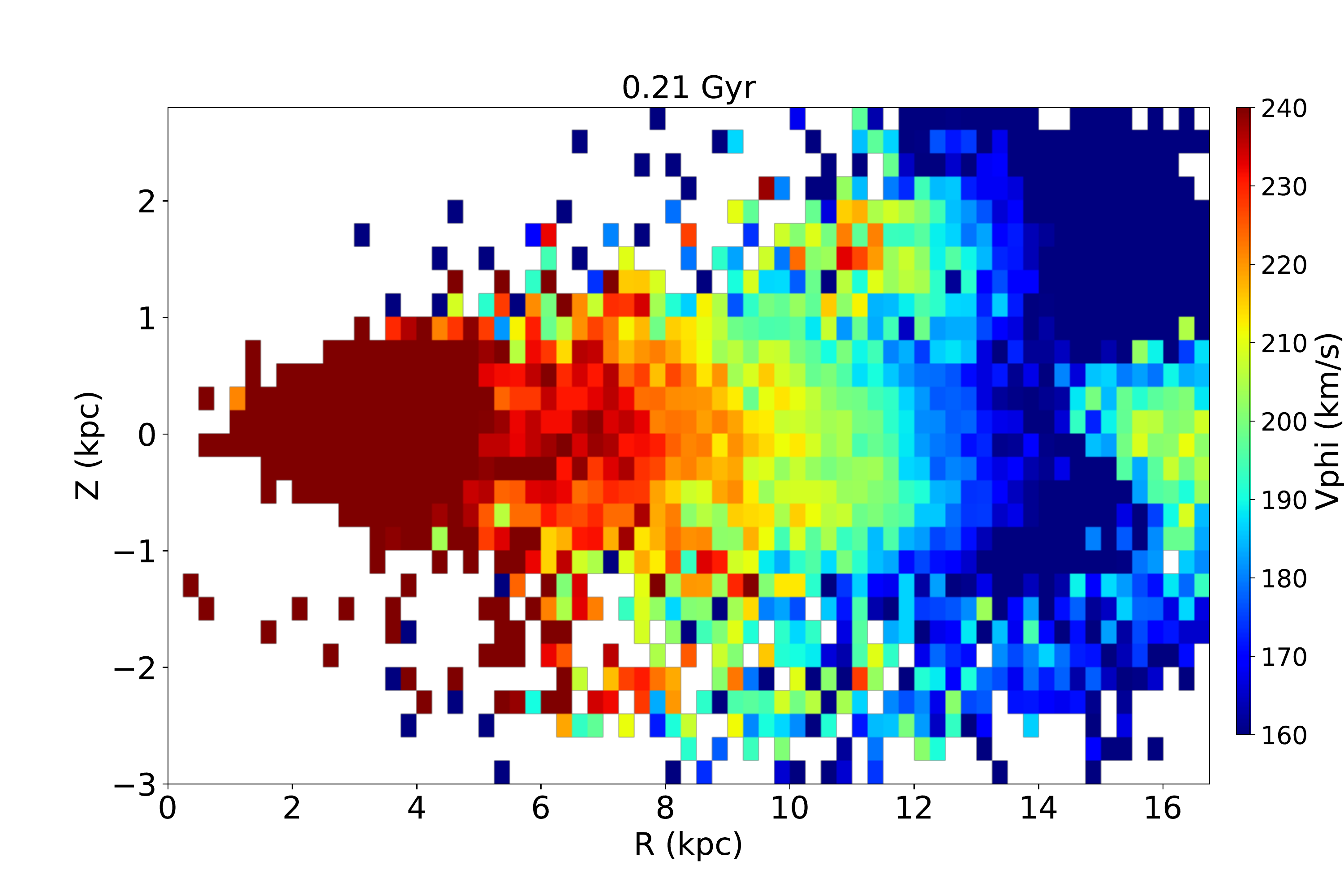}\\
      \includegraphics[width=25cm]{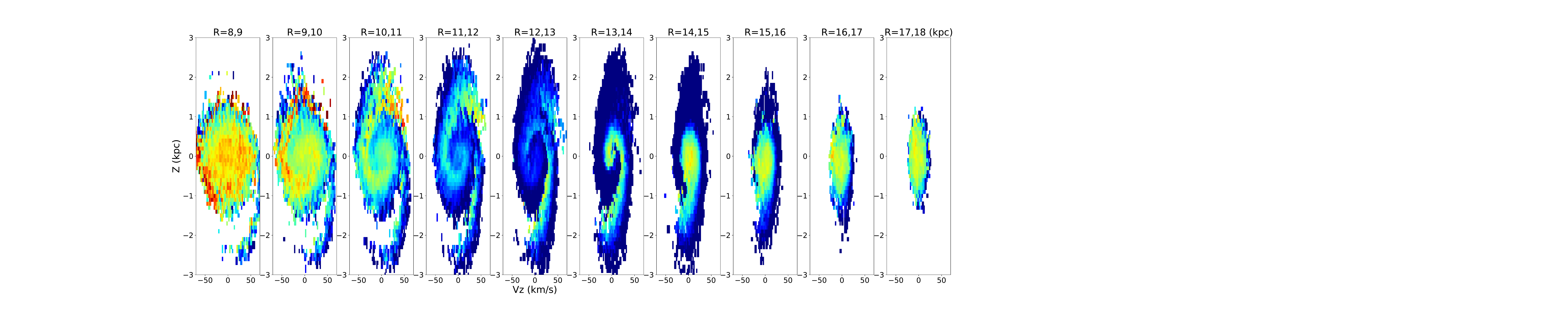}
    \caption{The upper panel shows that the median($V_\phi$) distribution in $R-Z$ space in the range of $315^\circ-20^\circ<\phi<315^\circ+20^\circ$ when $t=0.21$ Gyr after the impact. The lower panel shows the $V_\phi$ phase spiral in Z-$V_Z$ space of each $R$ bin. The phase spiral map is rotated unclockwise and lined up in the sequence of increasing $R$.}
    \label{ZVzmap_vertical_Vphi_0p22_315}
\end{figure*}

We validate the {\it Gaia} proper motions further than $R=20$ kpc, where $R$ is the distance from the Galactic center. The distance and radial velocities of the LAMOST K giants are validated to similar Galactocentric distance. The combination of {\it Gaia} proper motions, LAMOST radial velocities, and distances calculated from LAMOST spectra allow us to study the velocity field within the range $6<R<20$ kpc.  
Because LAMOST K giants are mostly observed within $\phi=\pm20^\circ$ of the anticenter direction, it is most convenient to project velocities in $R-Z$ space, while slicing the data in $R$. We study the velocity distribution in the $R-Z$ plane, and compare it with projections in $Z-V_Z$ phase space. From the data we learn:

1)There are plenty of features in the map of the $V_\phi$ distribution in the $R-Z$ plane. The main characteristics are: (i) The mid-plane as determined by the location of the minimum standard deviation $V_\phi$ is oscillating in the $Z$ direction. (ii) The outer disk bends southward. (iii) The oscillation in the centroid of the phase spiral traces the large scale bulk motion of the disc (see Figure~\ref{ZVz_Vphi_vertical}).  (iv) The high $V_\phi$ locus is trifurcated after $R=13$ kpc.

2)The $V_\phi$ Spiral pattens in the $Z-V_Z$ plane extend from $R=7$ to 15 kpc. Spirals are more tightly wound at smaller $R$ and more loosely wound at larger $R$. In addition, spirals more elongated along $Z$ and are more squashed along $V_Z$ with increasing $R$. The simulations predict the same trends due to decreasing of self gravity and longer dynamical time scale with increasing $R$ \citep{ 2019MNRAS.485.3134L,2019MNRAS.486.1167B}. In the observational data, after $R=15$ kpc, there are no complete spirals and the $Z-V_Z$ phase space is full of ``hatched chunk." Similar to the prediction of \citet{2019MNRAS.485.3134L}, the imprint of multiple passages of the dwarf galaxy through the disk are present in the stars of the outer disk. 

3)The most significant bulk motion is consistent with the 1D projection of the $V_\phi$ phase space spiral, integrated along $V_Z$ in each radial bin.

4) The observed flaring of the disk is substantial. The flare in the outer disk is even thicker than the thick disk in the Solar neighborhood. From Fig~\ref{ZVz_Vphi_vertical}, the boundary of the flare is defined by the extension along $Z$ of the phase space spiral. Since the Sgr dSph was found to induce the phase space spiral, this suggests that  the recent impact of the Sgr dSph also contributes to the flare.

We can't rule out other possible mechanisms(such as \citet{2002A&A...394...89N,2012A&A...548A.127M}) to produce a flare except the explanation of Saggitarius passage.
 
5) The test particle simulation and N-body simulation can reproduce the above observational characteristics qualitatively, including the phase spiral and bulk motions.

In summary, we observed the one-to-one relationship between the phase spiral and the most significant bulk motion of the disk stars. Because the most likely origin of the phase spiral is the last impact of the Sgr dSph, the bulk motion of the disk stars is likely  the result of the same impact.

\acknowledgments
This work is supported by National Key R\&D Program of 
China No. 2019YFA0405500.
C.L. thanks the National Natural Science Foundation of China (NSFC) with grant No. 11835057.
H.F.W. is supported by the LAMOST Fellow project, National Key Basic R\&D Program of China via 2019YFA0405500 and funded by China Postdoctoral Science Foundation via grant 2019M653504 and 2020T130563, Yunnan province postdoctoral Directed culture Foundation, and the Cultivation Project for LAMOST Scientific Payoff and Research Achievement of CAMS-CAS and the National Natural Science Fondation of China grant number 12003027. HJN is supported by the US National Science Foundation grant AST 19-08653. X.F. acknowledges the support of China Postdoctoral Science Foundation No. 2020M670023, the National Natural Science Foundation of China under grant number 11973001, and the National Key R\&D Program of China No. 2019YFA0405504. 
This work was supported in part by World Premier International Research Center Initiative (WPI Initiative), MEXT, Japan. 
  
Guoshoujing Telescope (the Large Sky Area Multi-Object Fiber Spectroscopic Telescope 
LAMOST) is a National Major Scientific Project built by the Chinese Academy of Sciences.
 Funding for the project has been provided by the National Development and Reform 
 Commission. LAMOST is operated and managed by the National Astronomical Observatories, 
 Chinese Academy of Sciences.
 
This work has made use of data from the European Space Agency (ESA)
mission {\it Gaia} (\url{https://www.cosmos.esa.int/gaia}), processed by
the {\it Gaia} Data Processing and Analysis Consortium (DPAC,
\url{https://www.cosmos.esa.int/web/gaia/dpac/consortium}). Funding
for the DPAC has been provided by national institutions, in particular
the institutions participating in the {\it Gaia} Multilateral Agreement. 

This work benefited from the International Space Science Institute (ISSI/ISSI-BJ) in Bern and Bei- jing, thanks to the funding of the team Chemical abundances in the ISM: the litmus test of stellar IMF variations in galaxies across cosmic time (PI D. Romano and Z-Y. Zhang).
 
\clearpage

\setcounter{figure}{0}    
\renewcommand{\thefigure}{A\arabic{figure}}
 \appendix
  

 
\section{Detailed results of the test particle simulation}
\begin{figure}
    \includegraphics[width=15cm]{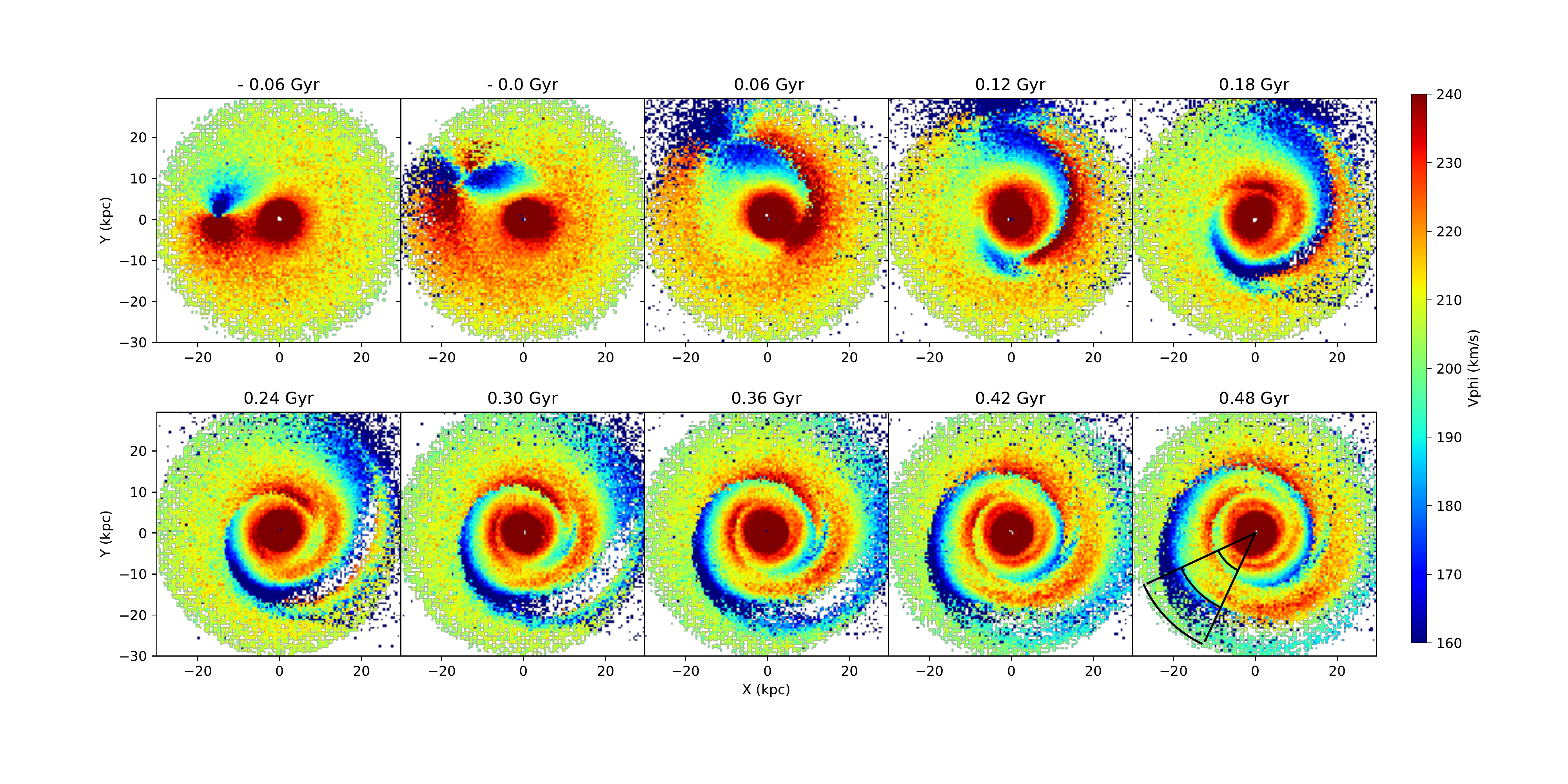}
    \caption{The median($V_\phi$) distribution on the $X-Y$ plane in a series of snapshots from the test particle simulation. The time after the impact is labeled on top of each panel. The wedge in the lower right panel shows the direction of $315^\circ-20^\circ<\phi<315^\circ+20^\circ$. Arcs at 10 kpc, 20 kpc, and 30 kpc from the Galactic center are shown in the wedge.}
    \label{testpartical_XY_Vphi}
\end{figure}

\begin{figure*}
    \includegraphics[width=15cm]{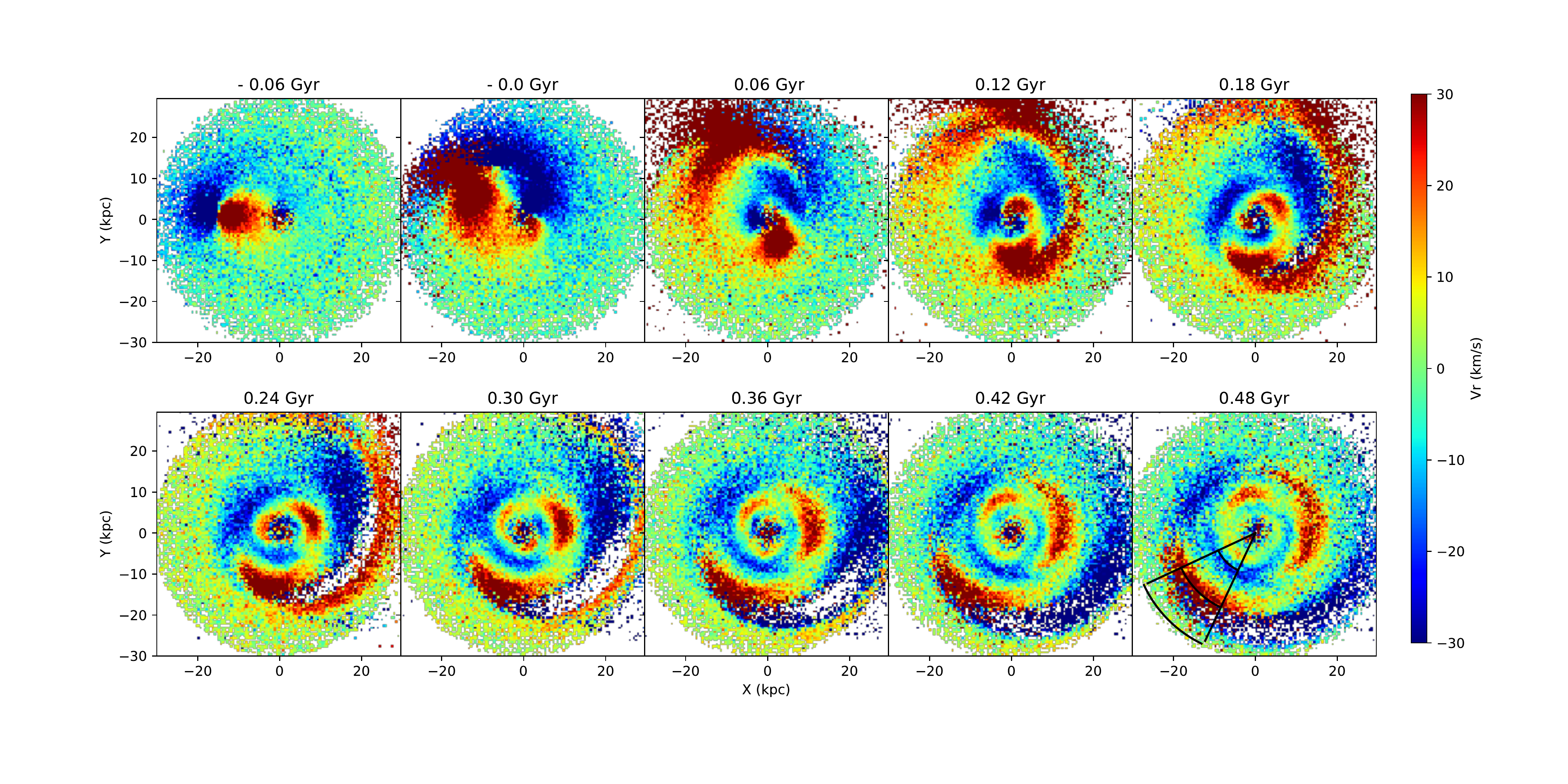}
    \caption{The $V_r$ distribution on the $X-Y$ plane in the test particle simulation. The time after the impact is labeled on top of each panel. The wedge in the lower right panel shows the direction of $315^\circ-20^\circ<\phi<315^\circ+20^\circ$. Arcs at 10 kpc, 20 kpc, and 30 kpc from the Galactic center are shown in the wedge.}
    \label{testpartical_XY_Vr}
\end{figure*}

\begin{figure*}
    \includegraphics[width=15cm]{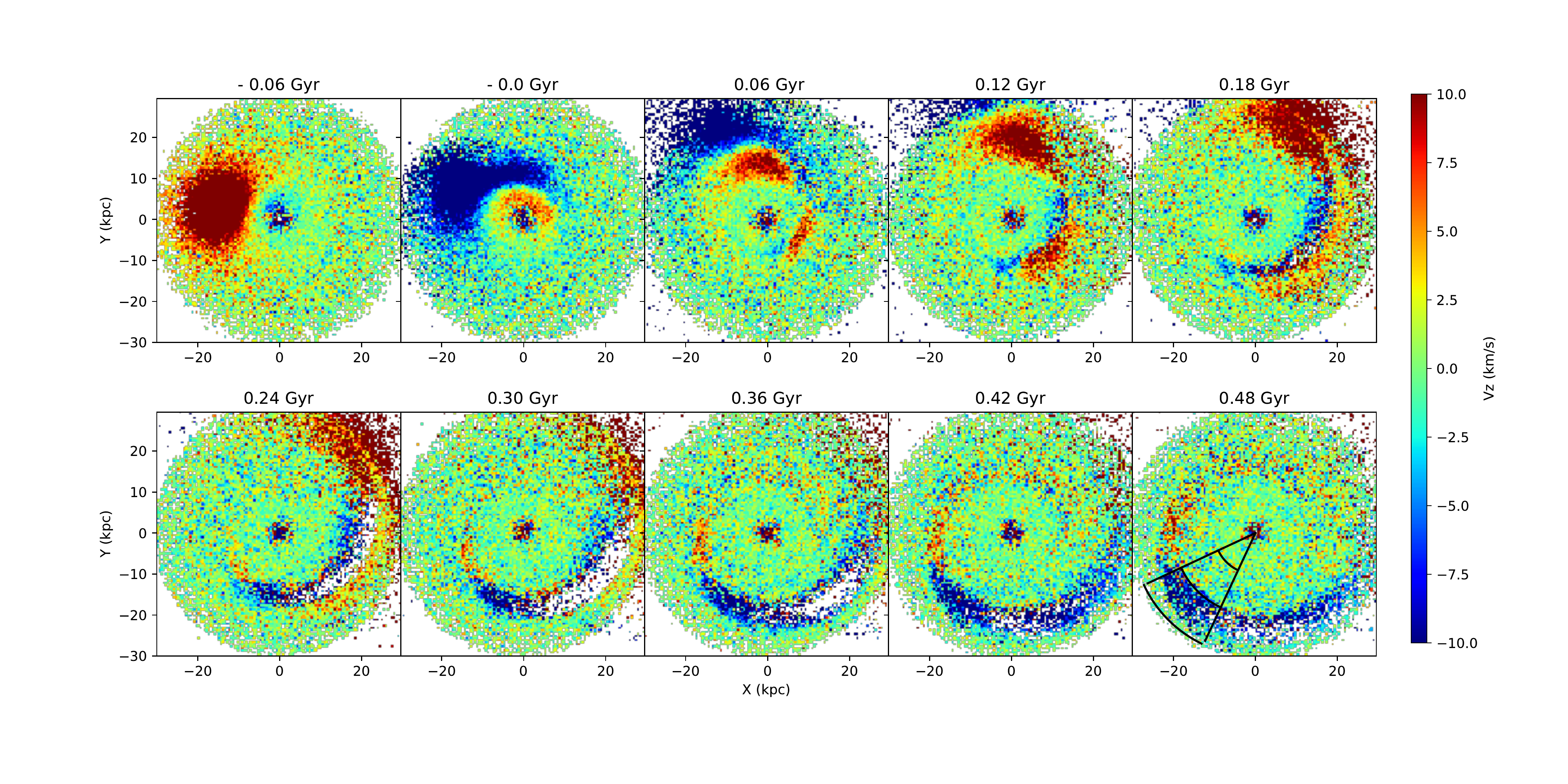}
    \caption{The median($V_Z$) distribution on the $X-Y$ plane in the test particle simulation. The time after the impact is labeled on top of each panel. The wedge in the lower right panel shows the direction of $315^\circ-20^\circ<\phi<315^\circ+20^\circ$. Arcs at 10 kpc, 20 kpc, and 30 kpc from the Galactic center are shown in the wedge.}
    \label{testpartical_XY_Vz}
\end{figure*}

The Figure~\ref{testpartical_XY_Vphi}, ~\ref{testpartical_XY_Vr}, and ~\ref{testpartical_XY_Vz} show the median($V_\phi$), median($V_R$), and median($V_Z$) distributions, respectively, of the simulation results in $X-Y$ plane. The Sun is located at $(X,Y)=(-8,0)$ kpc. The disk stars rotate clockwise. The azimuth angle ($\phi$) is defined so that $\phi=0$ in the direction from the Galactic center to the Sun and $\phi$ increases in the clockwise direction.
The toy model is simplistic, but allows us to study the effect of the impact itself, without complication from other physical effects. 

The motion of stars on the $X-Y$ plane are described by median($V_\phi$) and median($V_R$). From Figures~\ref{testpartical_XY_Vphi} and ~\ref{testpartical_XY_Vr}, we see that
the disk feels the disturbance right after the impact begins.  
The stars are accelerated towards the intruder; stars moving towards the intruder develop a faster speed and stars moving away from the intruder are slowed. Inside the radius of the position of impact, the median$(V_R)>0$. Outside that radius, the median$(V_R)<0$. 
  After the impact, the high median($V_\phi$) accelerated stars and the low median($V_\phi$) of decelerated stars are dragged into rings. The high median($V_\phi$) stars overtake the low median($V_\phi$) stars at about 0.02 Gyr. Then the high median($V_\phi$) stars are separated into two parts with median$(V_R)>0$ and median$(V_R)<0$. The part with median$(V_R)<0$ moves inward and rolls up, and the part with median$(V_R)>0$ becomes a ring at about 0.1 Gyr.  The low median$(V_\phi)$ forms a ring in between two high median$(V_\phi)$ rings at about 0.18 Gyr. The rings generally propagate outwards.   

The distribution of  median($V_Z$)  on the $X-Y$ plane of Figure~\ref{testpartical_XY_Vz} shows us that the stars around the point of impact have positive median$(V_Z)$  when the intruder is on the north side of the disk. The median$(V_Z)$  of stars around the point of impact is negative while the intruder is on the south side of the disk. The stars that gain an extra $V_Z$ deviate from their original planar orbit to oscillate above and below the disk. The influenced stars are then dragged to rings as the disk rotates, and the rings propagate outwards.

\begin{figure*}
    \includegraphics[width=15cm]{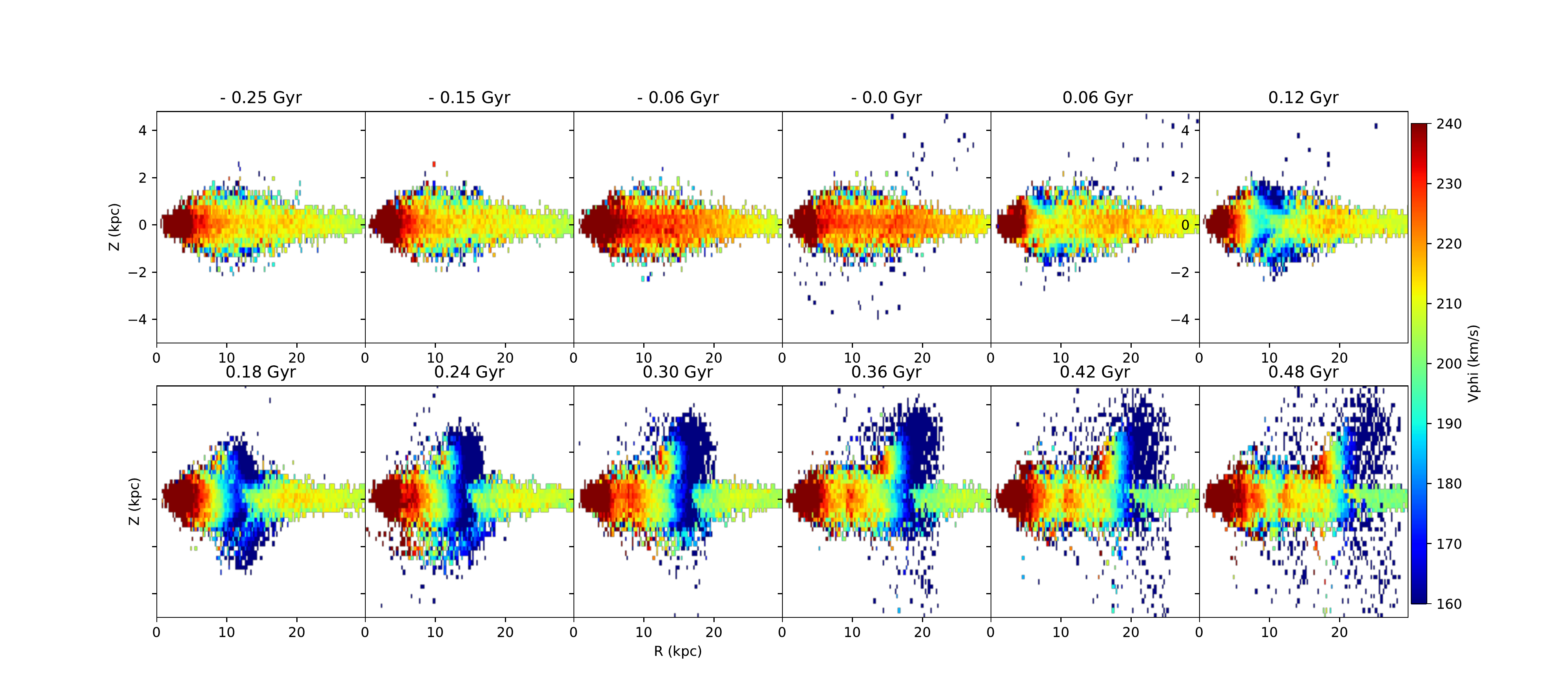}
    \caption{The median($V_\phi$) distribution of data with azimuthal angle $315^\circ-20^\circ<\phi<315^\circ+20^\circ$ in the $R-Z$ plane in the test particle simulation. The time after impact is labeled on top of each panel. }
    \label{testpartical_RZ_Vphi}
\end{figure*}

\begin{figure*}
    \includegraphics[width=15cm]{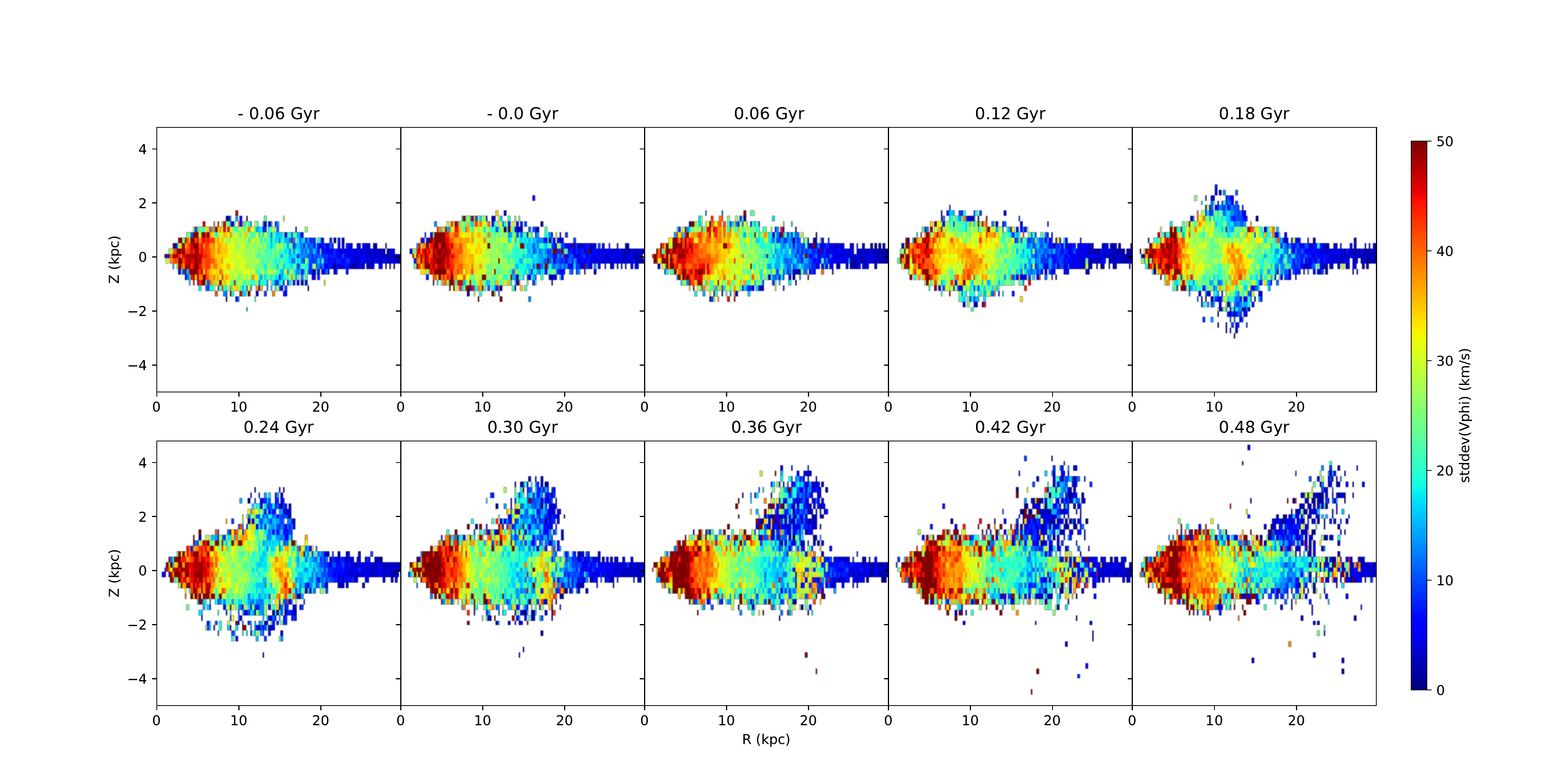}
    \caption{The standard deviation of the median($V_\phi$) distribution of data in the range of $315^\circ-20^\circ<\phi<315^\circ+20^\circ$ in the $R-Z$ plane in the test particle simulation. The time after impact is labeled on top of each panel. }
    \label{testpartical_RZ_stdVphi}
\end{figure*}

\begin{figure*}
    \includegraphics[width=15cm]{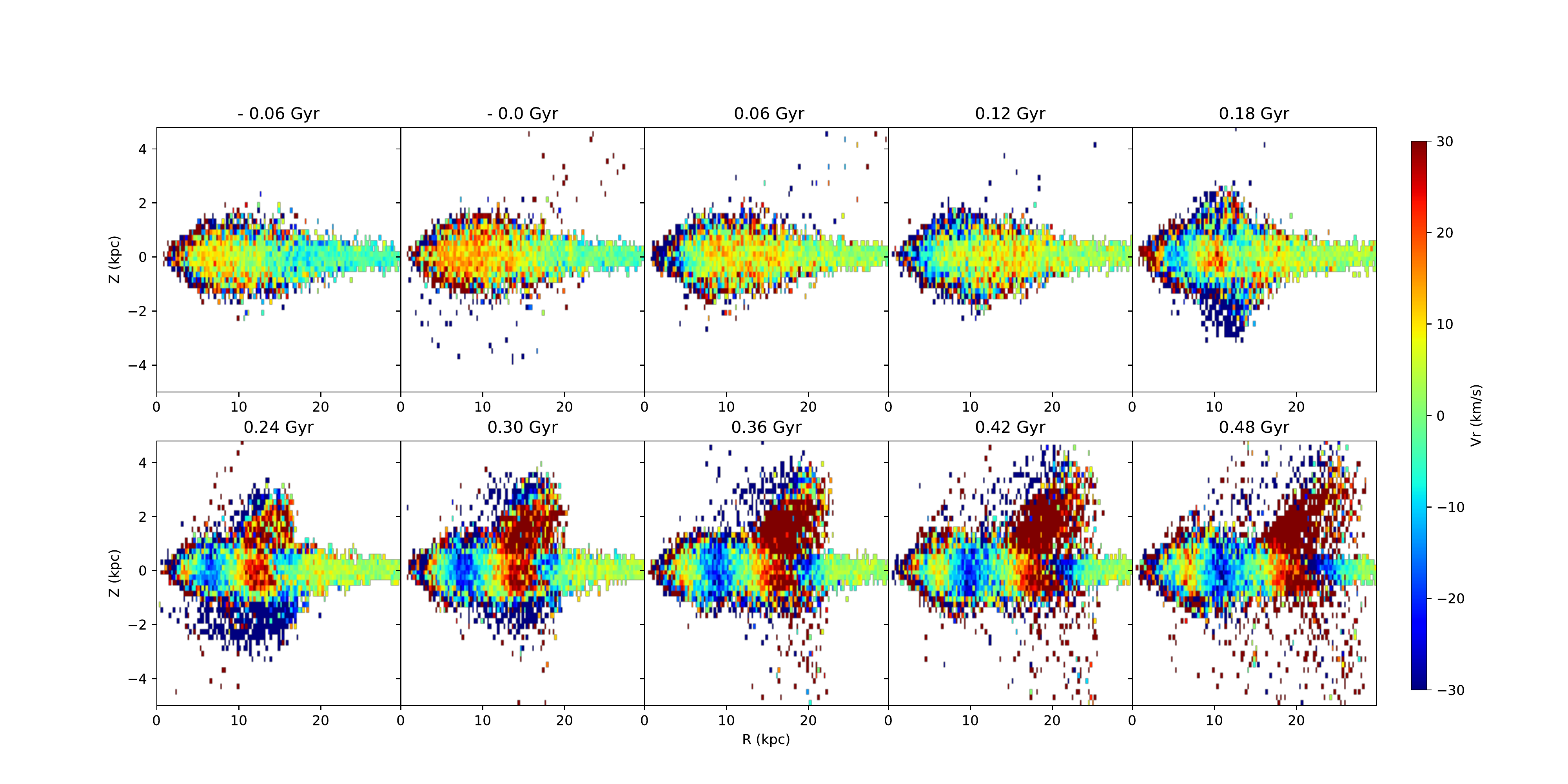}
    \caption{The median($V_r$) distribution of data in the range of $315^\circ-20^\circ<\phi<315^\circ+20^\circ$ on the $R-Z$ plane in the test particle simulation. The time after impact is labeled on top of each panel. }
    \label{testpartical_RZ_Vr}
\end{figure*}

\begin{figure*}
    \includegraphics[width=15cm]{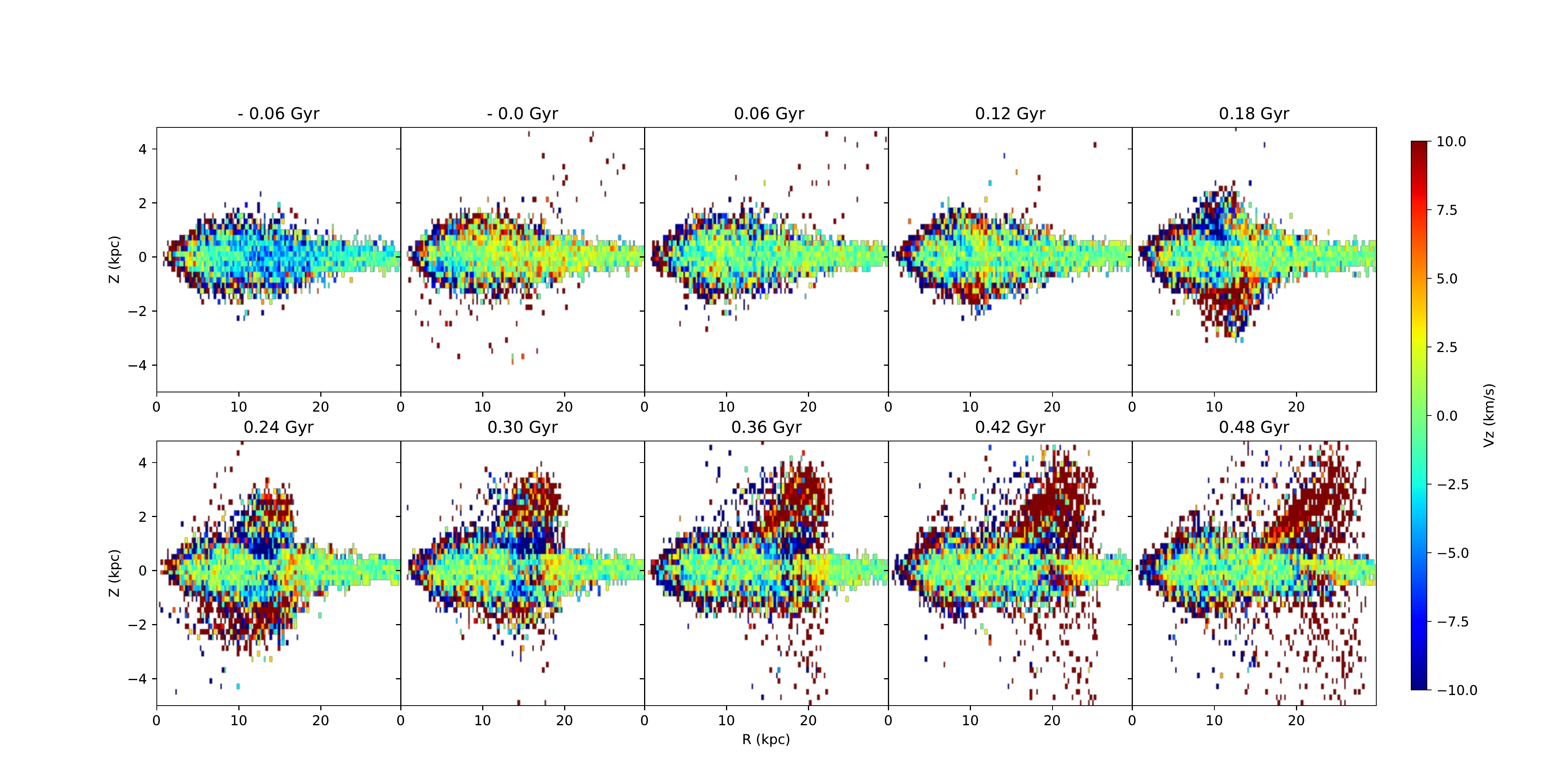}
    \caption{The median($V_z$) distribution of data in the range of $315^\circ-20^\circ<\phi<315^\circ+20^\circ$ on the $R-Z$ plane in the test particle simulation. The time after impact is labeled on top of each panel. }
    \label{testpartical_RZ_Vz}
\end{figure*}

Figures~\ref{testpartical_XY_Vphi}, ~\ref{testpartical_XY_Vr}, and ~\ref{testpartical_XY_Vz} show that the entire disk of stars is oscillating after several hundred million years. We observed snapshots in the direction of $\phi=0^\circ, 180^\circ, 45^\circ,135^\circ, 225^\circ,$ and $315^\circ$.  The stars in the wedge of $315^\circ-20^\circ<\phi<315^\circ+20^\circ$ were selected to show the detailed of kinematic features in $R-Z$ space (Figure~\ref{testpartical_RZ_Vphi}, ~\ref{testpartical_RZ_stdVphi},  ~\ref{testpartical_RZ_Vr}, and ~\ref{testpartical_RZ_Vz}). This aximuthal angle was selected because the phase space spiral in this direction, several hundred million years after the impact, is most similar to the observed one. From the snapshots, we see the evolution of kinematic substructures induced by the impact of the passing dwarf galaxy. The position of the wedge is shown in the last panel of Figures~\ref{testpartical_XY_Vphi}, ~\ref{testpartical_XY_Vr}, and ~\ref{testpartical_XY_Vz}. 

\end{document}